\documentclass[amsmath, amssymb, floatfix, reprint, twocolumn, notitlepage, 10pt]{revtex4-1}
\usepackage[T1]{fontenc}
\usepackage[utf8]{inputenc}
\usepackage{microtype,bm,bbm,graphicx,booktabs,times}
\usepackage[usenames,dvipsnames]{xcolor}
\usepackage[normalem]{ulem}
\usepackage{setspace}
\usepackage{tikz-cd}

\usepackage{hyperref}
\usepackage{subfiles} 
\hypersetup{colorlinks,allcolors=NavyBlue,linktoc=all}
\usepackage[capitalize]{cleveref}
\crefformat{section}{Sec.~#2#1#3}
\usepackage{tikz}
\usetikzlibrary{shapes}
\usepackage{amsmath}
\usepackage{amsfonts}
\usepackage{amssymb}
\usepackage{amsthm}

\newcommand{\change}{}
\newcommand{\changetwo}{}

\newtheorem*{oproblem}{Open problem}
\newcommand{\Tot}{\operatorname{Tot}}

\newcommand{\tr}{\operatorname{tr}}
\newcommand{\poly}{\operatorname{poly}}
\newcommand{\polylog}{\operatorname{polylog}}
\newcommand{\sys}{\operatorname{sys}}
\newcommand{\vol}{\operatorname{vol}}
\newcommand{\area}{\operatorname{area}}
\newcommand{\rk}{\operatorname{rk}}
\newcommand{\im}{\operatorname{im}}
\newcommand{\id}{\operatorname{id}}

\begin{document}
\title{Quantum LDPC Codes}
\author{Nikolas P. Breuckmann}
\email{n.breuckmann@ucl.ac.uk}
\affiliation{Department of Computer Science, University College London, WC1E 6BT London, United Kingdom}
\author{Jens Niklas Eberhardt}
\email{mail@jenseberhardt.com}
\affiliation{Mathematical Institute, University of Bonn, Germany}
\date{\today}

\begin{abstract}
Quantum error correction is an indispensable ingredient for scalable quantum computing.
In this Perspective we discuss a particular class of quantum codes called \emph{quantum low-density parity-check (LDPC) codes}.
The codes we discuss are alternatives to the surface code, which is the currently leading candidate to implement quantum fault tolerance.
We introduce the zoo of quantum LDPC codes and discuss their potential for making quantum computers robust against noise.
In particular, we explain recent advances in the theory of quantum LDPC codes related to certain product constructions and discuss open problems in the field.
\end{abstract}

\maketitle

\section{Introduction}
Whenever quantum information is stored or manipulated errors are bound to occur.
While there has been tremendous progress towards the realization of quantum processors~\cite{arute2019quantum,zhong2020quantum,jurcevic2021demonstration}, the implementation of error-corrected quantum memories and the demonstration of scalable fault-tolerant quantum computations remain formidable challenges.

One reason why error correction takes a prominent role in the realization of quantum computation is that qubits are inherently more fragile than classical bits.
Another, more subtle reason, is that quantum computers mix the analog (amplitudes) with the discrete (measurements).
Analog computation, which is based on the manipulation of continuous variables rather  than bits, should serve as a cautionary tale:
While it is well-known that, in theory, certain analog computers are vastly more powerful than even quantum computers~\cite{schonhage1979power}, they remain fictional devices since {\changetwo there are no known error correction schemes which apply to the relevant analog computing architectures}.
It is  {\changetwo thus unknown how} to scale them to a size relevant for solving practically relevant tasks.
This fact was already pointed out by early critics of quantum computers, who considered them to be nothing but analog computers in disguise~\cite{landauer1996physical,aaronson2013quantum}.
Shor countered this criticism by introducing the first \emph{quantum error correcting code} in his foundational paper~\cite{shor1995scheme}, showing that it is possible to protect quantum information.
This established quantum error correction as a field and started the pursuit to find optimal quantum error correcting codes.

Shortly after Shor's work, Kitaev's toric code~\cite{kitaev1997quantum,kitaev2003fault} and the related planar \emph{surface code}~\cite{freedman2001projective,bravyi1998quantum} were put forward. 
Surface codes are currently the leading approach for fault-tolerant quantum computation due to their high error correction threshold and planar layout. 
They exist for a variable number \emph{physical qubits} $n,$ and encode a single \emph{logical qubit} $k = 1$. 
Their \emph{distance}~$d$, a measure for the error-correcting capability, scales with~$\sqrt{n}.$ 
It is likely that we will soon see the first realization of small instances of surface codes, see~\cite{chen2021exponential} for recent results and a historical overview of experimental progress.

However, the surface code family does not compare favourably to the best known families of classical codes, where~$k$ and~$d$ scale linearly with~$n.$
While there exist quantum codes which match the properties of classical codes~\cite{calderbank1996good,ashikhmin2001asymptotically}, they have a significant technical draw-back: 
the parity checks, which have to be measured to infer the error, involve a growing number of physical qubits {\changetwo per logical qubit.}
This is an issue as arbitrarily large checks can not be reliably facilitated {\changetwo without additional fault-tolerance constructions}.
Further, the measurement of the checks can not be parallelized, leading to a build-up of errors due to idling qubits. Finally, many  decoding algorithms are based on the assumption that the parity checks are sparse.

Similar issues arose in classical coding theory and were solved by \emph{low-density parity check codes} (LDPC codes) where the number of bits involved in each check and the number of checks acting on each bit are bound by a constant for all members of the code family. 
LDPC codes have been very successful in the classical setting as they approach upper bounds due to Shannon on the amount of information that can be reliably transferred through a noisy channel~\cite{gallager1962low,mackay1996near}.
Many modern technologies such as WiFi, DVB-T and 5G are error corrected by LDPC codes~\cite{iliev2008application}.

It is therefore natural to consider \emph{quantum} LDPC codes which are defined in the same way.
While LDPC codes have been subject of intense study in classical coding theory for the last decades, their quantum analogues have only recently become a focus of attention.
Much of the interest in LPDC quantum codes was spurred by Gottesman's remarkable result in 2013 showing that quantum LDPC codes with a constant encoding rate can reduce the overhead of fault-tolerant quantum computation to be \emph{constant}~\cite{gottesman2014overhead}.
This is in contrast with other quantum fault tolerance schemes where, in order to perform a longer computation, it is necessary to suppress errors further, which requires larger codes and thus a growing number of physical qubits~\cite{kitaev1997quantum,knill1998resilient,aharonov2008fault}. 
{\change Moreover, one might hope that quantum LDPC codes can approach channel capacity similarly to (classical) LDPC codes.
We will not touch upon the challenging field of quantum channel capacities here, see~\cite{wilde2013quantum,gyongyosi2018survey,delfosse2012upper} for references.}

Classical coding theory is spoiled with constructions of LDPC codes with good properties.
In fact, taking random codes gives LDPC codes with constant encoding rate~$k/n$ and linear distance $d \propto n$ with high probability~\cite{gallager1962low,richardson2008modern}.
In comparison, it is much harder to construct quantum LDPC codes and it is still a major open problem whether quantum LDPC codes exist which rival the parameters of their classical counterparts~\cite{mackay2004sparse}.
{\change
For example, surface codes and color codes, which are LDPC and amongst the most studied quantum codes, only encode a low constant number of qubits and have $d\propto \sqrt{n}$ at best.
There has been a lot of recent progress towards better LDPC codes and several families of quantum LDPC codes have been constructed that significantly outperform surface codes and color codes in terms of their asymptotic parameters.
This is a strong indication for their potential, although they have not received the same amount of attention as the surface code as of now.
}

{\change 
Driven by this recent success, this perspective surveys the exciting and emerging field of quantum LDPC codes.
We discuss the construction and analysis of these novel codes which involve ideas and challenges which go far beyond the theory of surface and color codes. The field still has many open questions of central importance that we highlight throughout the text.
}

First, we introduce some background on stabilizer and CSS codes as well as their relation to homological algebra and geometry (\cref{sec:background}) for the convenience of the reader.

Next, we discuss constructions of quantum LDPC codes using the geometry of manifolds (\cref{sec:geometric_constructions}).
We discuss various product constructions of quantum LDPC codes inspired by the topological notions of cartesian products and fiber bundles which  are at the heart of recent breakthrough results (\cref{sec:product_constructions}).

{\change Somewhat paradoxically, we will also review quantum codes which are strictly speaking \emph{not} LDPC but also yield high encoding rates and require some sort of non-locality (\cref{sec:nonLDPC}).
We will not discuss surface and color codes in much detail since they are already covered extensively in the literature, see~\cite{lidar2013quantum,dennis2002topological,kubica2018abcs}.}

Besides introducing the zoo of quantum LDPC codes we discuss challenges and opportunities regarding their use for quantum error correction (\cref{sec:challandopp}).
For example, we will address decoding algorithms and the challenges in hardware implementation.
We do refrain from directly comparing different codes in terms of their thresholds due to the wide variety of error models and assumptions going into numerical simulations, but we do refer the reader to the relevant literature.
We also discuss applications of quantum LDPC codes outside of quantum error correction and quantum fault tolerance (\cref{sec:applicationsoutside}).

\subsection*{Notation and Conventions}
All vector spaces, unless otherwise mentioned, are over the field with two elements $\mathbb{F}_2.$
The notation $[n,k,d]$ describes the parameters of a classical binary code: number of bits~$n$, number of encoded bits~$k$ and minimum distance~$d$.
Similarly, we use the notation $[[n,k,d]]$ for quantum codes.

Often, the exact relation between the code parameters is not known.
However, it is sometimes possible to make asymptotic statements for which we need the following notation.
For two positive functions~$f$ and~$g$ we write $f\in O(g)$ if $\limsup_{n\rightarrow\infty} f(n)/g(n) < \infty$, $f\in o(g)$ if $\lim_{n\rightarrow\infty} f(n)/g(n) = 0$, $f\in \Omega(g)$ if $\liminf_{n\rightarrow\infty} f(n)/g(n) > 0$ and $f\in \Theta(g)$ if $f\in O(g)$ and $f\in \Omega(g)$.
The above symbols can be interpreted as sets of functions. 
However, whenever convenient we will abuse notation and write expressions such as $f \leq O(g)$ with the obvious interpretation.

{
	\renewcommand{\arraystretch}{1.2}%
	\begin{table}
		\centering
		\bgroup
		\def\arraystretch{1.5}%
		\resizebox{1\columnwidth}{!}{%
			\begin{tabular}{l r r r r r}
				\hline
				\hline
				Name & $k$ & $d$ & reference \\
				\hline
				2D hyperbolic codes & $\Theta\left(n\right)$ & $\Theta\left(\log(n)\right)$ & \Cref{sec:hyperbolic} \\
				4D hyperbolic codes & $\Theta\left(n\right)$ & $\Omega\left(\sqrt[10]{n}\right)$ & \Cref{sec:hyperbolic} \\
				Freedman--Meyer--Luo & $2$ & $\Omega\left(\sqrt[4]{\log(n)} \sqrt{n}\right)$ & \Cref{sec:freedman_meyer_luo} \\
				TP (good classical codes) & $\Theta\left(n\right)$ & $\Theta\left(\sqrt{n}\right)$ & \Cref{sec:tensor_product} \\
				TP (Ramanujan complexes) & $\Theta\left(\sqrt{n}\right)$ & $\Omega\left(\operatorname{polylog(n)} \sqrt{n}\right)$ & \Cref{sec:tensor_product} \\
				Fibre bundle codes & $\Theta\left(n^{3/5}/ \operatorname{polylog}(n)\right)$ & $\Omega\left(n^{3/5}/ \operatorname{polylog}(n)\right)$ & \Cref{sec:fiber_bundle} \\
				Lifted product codes & $\Theta\left(n^\alpha \log(n)\right)$ & $\Omega\left(n^{1-\alpha}/ \log(n)\right)$ & \Cref{sec:lifted_product} \\
				Balanced product codes & $\Theta\left(n^{4/5}\right)$ & $\Omega\left(n^{3/5}\right)$ & \Cref{sec:balanced_product} \\
				\hline
				\hline
			\end{tabular}
		}
		\egroup
		\caption{The best proven parameters of quantum LDPC codes discussed in this manuscript. Some entries refer to a whole method of constructing codes. In these cases we cite the best known proven parameters of a family quantum LDPC codes constructed using this method.}\label{tab:parameters}
	\end{table}
}

\section{Background}\label{sec:background}
One exciting aspect of the theory of quantum codes is the fact that it draws from a diverse mathematical and physical background.
It combines research from classical coding theory, systolic geometry, homology and combinatorics.
This manifests itself in several different perspectives that people from different areas have on quantum LDPC codes.
We will briefly survey these perspectives here for the convenience of the reader. 
\subsection{Quantum codes}
For background on general quantum codes we refer to the review by Terhal~\cite{terhal2015quantum} and Preskill's lecture notes~\cite{preskill1998lecture}. %
In this text, we will focus on stabilizer codes, which are the most studied class of quantum codes, see Gottesman's PhD thesis~\cite{gottesman1997stabilizer}.

\subsubsection{Stabilizer and CSS codes}
An $[[n,k,d]]$ stabilizer quantum code is defined by a commutative group~$S$ which is a subgroup of the Pauli group acting on the state space of~$n$ \emph{physical qubits} $(\mathbb{C}^2)^{\otimes n}$ {\change not containing~$-I$}.
The group~$S$ has $n-k$ independent generators, called \emph{stabilizer checks}.
The code subspace is defined as the $+1$-eigenspace of~$S$ and can be interpreted as the state space of~$k$ \emph{logical qubits}.
The {\change non-trivial} \emph{logical operators} on the code space correspond to the elements of the Pauli group which commute with~$S,$ but are not in~$S$ themselves.
The \emph{distance}~$d$ is the smallest number of physical qubits in the support of a non-trivial logical operator.
{\change
A slight generalization of stabilizer codes are \emph{subsystem codes} where only some logical degrees of freedom are used while the others are downgraded to become \emph{gauge qubits} and their corresponding logical operators are called \emph{gauge operators}. 
}

A CSS code is defined by a pair of classical linear binary codes $C_X,C_Z\subset\mathbb{F}^n_2$ such that the orthogonality condition $C_X\subset C_Z^\perp$ is satisfied, see~\cite{calderbank1998quantum}.
We assume that the codes~$C_X$ and~$C_Z$ are given by their parity check matrices $H_X$ and $H_Z.$
A CSS code defines a stabilizer code, where the stabilizer group is generated by the \emph{stabilizer checks} $X^c = \prod_{i=1}^n X_i^{c_i}$ where~$c$ is a row of~$H_X$ and $Z^d = \prod_{i=1}^n Z_i^{d_i}$ where~$d$ is a row of~$H_Z$.
The commutativity of the stabilizer group is ensured by the orthogonality condition $C_X\subset C_Z^\perp$ which is equivalent to
\begin{align}\label{eqn:CSS_commutativity_constraint}
H_Z H_X^{\tr} = 0 \mod 2.
\end{align}
It is straightforward to express properties of the stabilizer code in terms of the CSS code.
We will mostly focus on CSS codes and note that this is a only a minor restriction since any $[[n,k,d]]$ stabilizer code can be mapped onto a $[[4n,2k,2d]]$ CSS code, see~\cite{bravyi2010majorana}.

\subsubsection{Quantum LDPC  codes}\label{sec:LDPC_definition}
Generally, we are not interested in individual quantum codes but rather in families of codes with growing number of physical qubits.
By abuse of language we will often simply speak of \emph{a} code when we actually mean a family. 

A \emph{low-density parity-check code} (LDPC code) is a family of stabilizer codes such that the number of qubits participating in each check operator and the number of stabilizer checks that each qubit participates in are both bounded by a constant.
For CSS codes this means that the Hamming weight of each row and column of~$H_X$ and~$H_Z$ is bounded by a constant.

More generally, one can define quantum LDPC codes to include codes which are defined by a set of commutative projectors in the obvious way, see \cref{sec:spacetimecircuit}.
We consider subsystem codes to be LDPC if their stabilizer checks fulfil the LDPC condition.

A major open problem in quantum error correction is whether \emph{good} quantum LDPC codes exist.
Good is terminology from classical coding theory referring to the property of a code to have $k \in \Theta(n)$ and $d\in \Theta(n)$.
It turns out that taking a sparse parity check matrices at random defines good classical codes~\cite{richardson2008modern}, but taking two random  parity check matrices to define~$H_X$ and~$H_Z$ of a quantum code does not work, as \cref{eqn:CSS_commutativity_constraint} will not be satisfied.
Good quantum codes which are \emph{not} LDPC have been known since the early days of quantum error correction~\cite{calderbank1996good,ashikhmin2001asymptotically}.
However, it was only in 2020 that quantum LDPC codes have been constructed which have distances scaling as $d \geq \Omega(\sqrt{n} \polylog(n))$.
Recently, there has been rapid progress on increasing the distance of quantum LDPC codes (see \cref{sec:product_constructions}).

\subsection{Perspectives on CSS codes}
Besides their description in terms of parity check matrices, there are other useful representations of CSS codes, which we briefly collect in the following sections.

\subsubsection{Chain complexes}\label{sec:chain_complexes}
Soon after quantum codes were introduced by Shor, it was discovered that they can be constructed using tools from homological algebra \cite{kitaev2003fault,freedman2001projective,bravyi1998quantum,bombin2007homological}.
Let us briefly show how this homological description is related to our earlier definition of CSS codes:
We consider chain complexes $C$ of length $n+1$ which are collections of linear maps~$\partial_i$ called \emph{boundary operators} and $\mathbb{F}_2$-vector spaces~$C_i$
\begin{center}
    \begin{tikzcd}
C=(C_{n} \arrow[r, "\partial_{n}"] & \cdots \arrow[r, "\partial_2"] & C_1 \arrow[r, "\partial_1"] & C_0)
\end{tikzcd}
\end{center}
fulfilling {\change $\partial_{i}\partial_{i+1}=0.$}
{\change Moreover, we assume that all vector spaces $C_i$ are equipped with a basis such that the boundary operators $\partial_i$ can be interpreted as matrices. 

Chain complexes come from algebraic topology, see \cref{sec:geometric_construction}, and are hence described in topological terms. For example, the basis vectors of~$C_i$ are called $i$-cells and elements $i$-chains. An $i$-cycle is an $i$-chain with trivial boundary so an element in $\ker(\partial_i)$, whereas an $i$-boundary is an $i$-chain in the image of the boundary operator so an element in $\im(\partial_{i+1})$. The $i$-th homology $H_i(C)=\ker(\partial_i)/\im(\partial_{i+1})$ of $C$ is the vector space of $i$-cycles modulo $i$-boundaries. Dually, one defines $i$-cocycles, $i$-coboundaries and the $i$-th cohomology $H^i(X)=\ker(\partial^{tr}_{i+1})/\im(\partial^{tr}_i).$
}

In this homological language, a classical code corresponds to a chain complex of length two via its parity check matrix, while a CSS code can be represented by a chain complex of length three
\begin{center}
\begin{tikzcd}
C=(C_2 \arrow[r, "\partial_2=H_Z^{tr}"] & C_1 \arrow[r, "\partial_1=H_X"] & C_0).
\end{tikzcd}
\end{center}
With this correspondence, the logical $Z$-operators correspond to the homology group $H_1(C)=\ker(\partial_1)/\im(\partial_2)$ and the logical $X$-operators correspond to the cohomology group $H^1(C)=\ker(\partial_2^{\tr})/\im(\partial_1^{\tr}).$ The number of logical qubits is $k=\dim H_1(C)=\dim H^1(C).$
The $Z$-distance $d_Z$ and $X$-distance $d_X$ is the minimum Hamming weight of all non-trivial homology and cohomology classes, respectively. 
The distance $d$ is the minimum of $d_X$ and $d_Z.$
Vice versa, a single chain complex can yield many CSS codes, by taking any two consecutive boundary operators.

This is a fruitful perspective that allows to import language and constructions from homological algebra to the theory of quantum codes.
We refer to \cref{app:chaincomplexes} for more details.

\subsubsection{Tanner graphs} \label{sec:tannerconstruction}
\begin{figure}[h]
    \centering
    \includegraphics{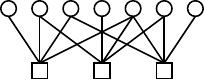}
    \caption{The Tanner graph of the $[7,4,3]$ Hamming code. Physical bits are represented by circles and checks by squares.}
    \label{fig:hammingcode74tanner}
\end{figure}

A linear binary code $C$ can be represented by a \emph{Tanner graph}. 
The Tanner graph is a bipartite graph where each side of the partition corresponds to the bits and checks, respectively.
Bits are connected to the checks in which they appear, see Figure \ref{fig:hammingcode74tanner}.

\begin{figure}[h]
    \centering
    \includegraphics{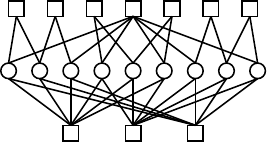}
    \caption{The Tanner graph of the Shor code. Physical qubits are represented by circles, $X$-checks and $Z$-checks by the bottom and top squares, respectively. {\changetwo We note that we chose a linearly dependent set of stabilizer checks in order to establish a connection to a geometric interpretation of Shor's code, see \Cref{fig:shorcodesimplicial}.}}
    \label{fig:shorcodetanner}
\end{figure}

Analogously, a quantum CSS code can be described by a Tanner graph with three layers, representing $X$-checks, physical qubits and $Z$-checks, see \cref{fig:shorcodetanner}. 
{\change A check acts on the qubits incident in the Tanner graph.}
Accordingly, the incidence matrices between the layers are given by the parity check matrices $H_X$ and $H_Z,$ respectively.
The commutativity constraint in \cref{eqn:CSS_commutativity_constraint} translates to the condition that the intersection of the neighborhoods of each $X$-check and $Z$-check contains an even number of physical qubits.

\subsubsection{Manifolds and Cell complexes}\label{sec:geometric_construction}
The toric code is arguably the most well-known CSS quantum code {\change \cite{kitaev1997quantum,kitaev2003fault}}.
It is defined from a tessellation of a torus with square tiles, where edges correspond to physical qubits and the stabilizer checks to faces ($Z$-checks) and vertices ($X$-checks).
Each check acts on all its incident qubits/edges and the logical operators correspond to non-contractible loops.

\begin{figure}
    \centering
    \includegraphics{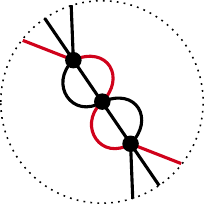}
    
    \caption{The Shor code~\cite{shor1995scheme} as a tessellation of the real projective plane, see~\cite{freedman2001projective}. 
    Antipodal points on the circle are identified. 
    The code has three $X$-checks, nine physical qubits and seven $Z$-checks represented by vertices, edges and faces, respectively. The red line represents a logical $Z$-operator.}
    \label{fig:shorcodesimplicial}
\end{figure}

{\change Interesting quantum codes can be derived from tessellations of surfaces or manifolds other than the torus. For example, Shor's $[[9,1,3]]$ code can be constructed from a tessellation of the real projective plane, the non-orientable cousin of the torus, see ~\cite{freedman2001projective}. In the same way as in the toric code, $X$-checks, physical qubits and $Z$-checks of Shor's $[[9,1,3]]$ code are associated with vertices, edges and faces of the tessellation and checks act on their incident edges.
This is shown in \cref{fig:shorcodesimplicial} where a tessellation of the projective plane with seven faces, nine edges and three vertices is depicted.
Note that pairs of antipodal points on the circle are identified.
Four of the faces are adjacent to two edges and hence give rise to $Z$-checks of weight two. One face is adjacent to six edges, giving rise to a $Z$-check of weight six.
Similarly, all $X$-checks, which are associated to the vertices, have weight six.
This should be compared to \cref{fig:shorcodetanner} representing the same code by a Tanner graph. From both representations, one can immediately extract the parity check matrices $H_X$ and $H_Z$ as incidence matrices.
}

This construction generalizes to tessellations of other surfaces and to higher-dimensional manifolds~$M$.
We will call the $i$-dimensional elements of the tessellation \emph{$i$-cells}, so that vertices are $0$-cells, edges are $1$-cells, faces are $2$-cells etc.
Given a tessellation of a $D$-dimensional manifold we identify $i$-cells of the tessellation ($0< i < D$) with qubits, $X$-checks with the $i-1$-cells and $Z$-checks with $i+1$-cells.
{\change Non-trivial logical $Z$-operators correspond to subsets of $i$-cells which have no boundary and do not arise as the boundary of a subset of $i+1$-cells.
In fact, the vector spaces spanned by $i$-cells together with the boundary operator form the so-called \emph{cellular chain complex}, see~\cite[Section 2.2.]{hatcher2002algebraic}.
This connects the homological perspective of \Cref{sec:chain_complexes} with the geometrical perspective. Moreover, there is also a relation to Tanner graphs by considering the Hasse diagram of the tessellation, see~\cite[Section 2.2.]{breuckmann2017homological}.
The logical $X$-operators are obtained in the same way by taking the dual tessellation or by considering cohomology classes.}

The distance %
is related to the $i$-\emph{systole}~$\sys_i(M)$ of $M$ which is the length/area/volume of the smallest non-contractible $i$-dimensional submanifolds of~$M$.
Note that this yields families of codes, by taking finer and finer tessellations. For reasonable tessellations these families are LDPC.

\section{Geometrical constructions}\label{sec:geometric_constructions}

Arguably the most famous quantum code is the toric code and its planar variant the surface code.
These two examples are in fact part of a much larger family of codes derived from geometrical objects, see \Cref{sec:geometric_construction}.
Importantly, the properties of these codes are determined by the geometric of the underlying space, so that tools from geometry become amenable to quantum error correction.

\subsection{Hyperbolic quantum codes}\label{sec:hyperbolic}
Quantum codes with a finite encoding rate $k/n \rightarrow R > 0$ for $n \rightarrow \infty$ naturally arise from manifolds of negative curvature, called \emph{hyperbolic manifolds}.
The reason for this lies in the \emph{Gau\ss --Bonnet--Chern theorem} which relates the geometry of a manifold to its topology~\cite{nakahara2003geometry}.
{\change More concretely, it shows that for hyperbolic manifolds of even dimension $D=2i$ the dimension of the homology group~$H_i$ grows linearly with the total volume of the manifold. Hence, the associated quantum code has a linear encoding rate.}
Therefore, any code derived from such a manifold (cf. \cref{sec:geometric_construction}) will have a constant encoding rate.

Compare this to the $L\times L$ toric code where the logical operators of minimum weight correspond to $1$-dimensional submanifolds (circles).
The number of these is two, regardless of $L$.
The same is true for higher dimensions, e.g. one can define a four-dimensional toric code~\cite{dennis2002topological} where the logical operators of minimum weight correspond to six $2$-dimensional tori.

\subsubsection{Hyperbolic Surface Codes}\label{sec:hyperbolicsurfacecodes}
Hyperbolic surface codes are the closest relatives of the toric code.
They are defined in exactly the same way as the toric code, except that the tessellations are derived from hyperbolic geometry.

If we consider a closed surface with a hyperbolic metric then the Gau\ss --Bonnet--Chern theorem~\cite{nakahara2003geometry} mentioned above can be used to derive an exact formula for the number of encoded logical qubits.
In particular, for regular tessellations based on regular polygons with~$r$ sides and~$s$ polygons meeting at each edge one can show that the number of logical qubits is given by $k = (1-2/r-2/s)\, n + 2$.
Note that the stabilizer check weight is~$r$ for $Z$-checks and~$s$ for $X$-checks, so that there is a trade-off between check-weight and encoding rate.
Hyperbolic surface codes exist with check-weights five and four for $X$-checks and $Z$-checks or vice versa~\cite{breuckmann2016constructions,breuckmann2017hyperbolic}.
Hyperbolic surface codes and their properties were discussed in~\cite{kim2007quantum,zemor2009cayley,delfosse2013tradeoffs}.
A general construction as well as a planar version were introduced in~\cite{breuckmann2016constructions}.
{\change In~\cite{delfosse2012upper} the authors utilize hyperbolic surface codes to obtain results in percolation theory.}

The distance of hyperbolic surface codes is logarithmic which suffices to prove that a threshold under minimum-weight decoding exists~\cite{kovalev2013fault}.
Many decoders that apply to the surface code can be used directly for hyperbolic surface codes, such as minimum-weight perfect matching~\cite{breuckmann2016constructions} and the union-find decoder~\cite{delfosse2017almost}.
However, this means that error suppression on the logical qubits for physical error rates below the threshold scales only polynomially with the system size.
Nevertheless, numerical simulations show that hyperbolic surface codes offer a reduction of physical qubits in the phenomenological noise model~\cite{breuckmann2016constructions,breuckmann2017hyperbolic} and gate-based noise model~\cite{higgott2020subsystem}.
Based on the symmetry of hyperbolic surface codes it is possible to find optimal measurement schedules of the check operators~\cite{higgott2020subsystem} and they are currently the only finite-rate quantum codes for which such schedules are known.

In~\cite{higgott2020subsystem} Higgott--Breuckmann show that hyperbolic surface codes can be turned into subsystem codes with weight-3 checks.
There also exist hyperbolic versions of color codes~\cite{delfosse2013tradeoffs,soares2018hyperbolic,vuillot2019quantum} which could simplify the implementation of logical gates.

\begin{figure}
    \centering
    \includegraphics[scale=0.11]{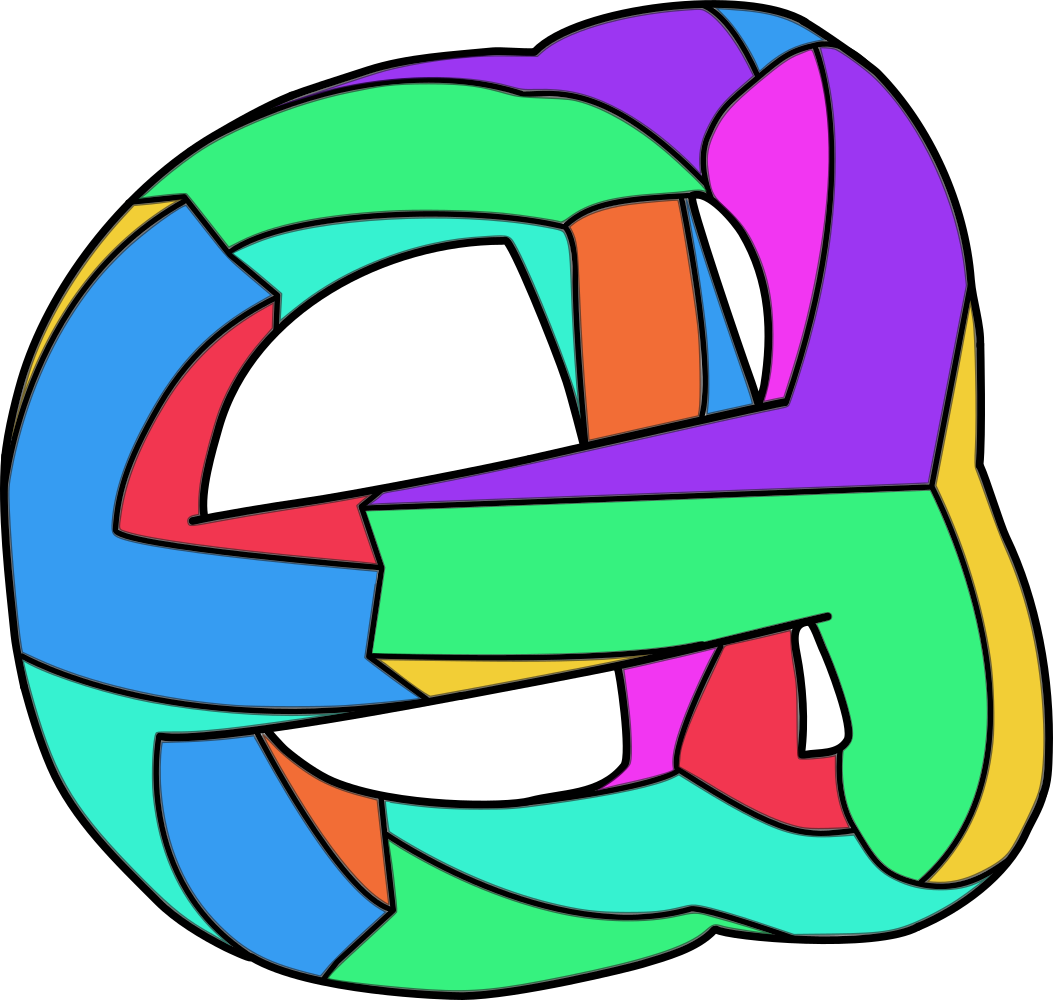}\hfil
    \includegraphics[scale=0.25]{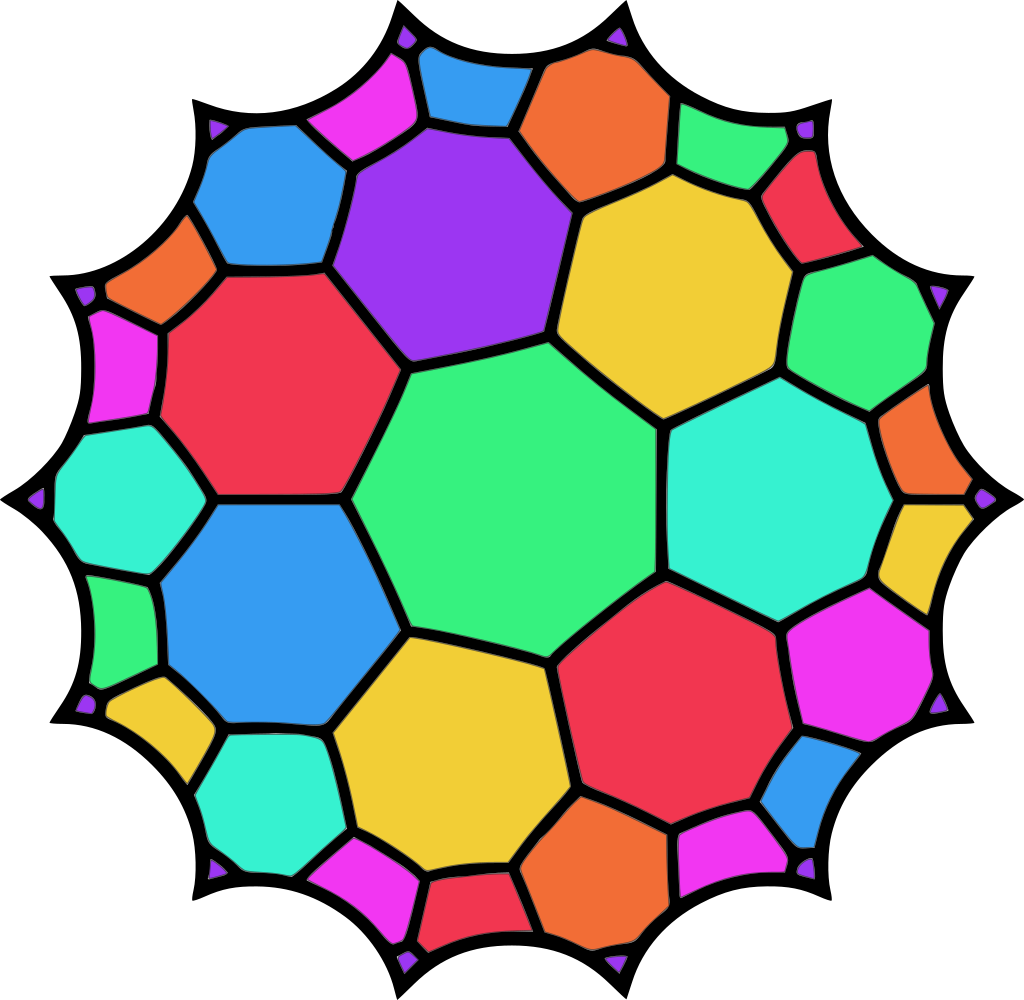}
    \caption{A hyperbolic surface of genus 3 tessellated by heptagons.
    It gives rise to a code with parameters $n = 84$, $k = 6$, $d_X=4$, $d_Z=8$.
    The colors have no intrinsic meaning and are only included to guide the eye.
    A weight-four $X$-operator goes through the following four faces on the right: 
    magenta (top), violet, green (middle), yellow (below) and back to the same magenta face (periodic boundary).
    A weight-eight $Z$-operator runs along the left-hand side of these faces.
    }
    \label{fig:klein_quartic}
\end{figure}

\subsubsection{Higher-Dimensional Hyperbolic Codes}
Lubotzky--Guth showed that codes derived from hyperbolic manifolds of dimension larger than two give quantum codes with distance scaling as $d\in \Theta(n^\alpha)$ for some $\alpha > 0$~\cite{guth2014quantum}.
In fact, they constructed families of 4D hyperbolic quantum codes such that $\alpha > 0.1$.
For arithmetic 4D hyperbolic manifolds the authors establish an upper bound of $\alpha < 0.3$.
However, it is an open problem whether these bounds hold for quantum codes derived from general 4D hyperbolic manifolds.

Hastings has proposed an efficient local decoding strategy for 4D hyperbolic codes~\cite{hastings2013decoding}.
However, despite having a distance scaling like $n^\alpha$, Hasting's decoder is only shown to correct errors up to size $\log(n)$.

The description of the codes by Lubotzky--Guth~\cite{guth2014quantum} is implicit. 
Londe--Leverrier~\cite{londe2017golden} consider a tessellation of 4D hyperbolic space by hypercubes giving rise to a family of codes with asymptotic encoding rate $R \geq 17/360$.
A construction based on a self-dual tessellation by 120-cells was given by Breuckmann--Londe~\cite{breuckmann2020single} giving an asymptotic encoding rate~$R \geq 13/72$.
Furthermore, they show how topological coverings can be used to reduce the size of these codes and perform simulations of the codes using a belief-propagation decoder which indicates that it has intrinsic robustness against measurement errors, see \cref{sec:decoding}.

\subsection{Freedman--Meyer--Luo codes}\label{sec:freedman_meyer_luo}
Hyperbolic geometry was used in earlier work by Freedman--Meyer--Luo~\cite{freedman2002z2} to construct a family of quantum codes with parameters $[[n,2,\Omega(\sqrt[4]{\log(n)}\sqrt{n})]]$ \footnote{{\change Note that there is an unfortunate typo in~\cite{freedman2002z2}, claiming distance scaling as $\Omega(\sqrt[2]{\log(n)}\sqrt{n})$.}}.
These codes held the record for distance scaling for around 20 years until the record was broken in 2020 by several works discussed in \cref{sec:product_constructions}.
{\change
The arguments used in~\cite{freedman2002z2} are quite involved and beyond the scope of this Perspective. However, we sketch the main ideas behind the construction in order to give an intuition behind the distance scaling for the interested reader.
}

\begin{figure}[h]
    \centering
    \includegraphics[width=\columnwidth]{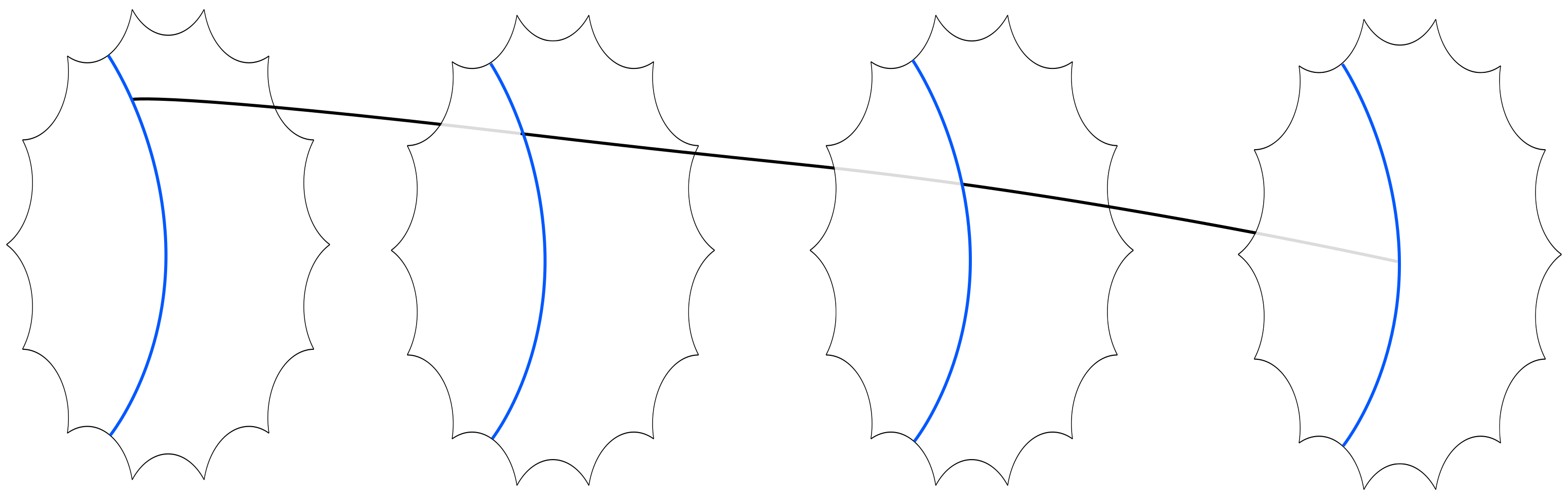}
    \caption{This figure shows a crucial step in the construction of Freedman--Meyer--Luo codes.
    The hyperbolic polygon is the fundamental domain of a hyperbolic surface~$\Sigma_g$ with genus~$g$ (cf.~\cref{fig:klein_quartic}).
    The blue line corresponds to a geodesic on the surface of length $\Theta(\log(g))$.
    All points on the surface are translated by a distance~$\Theta(\sqrt{\log(g)})$ along this geodesic.
    The thick black line traces out the position of a single point in the product $\Sigma_g \times [0,1]$.
    The two ends of the product, corresponding to coordinates $0$ and $1$ in the interval, are then identified so that we obtain $\Sigma_g \times S^1$ with a twist.
    }
    \label{fig:fml_code}
\end{figure}

{\changetwo
We will now sketch the construction of the underlying manifolds to give an intuition for the distance bound:
First, take a closed hyperbolic surface~$\Sigma_g$ of genus~$g$ and take the Cartesian product with the interval $[0,1]$ of unit length.
The length of the shortest non-contractible loop on~$\Sigma_g$ is called \emph{1-systole} and is denoted $\sys_1(\Sigma_g)$.
We identify the two ends of $\Sigma_g\times [0,1]$ with a twist of length $\sqrt{\sys_1(\Sigma_g)}$, so that the 1-systole of the resulting 3-dimensional manifold is $\sqrt{\sys_1(\Sigma_g)}$, see \Cref{fig:fml_code}.
All non-contractible loops of length $\sys_1(\Sigma_g)$ coming from $\Sigma_g$ are removed using surgeries, so that we obtain a 3-manifold which we denote by~$P_g$.
Since the interval $[0,1]$ has unit length, the 3-dimensional volume of~$P_g$ is proportional to the area of~$\Sigma_g$.
Further, due to the Gau\ss --Bonnet--Chern theorem (cf.~\Cref{sec:hyperbolic}) we have that $\area(\Sigma_g) = \Theta(g)$.
The final step in the construction of the manifolds is to take a loop~$S^1$ of length~$g/\sqrt{\log(g)}$ and take the Cartesian product with~$P_g$.
The resulting 4-manifold then has 4-dimensional volume $\vol_4(P_g \times S^1)=g^2/\sqrt{\log(g)}$.

By tessellating the manifolds uniformly we obtain a cell complex on which we can define a homological quantum code by identifying the 2-cells (faces) with qubits, 1-cells (edges) with $X$-checks and 3-cells (volumes) with $Z$-checks, as described in \Cref{sec:geometric_construction}.
The number of physical qubits derived from these manifolds have number of qubits~$n$ scaling with $\vol_4(P_g \times S^1)$.
The logical operators correspond to non-contractible surfaces inside the 4-manifold.
Their area is called the 2-systole and they have size~$\sys_2(P_g \times S^1)=\Theta(g)$, leading to the distance bound.
See~\cite{fetaya2011homological} for a nice review of the construction which covers all the details.}

\subsection{Haah's code}
Haah developed a general formalism which describes translation-invariant stabilizer codes which are local in $D$-dimensional Euclidean space \cite{haah2016algebraic}.
Arguably the most famous example of a quantum code constructed by Haah's method is \emph{Haah's cubic code} which is defined on a 3D cubic lattice of size $L\times L\times L$ with periodic boundary conditions and two qubits per site \cite{haah2011local}.
The number of encoded qubits in Haah's cubic code grows with~$L$ and hence with the number of physical qubits.
However, the exact number depends in a non-trivial way on~$L$.
For the distance of Haah's cubic code only the bounds $\Omega(\sqrt[3]{n}) \leq d \leq O(n^{2/3})$ are known \cite{haah2011local,panteleev2020quantum}.
An interesting feature of Haah's cubic code is the fact that its logical operators are fractals.
Although we classify Haah's code as a geometric code here, as it is defined on a cubic lattice, it can also be understood as a special case of a product construction discussed in \cref{sec:product_constructions}.

\begin{oproblem}
What is the asymptotic distance-scaling of Haah's cubic code?
\end{oproblem}

Haah's code is a candidate for a self-correcting quantum memory.
By defining a local Hamiltonian which has the parity checks as energy-penalty terms, we obtain a physical system with the quantum code as its ground state.
A self-correcting memory would be such a system which is inherently robust against thermal noise, without the need for an active decoding procedure.
As this is not a focus of this text, we refer the reader to \cite{brown2016quantum} for more background.

\subsection{Bounds on parameters}
{\change 
While geometry is a useful tool for the construction of quantum LDPC codes it also comes with restrictions.
Fetaya~\cite{fetaya2012bounding} showed that any code derived from the tessellation of a surface, either closed or with boundary, must have its distance bounded as $d^2 \leq O(n)$.
Delfosse~\cite{delfosse2013tradeoffs} extended Fetaya's result to the bound $k d^2 \leq O(\log^2(k)\, n)$.

For higher-dimensions, Bravyi--Poulin--Terhal showed that for any $[[n,k,d]]$ stabilizer code on a $D$-dimensional Euclidean lattice $k d^\alpha \leq O(n)$ where $\alpha = 2/(D-1)$~\cite{bravyi2010tradeoffs}.
We note that the Bravyi--Poulin--Terhal bound does not extend to non-Euclidean lattices, as for $D=2$ the bound is violated by hyperbolic surface codes \cref{sec:hyperbolicsurfacecodes}.
A class of codes conjectured to satisfy the Bravyi--Poulin--Terhal in 3D was introduced by Devakul--Williamson~\cite{devakul2021fractalizing}.

\begin{oproblem}
Can the above bounds be modified or extended to non-Euclidean lattices in higher dimensions?
\end{oproblem}
This problem seems challenging as it relates to deep questions in a sub-field of mathematics called \emph{systolic geometry}, which analyzes the volume-scaling of non-contractible sub-manifolds~\cite{katz2007systolic}.
}

\section{Product constructions}\label{sec:product_constructions}

Classical coding theory is a long established field and it would be desirable to transfer results into quantum coding theory.
In this section we describe various \emph{product constructions} which allow to build quantum codes from classical codes and/or quantum codes.
{\change These constructions} are at the heart of recent breakthrough results in the theory of quantum LDPC codes.

The first class of examples are incarnations of the \emph{tensor product} of chain complexes from homological algebra. 
There are the \emph{hypergraph product} (HP) codes by Tillich and Z\'emor~\cite{tillich2013quantum} and the \emph{homological product} codes of Hastings--Bravyi~\cite{bravyi2014homological}, constructing a quantum code from two classical codes. 
The \emph{distance balancing} of Hastings~\cite{hastings2016weight} and Evra--Kaufman--Zémor~\cite{evra_decodable_2020} is achieved by taking tensor products  of quantum codes with classical codes, while the codes of Kaufman--Tessler~\cite{kaufman2020new} employ iterated tensor products of quantum codes.

There are multiple improvements and generalizations of these product construction. Hastings--Haah-O'Donnell~\cite{hastings2020fiber} define fiber bundle codes that introduce a twist in the tensor product in order to increase the distance. Another approach is found in the \emph{generalized hypergraph product} and \emph{lifted product} of  Panteleev--Kalachev \cite{panteleev2019degenerate,panteleev2020quantum} as well as the \emph{balanced product} of Breuckmann--Eberhardt \cite{breuckmann2020balanced}. All these are very closely related to each other, see \cref{app:chaincomplexes}.
In 2020, the distance record of Freedman--Meyer--Luo from 2002, see Section \ref{sec:freedman_meyer_luo}, was broken multiple times utilising these product constructions.  
{\change These are breakthrough results, as they surpass the the $\operatorname{polylog}(n) \sqrt{n}$-distance barrier which was by some believed to be unsurpassable.
While many of the constructions do not have a constant encoding rate and hence do not satisfy the assumptions of Gottesman's constant overhead theorem \cite{gottesman2014overhead}, they constitute an important step in the pursuit of good quantum LDPC codes.
Independent of asymptotic results these constructions provide tools to construct concrete examples of codes worth studying, as done in \cite{panteleev2019degenerate}.
}

\subsection{Tensor/hypergraph products}\label{sec:tensor_product}
\begin{figure}
    \centering
    \includegraphics{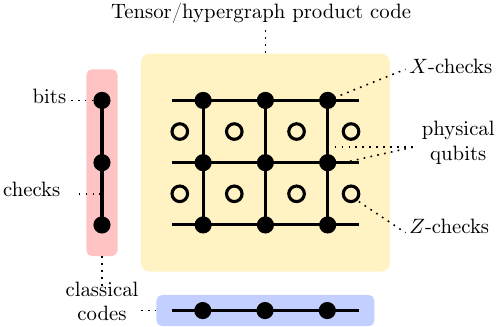}
    \caption{The tensor/hypergraph product of two repetition codes yields a surface code. The physical qubits of the quantum code are represented by edges and correspond to pairs of bits (vertical) or pairs of checks (horizontal) in the two repetition codes. We note that in the hypergraph product construction one of the classical codes is transposed, which is not depicted here.}
    \label{fig:hypergraphproduct}
\end{figure}
The tensor product of vector spaces extends to a notion of tensor product of chain complexes. This is a classical construction in homological algebra closely related to the Cartesian product of topological spaces, see \cite[Section 2.7]{weibel_introduction_1994}.

In particular, new quantum codes can be constructed by tensor products of classical codes and/or quantum codes using tensor products.

The first product construction in this spirit is the \emph{hypergraph product} introduced by Tillich--Z\'emor \cite{tillich2013quantum} in 2009.
The hypergraph product constructs a $[[n_1 n_2 + r_1 r_2,\, k_1 k_2,\, \min \lbrace d_1, d_2 \rbrace]]$ quantum code from two classical $[n_i,k_i,d_i]$-codes with $r_i$ linearly independent checks for $i=1,2$. Its stabilizer checks are a combination of the physical bits and parity checks of the classical codes, see \cref{fig:hypergraphproduct}.
By taking the hypergraph product of suitable classical LDPC codes, the authors were the first to achieve quantum LDPC codes with \emph{constant encoding rate} and distance $d\in\Theta(\sqrt{n}).$
Hypergraph products were used to define \emph{quantum expander codes}, see~\cite{leverrier_quantum_2015, fawzi_constant_2018}, using Sipser--Spielman's expander codes~\cite{sipser_expander_1996}. 
Further, there are constant factor improvements~\cite{kovalev_improved_2012} and higher dimensional generalisations~\cite{zeng_higher-dimensional_2019} of hypergraph products.

While Tillich--Z\'emor's definition is of combinatorial nature and utilising Tanner graphs, see \cref{sec:tannerconstruction}, it is equivalent to taking the tensor product of two chain complexes induced by the classical codes, see \cref{sec:chain_complexes}.
This perspective was, for example, taken by Hasting--Bravyi~\cite{bravyi2014homological} in 2011 with their \emph{homological product codes} and Audoux--Couvreur's work on tensor products of CSS codes~\cite{audoux_tensor_2019}. The tensor product has the advantage over hypergraph products of being defined for arbitrary chain complexes and not just classical codes which gives immediate higher dimensional generalisations. For more details see \cref{app:tensorproduct}.

In April 2020, generalizing a construction of Hastings~\cite{hastings2016weight}, Evra--Kaufman--Zémor~\cite{evra_decodable_2020} introduced a \emph{distance balancing} procedure for quantum codes utilising tensor products. They showed that the tensor product of a $[[n,k,d_X,d_Z]]$ quantum code with $r_X$ $X$-checks and a classical $[m,l,d]$  code with~$r$ checks yields a $[[nm+r_Xr,kl, d_X, d_Zd]]$ quantum code.
Armed with this new tool, the authors were the first to break Freedman--Meyer--Luo's distance record, see \cref{sec:freedman_meyer_luo}. 
The authors consider Ramanujan complexes, a higher dimensional generalisation of Ramanujan graphs, see \cite{lubotzky_high_2017}. 
Using Ramanujan complexes directly would yield quantum codes with distances $d_X\in \Theta(\log(n))$ and $d_Z\in \Theta(n).$
By applying distance balancing to these, Evra--Kaufman--Zémor construct codes with distance $d\in O(\sqrt{n}\log(n)).$

Just four months later, in August 2020, Kaufman--Tessler~\cite{kaufman2020new} set a new record with $d\in O(\sqrt{n}\log(n)^m)$ for arbitrary positive integers $m$, by making use of iterated tensor products of Ramanujan complexes.

\subsection{Fiber Bundles}\label{sec:fiber_bundle}

One month after Kaufman--Tessler's results \cite{kaufman2020new}, in September 2020, the $\polylog(n) \sqrt{n}$ distance barrier was broken by Hastings--Haah--O'Donnell's \emph{fiber bundle codes} \cite{hastings2020fiber}, a generalisation of the tensor product.

\begin{figure}
    \centering
    \includegraphics{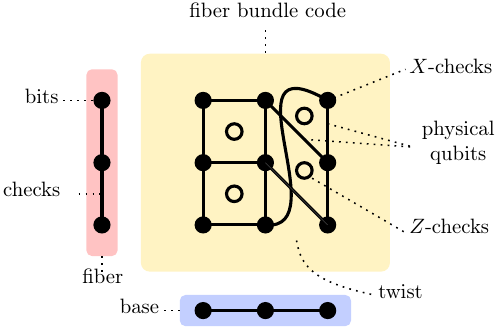}
    \caption{A fiber bundle code obtained from two repetition codes. Here, we assume periodic boundary conditions in the vertical direction.}
    \label{fig:fiberbundlecodes}
\end{figure}

Much like tensor products, fiber bundle codes are constructed from two classical codes, referred to as base and fiber code. 
While the number of physical qubits, logical qubits and checks is the same as in the tensor product, certain twists are introduced in the checks of the fiber, permuting the position of the qubits in the direction of the fiber. 
The twists are determined by a collection of automorphisms of the fiber code that are specified for every pair of bit and incident check of the base code. 
This can result in an increased the distance of resulting code.
{\change 
A trivial example is the \emph{twisted toric code} where the lattice is displaced along the vertical direction. Due to the twist the non-trivial horizontal loop representing a logical $Z$-operator has to take a `detour' to close up on itself.

The concept is derived from the topological notion of fiber bundles~\cite{nakahara2003geometry} and fiber bundle codes can be visualised as the fiber code varying non-trivially over the base code, see \cref{fig:fiberbundlecodes}, much like the M\"obius strip where the unit interval (fiber) varies non-trivially along the circle (base) or in the Klein bottle where a circle (fiber) varies over another circle (base).
For more details, see \cref{app:fiberbundle}.
}

{\change Hastings--Haah--O'Donnell applied the fiber bundle construction to a random classical code as base, a cyclic repetition code as fiber and a random choice of twists.}
By homological and probabilistic arguments, they show that this yields families of quantum LDPC codes with $k\in \Theta(n^{3/5}/\polylog(n))$ logical qubits and distance $d\in \Omega(n^{3/5}/\polylog(n)),$ a big step in the endeavor towards good quantum LDPC codes.

\subsection{Lifted Products}\label{sec:lifted_product}
Panteleev--Kalachev \cite{panteleev2020quantum} managed to break Hastings--Haah-O'Donnell's record just two months later, in December 2020, employing  a different improvement of tensor product codes.
Already in 2019 \cite{panteleev2019degenerate} the authors introduced \emph{generalized hypergraph product codes} (GHC) which they later renamed to \emph{lifted products codes} (LP). 
The LP construction allows to decrease the number of physical qubits in the tensor product by a reduction of symmetry. 

{\change Specifically, the construction takes as input two classical codes given by their parity check matrices. It assumes that the matrices admit a block decomposition into pairwise commuting submatrices of size $\ell \times \ell$ for some~$\ell$.
First, the hypergraph/tensor of the classical codes is taken. Then the parity check matrices of the resulting quantum code inherit a natural block decomposition into submatrices of size $\ell^2 \times \ell^2$. 
Next, the lifted product is obtained by collapsing each of these submatrices of size $\ell^2 \times \ell^2$ to matrices of size $\ell\times\ell$ via summing over all of their $\ell\times\ell$-sized blocks. 
The pairwise commutativity assumption ensures that the result defines a CSS code. 

The lifted product reduces the number of physical qubits by a factor of~$\ell$ in comparison to the hypergraph/tensor product. See Appendix \hyperref[app:liftedproducts]{A 5} for more details.

In~\cite{panteleev2020quantum}, Panteleev--Kalachev study a special case of lifted product codes constructed from coverings of Sipser--Spielman expander codes, see~\cite{sipser_expander_1996}. 
Their quantum codes are constructed from an $s$-regular expander graph (the \emph{base}), a cyclic covering of this graph of degree $\ell$ (the \emph{lift}) and a classical \emph{local} code on~$s$ bits. By grouping together vertices/edges in the lift graph that map to the same vertex/edge in the base graph, the incidence matrix of the lift graph has a natural block decomposition into submatrices of size $\ell\times\ell.$ The submatrices all commute since they correspond to a cyclic permutation. Next, the lift graph is combined with the local code, using the construction of Sipser--Spielman. The parity resulting check matrix inherits a block decomposition. In a last step the lifted product of this matrix with the adjacency matrix of the cycle graph of size~$\ell$ (the parity check matrix of cyclic repetition code) is formed.}

Most remarkably, they establish tight distance bounds for these lifted product codes assuming (co-)expansion properties of the associated classical expander code.
By a random choice of graph, cover and local code, they construct quantum LDPC codes with logical bits in $\Theta(n^\alpha\log(n))$ and distance in  $\Omega(n^{1-\alpha/2}\log(n))$ for any $0\leq \alpha <1.$ 
Thereby they achieve the first quantum LDPC  codes with almost linear distance.

\subsection{Balanced Products}\label{sec:balanced_product}
Almost simultaneously to Panteleev--Kalachev's work, Breuckmann--Eberhardt  introduced \emph{balanced product codes}~\cite{breuckmann2020balanced}. 
Similarly to lifted products, the balanced product construction is based on a reduction of symmetry in the hypergraph/tensor product. 
The balanced product is defined for two classical codes with a common symmetry group and arises by modding out the action of the group on their hypergraph/tensor product. 
The concept is derived from the balanced product of topological spaces, a classical construction in topology which is commonly used to construct fiber bundles from principal bundles~\cite{nakahara2003geometry}. 

\begin{figure}
    \centering
    \includegraphics{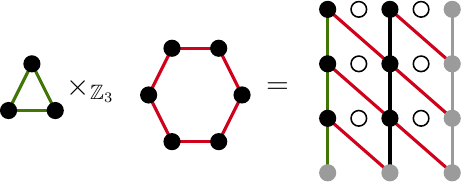}
    \caption{The balanced product of the length-3 and length-6 cyclic graphs over $\mathbb{Z}_3$. 
    This gives a $3\times 2$ twisted toric code, which is a $[[12,2,3]]$ quantum code.
    Grey edges and vertices are only included to visualize the periodic boundaries.
    This code has an equivalent interpretation as a fiber bundle with a length-2 cycle graph as base and a length-3 cycle graph as fiber.
    It is also a lifted product of the boundary operators~$\partial_1$ of the two cycle graphs.}
    \label{fig:balancedproduct}
\end{figure}

{\change We will now explain the construction in more detail.
To begin, assume we are given a classical code with a group~$H$ acting on the bits and checks via permutation such that the incidence relation between bits and checks is preserved.
One can form the quotient code by identifying bits and checks which lie in the same orbit of the action, thereby reducing the length of the code.
Now, given two classical codes~$C$ and~$D$ on which a group~$H$ acts, one can form the balanced product~$C\otimes_H D$ as follows.
First, the tensor/hypergraph product code~$C\otimes D$ is formed. There is an induced action of~$H$ on the physical qubits and checks of this quantum code. The balanced product~$C\otimes_H D$ is obtained by identifying physical qubits and checks in the same orbit, much like in the classical quotient code. We note that the balanced product construction also works in greater generality. For example, one can form the balanced product of two quantum codes or chain complexes. See \cref{app:balancedproducts} for more details.

In  \cref{fig:balancedproduct} the balanced of two repetition codes/cycle graphs of size $3$ and $6$ is visualised. 
Here~$\mathbb{Z}_3,$ the cyclic group of order $3,$ acts on both codes via rotation by~$120^\circ$. Their hypergraph/tensor product is a $3\times 6$ toric code on which the group $\mathbb{Z}_3$ acts via rotating the torus in both directions by $120^\circ$ simultaneously. 
Now all  $X$-checks/vertices, physical qubits/edges and $Z$-checks/faces which lie in the same orbit of the action are identified. The resulting balanced product code is a  $3\times 2$ twisted toric code with parameters $[[12,2,3]].$ 
}

Breuckmann--Eberhardt applied the balanced product to Sipser-Spielman codes and repetition codes with cyclic symmetry. The Sipser-Spielman codes are derived from Lubotzky--Phillips--Sarnak's expander graphs which are Cayley graphs of projective linear groups over a finite field $\operatorname{PSL}(2,q),$ see \cite{lubotzky_ramanujan_1988}.
The resulting codes are non-random and achieve a number of logical qubits~$k\in \Theta(n^{4/5})$ and distance~$d\in\Omega(n^{3/5}).$

{\change 
\subsection{The Possibility of Good Quantum LDPC Codes}
The fiber, lifted and balanced product open to the door to many new code families. 
They give hope that it possible to construct \emph{good} quantum LDPC codes, so codes with constant rate and linear distance.
However, we believe that it is not sufficient to only consider cyclic (or more generally abelian) twists/symmetries, such as those used in the previous three sections.
This is supported by arguments of Panteleev--Kalachev \cite[Section~IV~C]{panteleev2020quantum} where they show that their bounds $k\in \Theta(n^\alpha\log(n))$ and  $d\in \Omega(n^{1-\alpha/2}\log(n))$ for $0\leq \alpha <1$ are in some sense optimal in these cases.

One promising idea to circumvent these bounds is to take non-abelian twists.
This is possible in the balanced product construction by taking a non-abelian group $H$. For example, one could take the balanced product of two copies of a Sipser-Spielman code derived from a Cayley graph of a non-abelian group such as $\operatorname{PSL}(2,q)$. 
While one can show that this yields codes with linear rate, it is not yet clear how to determine their distance.
\begin{oproblem}
Can the balanced product of two expander codes derived from Cayley graphs be used to construct \emph{good} LDPC codes \cite[Section VI]{breuckmann2020balanced}?
\end{oproblem}
Balanced products codes have the advantage of being symmetric in their two input factors. 
The construction can easily be used to construct codes which are isomorphic to their dual and hence rendering any distance balancing unnecessary.
}

\subsection{XYZ Products}
Leverrier--Alpers--Vuillot~\cite{leverrier2020quantum} consider \emph{XYZ product codes}, a product construction of stabilizer codes which are not CSS codes. 
The idea was first suggested in~\cite{maurice2014codes} and generalizes a 3D non-CSS code due to Chamon~\cite{chamon2005quantum,bravyi2011topological}.
An XYZ product code is defined by a tensor product of three classical codes.
Each check contains Pauli-$X$, Pauli-$Y$ and Pauli-$Z$ operators.
Leverrier--Alpers--Vuillot argue that the distance of XYZ product codes could be as high as $\Theta(n^{2/3})$ as logical operators have natural representations as ``2D objects'' in the product.
However, there are no known general lower bounds on the distance.
Bravyi--Leemhuis--Terhal~\cite{bravyi2011topological} show that for the Chamon code, which is the XYZ product of three repetition codes with block lengths $n_1$,$n_2$ and $n_3$, the number of encoded qubits is given by $4 \gcd (n_1,n_2,n_3)$.
Recently, a 2D version of the Chamon code was found to perform remarkably well in numerical simulations when the noise is biased~\cite{bonilla2020xzzx}.

\begin{oproblem}
What is the minimum distance and performance of quantum XYZ product codes?
\end{oproblem}

\section{Other Constructions Related to LDPC~Codes}\label{sec:nonLDPC}
In this section we describe other families of quantum codes which are not LDPC under our strict definition, see \cref{sec:LDPC_definition}.
However, they are sometimes called LDPC codes in the literature and are based on very interesting ideas which is why we include them here.

\subsection{Bravyi--Hastings Codes}
In~\cite{bravyi2014homological} Bravyi--Hastings apply the tensor product construction, see \cref{sec:tensor_product}, to two random, non-LDPC CSS codes with check weights~$\Theta(n)$.
They show that with high probability the resulting codes have parameters $[[n,\Theta(n),\Theta(n)]]$, i.e. they are good codes.
However, the codes are not LDPC as the check weights are in $\Theta(\sqrt{n})$.
This improves on the earlier result by Calderbank--Shor \cite{calderbank1996good} who constructed good quantum codes with check weight~$\Theta(n)$.
Note that the square-root of the check weight comes from the fact that Bravyi--Hastings are taking the product of two codes with linear check weight and that the check weights are additive in the product.
This immediately suggests that the weight could be further suppressed by taking iterated products.

A related construction due to Hastings \cite{hastings2016quantum}, under the assumption of a conjecture in geometry, achieves distance $d\in \Omega(n^{1-\epsilon})$ for arbitrary $\epsilon > 0$ and with logarithmic stabilizer weight.

\begin{oproblem}
Does the iterated product of random codes provide a code family of good codes with stabilizer check weight scaling arbitrary low?
\end{oproblem}

\subsection{Bravyi--Bacon--Shor Codes}
Bravyi--Bacon--Shor codes are generalizations of Bacon--Shor subsystem codes defined in~\cite{bravyi2011subsystem} and studied by Yoder~\cite{yoder_bbs,li2020numerical}.
They are defined from a binary matrix $A\in \mathbb{F}_2^{m_1\times m_2}$ by placing physical qubits on a $m_1\times m_2$ square grid with a physical qubit placed on position $(i,j)$ if and only if $A_{i,j}=1$.
The gauge operators are generated by $XX$ interactions between any two consecutive qubits sharing a column and $ZZ$ interactions between any two consecutive qubits sharing a row.
The number of physical qubits $n$ of the resulting code is the number of non-zero entries in~$A.$  
Bravyi furthermore showed that the number of logical qubits is $k=\rk(A)$ and that the minimum distance is the minimum Hamming weight of the row- and column-span of~$A$, i.e. $d = \min_{c\in V} |c|$ where $V = (\im A \cup \im A^T ) \setminus \{0\}$.

In~\cite{yoder_bbs} Yoder considered taking two classical codes with parameters $[n_1,k,d_1]$ and $[n_2,k,d_2]$ with generating matrices~$G_1$ and~$G_2$ in order to define a Bravyi--Bacon--Shor Code based on the matrix $A=G_1^T Q G_2$ where $Q\in \mathbb{F}_2^{k\times k}$ is any full-rank matrix.
The resulting code then has between $\min \{ n_1 d_2, d_1 n_2 \}$ and $n_1 n_2$ physical qubits, $k$ logical qubits and distance $\min \{ d_1, d_2 \}$.
In particular, when the classical input codes have constant rate and linear distance then the resulting Bravyi--Bacon--Shor codes have optimal scaling for 2D subsystem codes.
Furthermore, the resulting code inherits a decoder from the classical codes used for the construction.

\begin{oproblem}
Can the Bravyi--Bacon--Shor Codes be extended to the novel product constructions discussed in \ref{sec:product_constructions}?
\end{oproblem}

\subsection{Subsystem Codes from Quantum Circuits}\label{sec:bacom_flammia_harrow_shi}
Bacon et al. showed in~\cite{bacon2017sparse} that it is possible to obtain quantum codes such that each physical check has weight~$O(1)$ with distance $\Theta(n^{1-\epsilon})$ where $\epsilon \in O(1/\sqrt{\log(n)})$.
Furthermore, restricting the code to be spatially local in $D$-dimensional Euclidean space, the authors show that a distance of $\Theta(n^{1-\epsilon-1/D})$ can be obtained.
The physical checks correspond to \emph{gauge operators} and not to the actual stabilizer checks.
The stabilizer checks can be written as products of the gauge operators, so that the outcome of stabilizer measurements can in principle be inferred from the measurements of the gauge operators.
The number of gauge factors of a stabilizer is not bounded, in other words, the actual stabilizer checks have unbounded weight, so that this code family is not LDPC.

Let us briefly sketch the main idea behind the construction.
The authors show that a quantum circuit can be mapped onto a quantum code by associating the gates with gauge operators which act on physical qubits positioned between the gates.
It can then be shown that if the mapping is applied to a suitable error-detection circuit of a stabilizer code then the resulting subsystem code has the same logical operators up to multiplication with gauge operators. 
The actual parameters stated earlier can be obtained by taking a quantum code with parameters $[[n_0,1, \Theta(n_0)]]$, which is guaranteed to exist by \cite{goodquantumcodes}, and concatenate it with itself a suitable number of times.

Note that in this construction the distribution of the stabilizer check weights is non-uniform, but logarithmically distributed.
Although it is unlikely that these codes can have a threshold it might still be worthwhile to find an efficient decoder to test whether the error suppression is competitive for relevant system sizes.
\begin{oproblem}
Can the codes of Bacon et~al.~\cite{bacon2017sparse} be efficiently decoded?
\end{oproblem}

\subsection{Approximate Codes from Spacetime Circuit Hamiltonians}\label{sec:spacetimecircuit}
An interesting approach was taken by Bohdanowicz et~al. in~\cite{bohdanowicz2019good}.
Similarly to the codes by Bacon et~al. discussed in \cref{sec:bacom_flammia_harrow_shi} they derive quantum codes from quantum circuits.
The parameters of their code are $[[n, \Omega(n/\polylog(n)), \Omega(n/\polylog(n))]]$.
They define their code as the ground-space of a local Hamiltonian where each term operates on~$9$ qubits and each qubit participates in $\polylog(n)$ many terms.
The codes are non-stabilizer codes, i.e. the terms of the Hamiltonian are not given by Pauli operators, so that many fault tolerance techniques developed for stabilizer codes do not apply.
For example, it is not clear how to measure the energy of each term of the Hamiltonian or how to process the information for a recovery.
Furthermore, the codes are \emph{approximate codes}, which means that the fidelity of the encoded state after a recovery is only~$1-\epsilon$, where $\epsilon\in o(1)$.

Their construction uses encoding circuits of good quantum codes of $\polylog$-depth which are guaranteed by \cite{brown2013short}.
This encoding circuit is mapped onto a local Hamiltonian which contains the valid computations of the circuit in its ground space~\cite{breuckmann2014space} and has a spectral gap which scales as $\Omega(1/n^\alpha)$ for some $\alpha > 0$.
The authors show that for arbitrary errors a recovery operation exists which restores the initial state with high fidelity.

{\change
A different approach to non-stabilizer codes was taken by Movassagh--Ouyang~\cite{movassagh2020constructing} who demonstrate how to map classical codes into the ground space of quantum spin chain Hamiltonians.
}

\begin{oproblem}
Can non-stabilizer codes and approximate codes give rise to practical and competitive fault tolerance schemes?
\end{oproblem}

\section{Challenges and Opportunities}\label{sec:challandopp}

\subsection{Reduction in Overhead}\label{sec:overhead_reduction}
A major achievement of fault-tolerant quantum computing is the \emph{threshold theorem}~\cite{kitaev1997quantum,knill1998resilient,aharonov2008fault} which shows that fault-tolerant quantum computation is possible with polylogarithmic overhead of physical qubits in the length of the computation.
A theorem due to Gottesman shows that it is even possible to perform quantum computation with only \emph{constant} overhead in resources~\cite{gottesman2014overhead}.

{\change
More precisely, Gottesman shows the following:
Assuming we have a family of LDPC codes with parameters $[[n_i,k_i,d_i]]$ such that (a)~it has a constant rate $\liminf_{i\rightarrow \infty} n_i/k_i = R > 0$, (b)~its elements are polynomially spaced, i.e. $0<n_i-n_{i-1} < n_i^\beta$ for some constant $\beta>0$, and (c)~there exists an efficient decoding algorithm which for suitably low noise parameters suppresses errors as $1/g(n_i)$ for $i \rightarrow \infty$ where~$g$ is some non-decreasing function.
Then any suitably large quantum circuit on $k$ qubits can be  approximated with arbitrary precision if the noise of the components is below a certain threshold parameter using at most~$\eta k/R$ physical qubits, where $\eta >1$ controls the threshold.
In particular, this result gives an exact upper bound compared to the earlier threshold theorems which can have very large constants hidden in the asymptotics~\cite{knill2005quantum}.
}

Fawzi--Grospellier--Leverrier showed that the assumptions of Gottesman's theorem can be satisfied~\cite{fawzi2018constant} using a hypergraph product code build from expander codes~\cite{leverrier2015quantum} decoded by a simple decoder that they call the \emph{small-set-flip decoder}.
It is very likely that other codes discussed in \cref{sec:geometric_constructions,sec:product_constructions} could fulfill the requirements of Gottesman's theorem as well.
The key to this is finding decoders which are sufficiently simple in order to be able to proof the required error suppression.
\begin{oproblem}
Which quantum LDPC codes can be used for Gottesman's constant overhead theorem?
\end{oproblem}

\subsection{Logical Operations}
Gottesman's theorem guarantees a constant overhead by performing operations sequentially with logical gates implemented using ancilla states.
However, it does require a minimum amount of logical qubits to become effective and this amount has yet to be determined.
Hence, there may be schemes which could potentially turn out to be more practical.
For an overview of the leading proposals of implementing operations on codes not discussed in the manuscript see~\cite{campbell2017roads}.

Bravyi--K\"onig showed that there is a trade-off between the implementability of constant-depth logical gates and the spacial locality in Euclidean space~\cite{bravyi2013classification}, see also~\cite{pastawski2015fault}.
A corollary of their result is that any code that is spatially local in two dimensions can only have constant-depth logical gates belonging to the Clifford group.
Therefore, in order to implement logical gates in codes like the surface code or two-dimensional color codes we need to execute circuits of depth scaling with the code size.

One could therefore expect that LDPC codes which are not bound by locality might offer an advantage.
Not much is known regarding logical gates for general LDPC codes.

Code deformations were considered for hyperbolic surface codes in order to perform CNOT gates~\cite{breuckmann2017hyperbolic}.
Krishna--Poulin~\cite{krishna2019fault} considered generalizations of code deformation techniques of the surface code to hypergraph product codes (cf.~\cref{sec:tensor_product}) in order to implement Clifford gates.
On the other hand, Burton--Browne~\cite{burton2020limitations} showed that it is not possible to obtain logical gates with circuits of depth one (transversal gates) outside of the Clifford group using hypergraph product codes.

A different approach is taken by Jochym--O'Connor~\cite{jochym2019fault} who showed that taking the tensor product of two suitable quantum codes with complementary sets of gates it is possible to perform the logical operations of either and thus obtain a fault-tolerant and universal set of gates.
Such a scheme may be an alternative to Gottesman's protocol~\cite{gottesman2014overhead} which achieves universality using ancillary states to obtain constant overhead.

\subsection{Decoding}\label{sec:decoding}
For a general stabilizer code it was shown by Iyer--Poulin that optimal decoding, i.e. maximizing the success probability of reversing the error, is \#P-complete~\cite{iyer2015hardness}.
However, it is often sufficient to consider a sub-optimal decoding algorithm, such as minimum-weight perfect matching for the surface code~\cite{dennis2002topological}.

Current decoding algorithms for the surface code or 2D~topological codes suffer from a large time complexity, although progress has been made in reducing the time complexity of decoding the surface code~\cite{delfosse2017almost,das2020scalable}.
Here, LDPC codes could offer an advantage.
First, the time complexity of decoding algorithms often depends on the number of physical bits. 
LDPC codes can achieve better encoding rates, offering the same level of protection, and consequently admit faster decoding.
For example, applying minimum-weight perfect matching to hyperbolic surface codes can yield significant performance improvements in comparison to 2D surface codes.
Second, LDPC codes offer simplified decoding algorithms, significantly decreasing the classical processing load and complexity compared to currently favored schemes. 
They can be implemented by simple logical gates and do not need complex processors and this would imply less heat dissipation into the system and could allow for the classical control hardware to be closer to the qubits.

A widely used decoding algorithm for classical codes is based on iterative message-passing on the Tanner graph and is called \emph{belief-propagation} (BP).
The BP decoder is very appealing due to its simplicity, which could benefit hardware implementations, as well as its versatility, as it can in principle be applied to arbitrary quantum LDPC codes.
Generally, BP does not work well when applied to Tanner graphs which contain small loops, a feature quantum codes necessarily have due to the commutativity constraint which introduces loops of length four (cf.~\cref{fig:shorcodetanner}).
Further, when applied to quantum codes BP tends to fail to converge as there exist many equivalent solutions up to the application of stabilizers.
These problems were addressed in~\cite{LEIFER20081899,poulin2008belief,poulin2008iterative,6145510,kuo2020refined,raveendran2019syndrome}.
In particular, Duclos-Cianci--Poulin combined BP with a renormalization decoder~\cite{duclos2010fast} and Panteleev--Kalachev combined BP with ordered statistics decoding, which showed good performance on a variety of quantum LDPC codes~\cite{panteleev2019degenerate}.
BP decoders were analyzed in numerical simulations for tensor products of classical codes~\cite{grospellier2018numerical,grospellier2020combining,roffe2020decoding} and to 4D hyperbolic codes~\cite{breuckmann2020single}.
As BP is widely used for classical codes one can draw from a wealth of literature.
For example, there has been rapid progress on efficient hardware implementations of BP~\cite{liang2011hardware,chen2002near}.

For classical codes it has been observed that expansion properties of the Tanner graph can lead to simple greedy decoding algorithms~\cite{sipser_expander_1996}.
Such greedy algorithms do not directly transfer to quantum codes.
However, Leverrier--Tillich--Z\'emor found a suitable generalization, called the small-set-flip decoder, which applies to tensor products of classical expander codes~\cite{leverrier2015quantum}.
Hastings showed that the expansion properties of 4D hyperbolic codes can be used for decoding using a local greedy procedure as well~\cite{hastings2013decoding}.

{\change
Delfosse--Londe--Beverland consider the union-find decoder, which was initially developed as an efficient decoder for the surface code, for decoding general quantum LDPC codes~\cite{delfosse2021toward}.}
Delfosse--Hastings combined the union-find decoder of the surface code~\cite{delfosse2017almost} with a look-up decoder of a small code of fixed size~\cite{delfosse2020union} applying it to the tensor product of both codes.
This raises the following question.
\begin{oproblem}
Is there a systematic approach to generalize decoders of classical codes to work for quantum codes based on their product?
\end{oproblem}
Bounds on the optimal decoding performance for tensor products of random classical codes were given in~\cite{kovalev2018numerical}.

A further potential advantage of LDPC codes over the surface code is \emph{single-shot decoding}~\cite{bombin2015single}.
As stabilizer check measurements are subject to noise they have to be repeated in order to build confidence \cite{dennis2002topological}.
Single-shot decoding refers to the property of some LDPC codes to exhibit robustness against such measurement errors, so that it is not necessary to repeat the stabilizer check measurements.
{\change
Numerical simulations of single-shot decoding were performed (under various assumptions and error models) for tensor product codes \cite{grospellier2021combining,quintavalle2020single} and 4D hyperbolic codes~\cite{breuckmann2020single}.
}

\subsection{Hardware Implementation}
A major concern often raised regarding the codes discussed in this manuscript is how they could be implemented in hardware. 
In the following section we discuss the main concerns and argue why we are optimistic about the potential of LDPC codes.

An important aspect of hardware implementation is that the maximal number of qubits involved in a stabilizer check should be low in order to keep the number of errors introduced down.
Although this number is constant for LDPC codes by definition, it can still be too high for practical purposes, although it is possible to reduce the stabilizer check weight using graph-based arguments \cite{hastings2016weight,hastingsreduction}.
Higgott--Breuckmann suggest an alternative construction by systematically breaking the stabilizer checks into smaller, so-called gauge checks, which do not commute, but from which the stabilizer check measurement can be inferred~\cite{higgott2020subsystem}.

However, the most obvious draw-back of LDPC codes comes with the question of how to lay out the physical qubits and their couplings in space.
As a proxy, we will discuss the layout of the Tanner graphs of codes.
Almost none of the Tanner graphs of the quantum LDPC codes discussed here are planar, with the exception of a planar variation of hyperbolic surface codes~\cite{breuckmann2016constructions}.
More severely, several codes discussed here do not have `nice' embeddings in Euclidean space, as their Tanner graphs have non-trivial expansion (although expanding graphs actually have been implemented in experiments~\cite{kollar2019hyperbolic}).

{\change A related concern is that some qubit hardware implementations and their couplings are only possible in a planar layout.}
While planar embeddings of the discussed quantum LDPC codes are generally not possible, it is possible to break up graphs into planar pieces which are then connected along a 1D line without intersections.
This can be done using \emph{book embeddings}~\cite{bilski1992embedding} where the vertices of the graph are arranged along a line (spine) and each edge is assigned a half-plane with the line as its boundary (page), such that no two edges on the same page intersect.
Clearly, the vertices do not have to be placed on the spine but can be pulled into the pages.
The number of pages should ideally not grow and it was shown in~\cite{dujmovic2015layouts} that there are indeed families of expander graphs even so that only three pages are sufficient for a book embedding.
However, the minimum number of pages for the Tanner graphs of codes discussed here are not known to us.

The viability of implementing quantum codes requiring non-local couplings depends on the hardware.
Currently, it is not settled which qubit architecture will succeed (see~\cite{qubitzoo} for an overview).
Hence, it is also not clear at this point in time how future quantum computing architectures will scale.
Although some proposals suggest that a large number of physical qubits may be placed in a single fridge~\cite{fowler2012surface},  it seems doubtful that arbitrary scaling inside a single fridge will be possible.
Other proposals pursue a modular architecture of interconnected modules linked by a photonic interface~\cite{monroe2013scaling,nigmatullin2016minimally,nickerson2014freely,rudolph2017optimistic}.
A modular approach would free us from spacial constraints, making LDPC codes competitive candidates for implementing quantum fault tolerance.
Other approaches to quantum computation, such as qubits coupled to a common cavity mode~\cite{PhysRevLett.75.3788,PhysRevA.94.053830}, even allow for direct, non-planar interactions between qubits.

In order to measure the stabilizer checks it is necessary to find a \emph{scheduling}, an ordering of the gates which couple the data qubits to an ancilla used for the measurement.
This ordering should not spread errors in order to be fault-tolerant and it should also be efficient to minimize the time of qubits idling.
Finding such circuits is a non-trivial task and, as far as we are aware, hyperbolic surface codes are the only finite-rate codes which have a known measurement schedule~\cite{conrad2018small,higgott2020subsystem}.
Finding such schedules will be challenging for random constructions.
\begin{oproblem}
Are there good measurement schedules for the LDPC codes discussed here?
\end{oproblem}

\section{Applications outside of QEC}\label{sec:applicationsoutside}
We have seen that quantum LDPC codes draw from many areas of mathematics, physics and computer science.
One could hope that quantum LDPC codes could in turn find use outside of quantum error correction and quantum fault tolerance.
Here we want to briefly highlight two examples where this is the case.

\subsection{Quantum Complexity Theory}
An important class in quantum complexity theory is QMA, an analogue of the classical complexity class NP, see \cite{kitaev2002classical,aharonov2002quantum}.
A prototypical QMA-complete problem is the $k$-local Hamiltonian problem (LocHam), see ~\cite{kitaev2002classical}.
It asks whether the ground state energy of a $k$-local Hamiltonian is either below~$a$ or above~$b$ where $b-a > 1/\poly(n)$ and can be seen as the quantum analogue of $3$-SAT.
One of the main achievements of classical complexity theory, the PCP theorem, also a admits a conjectural quantum version. The quantum PCP (qPCP) conjecture states that LocHam is equally hard when stated with a constant accuracy $b-a> \text{const.}$ instead of an inverse-polynomial accuracy, see \cite{arora1998probabilistic,dinur2007pcp}.

Hastings introduced the no low energy trivial state (NLTS) conjecture, a weakening of the qPCP conjecture~\cite{hastings2012trivial,freedman2013quantum}. 
It states that there a is family of local Hamiltonians acting on an increasing number of qubits such that the energy of any trivial state is below a universal constant.
The NLTS conjecture could be solved by construction quantum LDPC codes with linear distance for which there exist local Hamiltonians for which the energy of a quantum state is proportional to its distance from the ground-space of the Hamiltonian (quantum locally testable codes).
{\change
Towards solving the NLTS conjecture, Anshu--Nirkhe show that quantum LDPC codes with linear rate and polynomial distance has no trivial states of energy less than $o(n)$~\cite{anshu2020circuit}.
}
See~\cite{eldar_local_2017} for a zoo of the various complexity classes and their relation to quantum LDPC codes and~\cite{aharonov2013guest} for a comprehensive review of the qPCP and NLTS conjectures.

\subsection{Geometry}
In \cref{sec:geometric_constructions} we have seen that quantum LDPC codes can be constructed using tools from geometry.
More precisely, quantum codes can be defined from tessellations of manifolds such that the code properties are determined by the geometric properties.
Recently, Freedman--Hastings showed that the inverse is also possible~\cite{freedman2020building}.
Given a quantum LDPC code they construct manifolds of dimension $D=11$ such that geometric properties of the manifold are determined by the properties of the code.
Their work suggests that questions of systolic geometry can be answered using quantum LDPC code constructions.

\section{Conclusion and Outlook}
In this perspective, we gave an overview of the emerging field of quantum LDPC codes providing promising new approaches to quantum error correction.
Quantum LDPC codes use a plethora of techniques from mathematics, physics and computer science.
In particular, we showed how ideas from geometry and homological algebra shape the theory of quantum LDPC codes. 
The results discussed here make use of hyperbolic geometry, expander codes, algebraic topology, to name a few.  
The fast pace of new distance records in the last year suggests that one of the main goals of the field, the quest for quantum LDPC codes with constant encoding rate $k/n > \text{const.} > 0$ and linear distance $d \propto n$, may soon be in reach and that the next years may offer many exciting new developments.

Moreover, we discussed challenges and opportunities. 
In particular, the viability of quantum LDPC codes depends on future developments in hardware and
many problems in the implementation of scalable fault-tolerant quantum computation remain to be solved.
Low-latency classical control and fast decoding algorithms as well as inter-connectivity and wiring are challenging problems for the architecture of error-corrected post-NISQ quantum devices.
Quantum LDPC codes could play a decisive role in their realization.
Although the development of the surface code is ahead in many respects, quantum LDPC codes may well turn out to be better suited for the implementation of quantum computers in the mid- to long-term.

On a theoretical level, quantum LDPC codes may yield exciting applications in geometry, quantum complexity theory and potentially beyond, indicating that the flow of ideas can be reversed.

\vspace{1cm}
\emph{Acknowledgements}:
The authors would like thank the following people for helpful discussions: Matt Hastings, Gleb Kalachev, Anirudh Krishna, Alex Lubotzky, Pavel Panteleev, Joschua Ramette, Christophe Vuillot.
Special thanks to Barbara Terhal for valuable feedback on our manuscript.
NPB acknowledges support through the EPSRC Prosperity Partnership in Quantum Software for Simulation and Modelling (EP/S005021/1).

\section*{Appendix}
\subsection{Constructions for chain complexes}\label{app:chaincomplexes}
For the convenience of the reader, we describe various homological constructions, such as the different product codes from \cref{sec:tensor_product}, in greater detail.
\subsubsection{Chain complexes}

A chain complex $C=(C,\partial^C)$  of vector spaces over~$\mathbb{F}_2$ of length $n+1$ is a collection of vector spaces~$C_i$ and linear maps~$\partial^C_i$, called \emph{boundary operators},
\begin{center}
    \begin{tikzcd}
C=(C_{n} \arrow[r, "\partial^C_{n}"] & \cdots \arrow[r, "\partial^C_2"] & C_1 \arrow[r, "\partial^C_1"] & C_0)
\end{tikzcd}
\end{center}
fulfilling $\partial^C_{i+1}\partial^C_i=0.$
Often, the indices of boundary operators from the notation. For example, one simply writes $\partial^2=0.$ 

{\change To a chain complex $C$ one can associate the vector spaces
\begin{align*}
Z_i(C) &= \ker \partial_i\subset C_i\\
B_i(C) &= \im \partial_{i+1}\subset C_i\\
H_i(C) &=Z_i(C)/B_i(C)
\end{align*}
of  \emph{$i$-cycles}, \emph{$i$-boundaries} and the $i$-th \emph{homology}, respectively, whereas elements in $C_i$ are called \emph{$i$-chains}.

We assume that the spaces $C_i$ are equipped with bases of so called $i$-cells. This defines scalar products on each $C_i.$
We denote the linear dual of $C_i$ by  $C^i.$ Elements in $C^i$ are called cochains. The canonical basis allows to identify $C_i=C^i$ and one defines the vector spaces of
\begin{align*}
Z^i(C) &= \ker \partial_{i+1}^{tr}\subset C^i\\
B^i(C) &= \im \partial_{i}^{tr}\subset C^i\\
H^i(C) &=Z^i(C)/B^i(C)
\end{align*}
\emph{$i$-cocycles}, \emph{$i$-coboundaries} and the $i$-th \emph{cohomology} of the complex $C.$
The scalar product on $C_i$ and $C^i$ induces a well-defined and non-degenerate pairing of $H_i(C)$ and $H^i(C)$ since
$B^i(C)=Z_i(C)^\perp.$ This implies that $\dim H_i(C)=\dim H^i(C).$

For each $i,$ a quantum code can be extracted from the cell complex $C$ with parity check matrices $H_X$ and $H_Z$ corresponding to the operators $\partial_{i}$ and $\partial_{i+1}^{tr}.$ 
The $X$-checks, physical qubits and $Z$-checks then correspond to the $i+1$-, $i$- and $i-1$-cells of $C.$ 
The non-trivial logical $X$- and $Z$-operators correspond to the elements in $H^i(C)$ and $H_i(C),$ respectively.
See also \cref{sec:chain_complexes} for the relation of chain complexes and (quantum) codes.}

\subsubsection{Total Complex of Double Complexes}
\begin{figure}
    \centering
    \includegraphics{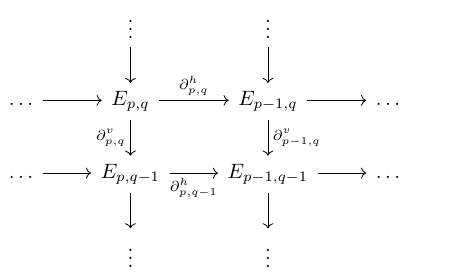}
    \caption{A part of a double complex.}
    \label{fig:doublecomplex}
\end{figure}

An interesting way of constructing chain complexes is by the total complex construction of a double complex. 
A double complex $E=(E_{\bullet,\bullet},\partial^{v},\partial^{h})$ is an array of vector spaces $E_{p,q}$ equipped with vertical and horizontal maps
\begin{align*}
\partial_{p,q}^{v}&:E_{p,q}\to E_{p,q-1}\text{ and }\\
\partial_{p,q}^{h}&:E_{p,q}\to E_{p-1,q}
\end{align*}
such that $\partial^{v}$ and $\partial^{h}$ are commuting boundary operators
\begin{align}\label{eqn:doublecomplex}
(\partial^{v})^{2}=(\partial^{h})^{2}=0 \text{ and } \partial^{v}\partial^{h}=\partial^{h}\partial^{v}.
\end{align}
It is convenient to visualise the double complex laid out on a two-dimensional grid
where each square is required to commute and composing two maps in the same direction yields zero, see \cref{fig:doublecomplex}.
To each double complex $E$, one can associate a chain complex $\Tot(E),$ called the \emph{total complex}, where the $n$-th degree is given by the direct sum over the $n$-th diagonal in $E,$ so
$$\Tot(E)_n=\bigoplus_{p+q=n}E_{p,q},$$
and the boundary operators of $\Tot(E)$ are the sum of all boundary operators passing from one diagonal to the next. The requirement that the boundary operators of $\Tot(E)$ square to zero directly follows from \cref{eqn:doublecomplex}.

We note that these concepts immediately generalize to higher dimensions. See \cref{fig:cubecomplex} for an example of a triple complex.

\subsubsection{Tensor product of chain complexes and hypergraph products}\label{app:tensorproduct}
\begin{figure}
    \centering
    \includegraphics{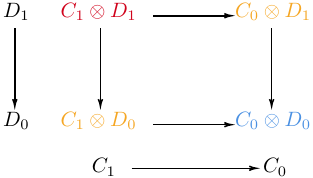}
    \caption{A double complex arising as the tensor product of two complexes of length two. The different color symbolize the different degrees in the total complex which is a chain complex of length three. The diagram should be compared with Figure \ref{fig:hypergraphproduct}.}
    \label{fig:tensordoublecomplextikz}
\end{figure}

\begin{figure}
    \centering
    \includegraphics{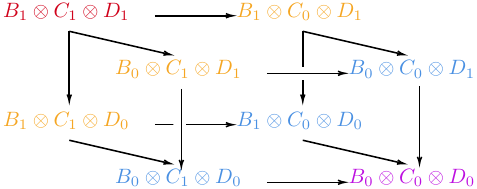}
    \caption{A triple complex arising as the tensor product of three complexes of length two. The different color symbolize the different degrees in the total complex which is a chain complex of length four.}
    \label{fig:cubecomplex}
\end{figure}
Let $C$ and $D$ be complexes of length $n$ and $m,$ respectively. The \emph{tensor product} $C\otimes D$ is a chain complex of length $n+m-1$ and can be seen as a generalization of the tensor product of vector spaces to complexes.

There is an elegant and quick definition of tensor product $C\otimes D$ in terms of double complexes. Namely, the  \emph{tensor product double complex} $C\boxtimes D$ is defined by
$$(C\boxtimes D)_{p,q}=C_p\otimes C_q$$ with boundary operators
$\partial^v=\partial^C\otimes \id_D \text{ and } \partial^h=\id_C\otimes \partial^D.$ Note, that after choosing a basis, the tensor product of two maps is given by the \emph{Kronecker product} of the corresponding matrices.
Then $C\otimes D=\Tot(C\boxtimes D)$ is the total complex of this double complex.

For example, if $C$ and $D$ be chain complexes of length $2$
then the tensor product $C\otimes D$ of $C$ and $D$ is a chain complex of length $3$
\begin{center}
\begin{tikzcd}
C_1\otimes D_1 \arrow[r] 
& C_0\otimes D_1\oplus C_1\otimes D_0 \arrow[r] & C_0\otimes D_0.
\end{tikzcd}
\end{center}
with boundary operators
\begin{align*}
\begin{pmatrix} \partial^C_1\otimes \id_{D_1}\\ \id_{C_1}\otimes\partial^D_1  \end{pmatrix}\text{ and }\begin{pmatrix}\id_{C_0}\otimes\partial^D_1, & \partial^C_1\otimes \id_{D_0}  \end{pmatrix},
\end{align*}
respectively. The relation between this direct definition and the definition via total complexes is visualised in \cref{fig:tensordoublecomplextikz}.

{\change The tensor product of chain complexes can be used to define the hypergraph product of two classical codes.
Assume that the codes have $n_i$ bits and $r_i$ checks for $i=1,2$ and parity check matrices $H$ and $H'.$ 
To the codes one associates two chain complexes $C$ and $D$ of length $2$ with boundary operators $\partial_C^1$ and $\partial_D^1$ represented by the matrices $H$ and $(H')^{tr}.$ 
Then the hypergraph product arises as the tensor product $C\otimes D.$ 
In particular, the parity check matrices $H_Z^{tr}$ and $H_X$ of the hypergraph product code are given by
\begin{align*}
\begin{pmatrix} H \otimes I_{r_2}\\ I_{n_1}\otimes (H')^{tr}  \end{pmatrix}\text{ and }\begin{pmatrix}I_{r_1}\otimes(H')^{tr}, & H\otimes I_{n_2}  \end{pmatrix}.
\end{align*}}

The homology of a tensor product is subject of the Künneth formula
$$H_n(C\otimes D)=\bigoplus_{p+q=n} H_p(C)\otimes H_q(D).$$
This allows to easily compute the number of logical qubits of a tensor product/hypergraph product quantum code.

Moreover, one can also take iterated tensor products of chain complexes, which correspond to higher dimensional complexes. See \cref{fig:cubecomplex} for an example of a tensor product of three complexes.

\subsubsection{Fiber Bundle Codes}\label{app:fiberbundle}
In topology, a fiber bundle is generalization of a product of two spaces, which allows for non-trivial twists. It consists of a projection map $\pi: E\to B$ from its \emph{total space} to its \emph{base}, such that the \emph{fibers} $F=\pi^{-1}(x)$ are isomorphic and $E$ is a product of the base and the fiber locally. A protopical example of a fiber bundle is the Klein bottle, which admits a map to a circle whose fiber is also a circle. The Klein bottle is a twisted version of a product of two circles, the torus.

Fiber bundle codes mimic this topological concept and were introduced by Hastings--Haah-O'Donnell~\cite{hastings2020fiber} to build quantum LDPC codes breaking the $\sqrt{n}\operatorname{polylog}(n)$ distance barrier.
The idea behind fiber bundle codes is to introduce a \emph{twist} in the boundary operators of tensor product codes, in order to increase the distance of the resulting code. 

Let $B$ and $F$ be two complexes of length two equipped with bases, we will refer to as  \emph{base} and \emph{fiber} complex respectively. Further, let 
$\varphi$ be a function (called \emph{twist}) that associates to any pair of incident basis vectors of $B_0$ and $B_1$ an automorphism of the fiber.

Then the fiber bundle code $B\otimes_\varphi F$ is a chain complex with the same underlying vector spaces as $B\otimes F$ but twisted boundary operators
\begin{align*}
\begin{pmatrix} \partial_\varphi\\ \id_{B_1}\otimes\partial^F_1  \end{pmatrix}\text{ and }\begin{pmatrix}\partial_\varphi & \partial^B_1\otimes \id_{F_0}  \end{pmatrix},
\end{align*}
where
\begin{align*}
\partial_{\varphi}(b^{1}\otimes f)=\sum_{b^{0}\in\partial^{B}b^{1}}b^{0}\otimes \varphi(b^{1},b^{0})(f)
\end{align*}
for basis vectors $b_i\in B_i$ and $f\in F_0$ or $f\in F_1.$ In particular, if $\varphi=1$ then the fiber bundle $B\otimes_\varphi F=B\otimes F$ specializes to the tensor product. In fact, the fiber bundle code $B\otimes_\varphi F$ can be interpreted as the total complex of the fiber bundle double complex $B\otimes_\varphi F$ with the obvious boundary operators.

Requiring that $\partial_1^B$ is surjective and some additional technical conditions~\cite{hastings2020fiber}, one can show that 
$$H^1(B\otimes_\varphi F)=H^1(B).$$
Hence the number of logical qubits in $B\otimes_\varphi F$ coincides with the number of encoded bits in the code associated the complex~$B.$
In \cite{hastings2020fiber} the construction was applied to a random code~$B$ as base, a repetition code with cyclic symmetry~$F$ as fiber and random twist $\varphi.$

\subsubsection{Lifted Product Codes}\label{app:liftedproducts}
Lifted product codes, introduced by Panteleev--Kalachev~\cite{panteleev2019degenerate,panteleev2020quantum} are based on the observation that the tensor product of vector spaces or Kronecker product of matrices extend to modules over algebras. This more general definition can be used to construct quantum codes.

Let $R\subset \mathbb{F}_2^{\ell\times \ell}$
be a commutative subalgebra of the ring of $\ell\times\ell$ matrices over $\mathbb{F}_2.$
Now let $A\in R^{n\times m}$ and~$B\in R^{k\times l}$ be matrices with entries in the algebra $R.$ Equivalently, $A$ and $B$ can be interpreted as matrices $\tilde{A}\in \mathbb{F}_2^{\ell n\times \ell m}$ and $\tilde{B}\in \mathbb{F}_2^{\ell k\times \ell l}$ whose blocks of size $\ell\times\ell$ are elements in the algebra $R.$ 

{\change Then the lifted product quantum code is defined in terms of the check matrices $H_X$ and $H_Z$ given by
$$\begin{pmatrix}
 I_m^R\otimes_R B &
 A\otimes_R I_l^R 
\end{pmatrix}\text{ and } 
\begin{pmatrix} A\otimes_R I_k^R & I_n^R\otimes_R B\end{pmatrix}$$
where $\otimes_R$ denotes the Kronecker product of matrices over $R,$ $I_q^R$ denotes the $q\times q$ identity matrix over $R$ and the resulting matrices are interpreted as matrices over $\mathbb{F}_2.$ 

Equivalently, the parity check matrices and $H_X$ and $H_Z$ can be obtained by first considering the matrices
$$\begin{pmatrix}
 I_{\ell m}\otimes \tilde{B} &
 \tilde{A}\otimes I_{\ell l} 
\end{pmatrix}\text{ and } 
\begin{pmatrix} \tilde{A}\otimes I_{\ell k} & I_{\ell n}\otimes \tilde{B}\end{pmatrix}$$
using the usual Kronecker product. The resulting matrices can be subdivided into submatrices of size $\ell^2\times\ell^2.$ Then, one collapses each of these these submatrices to matrices of size $\ell\times\ell$ by adding all of their $\ell\times\ell$ submatrices of size $\ell\times\ell$ together.}

Note that the number of $X$-, $Z$-checks and logical qubits in the lifted product is smaller by factor of $\ell$ then in the corresponding tensor/hypergraph product.

The lifted product can also be written as a tensor product of chain complexes. Here, one has to interpret the two classical codes as chain complexes of length two \emph{over the algebra} $R$ and use the tensor product over $R.$ This is closely related to the definition of balanced product codes, see \cref{app:balancedproducts}.

In \cite{panteleev2019degenerate, panteleev2020quantum} the lifted product construction is applied mostly in the case where $R$ is the algebra of circulant matrices, that is, the algebra generated by the cyclic shift matrix~$x$ of the $\ell$-cycle. 
Matrices with entries in $R$ can be for example constructed from the incidence matrix of graph with an $\ell$-fold cyclic covering, see \cite{agarwal_expansion_2019} and associated expander codes. Indeed, quantum LDPC codes with almost linear distance are obtain in~\cite{panteleev2020quantum} by taking the lifted product of an Sipser--Spielman code on a random cyclic covering of a random expander graph with the matrix $1+x$ of the repetition code.

\subsubsection{Balanced Product Codes}\label{app:balancedproducts}
The balanced product is a topological construction, which associates to two spaces $X,Y$ with right and left actions of a groups $G,$ respectively, a space $X\times_G Y.$ The space $X\times_G Y$ is defined as the quotient $(X\times Y)/G$ of the Cartesian product, where $G$ acts on $X\times Y$ via $g\cdot(x,y)=(xg^{-1},gy).$ 
The balanced product is often used to construct fiber bundles from principal bundles in physics and topology, see~\cite{nakahara2003geometry}. 
Namely, the natural projection $\pi: X\times_G Y\to X/G$ is a fiber bundle with base~$X/G$ and fiber~$Y,$ under some technical assumption.

Balanced product codes mimic this concept and were introduced by Breuckmann--Eberhardt~\cite{breuckmann2020balanced} to build quantum LDPC codes.

If $G$ is a group acting on a vector space $V$ and $W$ from the left and right, respectively, one can form the tensor product over $G$ via 
$$V\otimes_G W=V\otimes W/\langle v\cdot g\otimes w - v\otimes g\cdot w \rangle.$$
Similarly, to the tensor product for vector spaces, this definition extends to chain complexes. Let $C$ and $D$ be chain complexes with a right and left action of $G,$ respectively, that is In other words, $G$ acts on the individual spaces $C_i, D_i$ and commutes with all boundary operators. Then one can form the complex $C\otimes_G D$ which is the total complex of the double complex $C\boxtimes_G D.$

Under the assumption that $G$ is a finite group of odd order, there is a Künneth formula
$$H_n(C\otimes_G D)=\bigoplus_{p+q=n} H_p(C)\otimes_G H_q(D).$$
In the case that $G$ is a commutative group with a free action on each vector space $C_i, D_i$ the balanced product specializes to a lifted product.

The balanced product construction was applied by Breuckmann--Eberhardt to construct quantum LDPC codes from  highly symmetrical Sipser--Spielman codes and a repetition code with cyclic symmetry. To construct Sipser--Spielman codes, the authors used Cayley graphs of $\operatorname{PGL}(2,q)$, whose automorphism are exactly this group.
\subsubsection{Relation of Fiber Bundle and Balanced Product Codes}
As mentioned above, fiber bundles and balanced products are closely related concepts in topology. Similarly, balanced product, lifted product and fiber bundle codes are closely related, see \cref{fig:balancedproduct}. In fact, the code families breaking the $\sqrt{n}\polylog(n)$ distance barrier described in \cref{sec:tensor_product} can be interpreted in all three setups.

Let us illustrate the relationship in a topological example. Denote by $X=S^1$ the circle and let $G=\mathbb{Z}_{2}$ act on $X$ via a rotation by $\pi.$ The quotient space $X/G=S^{1}$ is also a circle and $\pi: X\to X/H$ a $2$-fold covering. Now let  $G$ act on another circle $Y=S^{1}$ by reflection along the $x$-axis.
Then the associated balanced product
$S^{1}\times_{\mathbb{Z}_{2}}S^{1}$
is a \emph{Klein bottle} which is a fiber bundle over the circle
$$\pi_{S^1}:S^{1}\times_{\mathbb{Z}_{2}}S^{1}\to S^{1}$$
with fiber $S^{1}.$ By choosing $G$-equivariant tessellations of~$X$ and~$Y,$ one obtains quantum codes, which can be interpreted as balanced product, lifted product and fiber bundle codes.
Similarly, it is often possible to relate similar such code constructions to each other.

\bibliography{library.bib}{}

\begin{thebibliography}{147}%
\makeatletter
\providecommand \@ifxundefined [1]{%
 \@ifx{#1\undefined}
}%
\providecommand \@ifnum [1]{%
 \ifnum #1\expandafter \@firstoftwo
 \else \expandafter \@secondoftwo
 \fi
}%
\providecommand \@ifx [1]{%
 \ifx #1\expandafter \@firstoftwo
 \else \expandafter \@secondoftwo
 \fi
}%
\providecommand \natexlab [1]{#1}%
\providecommand \enquote  [1]{``#1''}%
\providecommand \bibnamefont  [1]{#1}%
\providecommand \bibfnamefont [1]{#1}%
\providecommand \citenamefont [1]{#1}%
\providecommand \href@noop [0]{\@secondoftwo}%
\providecommand \href [0]{\begingroup \@sanitize@url \@href}%
\providecommand \@href[1]{\@@startlink{#1}\@@href}%
\providecommand \@@href[1]{\endgroup#1\@@endlink}%
\providecommand \@sanitize@url [0]{\catcode `\\12\catcode `\$12\catcode
  `\&12\catcode `\#12\catcode `\^12\catcode `\_12\catcode `\%12\relax}%
\providecommand \@@startlink[1]{}%
\providecommand \@@endlink[0]{}%
\providecommand \url  [0]{\begingroup\@sanitize@url \@url }%
\providecommand \@url [1]{\endgroup\@href {#1}{\urlprefix }}%
\providecommand \urlprefix  [0]{URL }%
\providecommand \Eprint [0]{\href }%
\providecommand \doibase [0]{http://dx.doi.org/}%
\providecommand \selectlanguage [0]{\@gobble}%
\providecommand \bibinfo  [0]{\@secondoftwo}%
\providecommand \bibfield  [0]{\@secondoftwo}%
\providecommand \translation [1]{[#1]}%
\providecommand \BibitemOpen [0]{}%
\providecommand \bibitemStop [0]{}%
\providecommand \bibitemNoStop [0]{.\EOS\space}%
\providecommand \EOS [0]{\spacefactor3000\relax}%
\providecommand \BibitemShut  [1]{\csname bibitem#1\endcsname}%
\let\auto@bib@innerbib\@empty
%</preamble>
\bibitem [{\citenamefont {Arute}\ \emph {et~al.}(2019)\citenamefont {Arute},
  \citenamefont {Arya}, \citenamefont {Babbush}, \citenamefont {Bacon},
  \citenamefont {Bardin}, \citenamefont {Barends}, \citenamefont {Biswas},
  \citenamefont {Boixo}, \citenamefont {Brandao}, \citenamefont {Buell} \emph
  {et~al.}}]{arute2019quantum}%
  \BibitemOpen
  \bibfield  {author} {\bibinfo {author} {\bibfnamefont {F.}~\bibnamefont
  {Arute}}, \bibinfo {author} {\bibfnamefont {K.}~\bibnamefont {Arya}},
  \bibinfo {author} {\bibfnamefont {R.}~\bibnamefont {Babbush}}, \bibinfo
  {author} {\bibfnamefont {D.}~\bibnamefont {Bacon}}, \bibinfo {author}
  {\bibfnamefont {J.~C.}\ \bibnamefont {Bardin}}, \bibinfo {author}
  {\bibfnamefont {R.}~\bibnamefont {Barends}}, \bibinfo {author} {\bibfnamefont
  {R.}~\bibnamefont {Biswas}}, \bibinfo {author} {\bibfnamefont
  {S.}~\bibnamefont {Boixo}}, \bibinfo {author} {\bibfnamefont {F.~G.}\
  \bibnamefont {Brandao}}, \bibinfo {author} {\bibfnamefont {D.~A.}\
  \bibnamefont {Buell}},  \emph {et~al.},\ }\href@noop {} {\bibfield  {journal}
  {\bibinfo  {journal} {Nature}\ }\textbf {\bibinfo {volume} {574}},\ \bibinfo
  {pages} {505} (\bibinfo {year} {2019})}\BibitemShut {NoStop}%
\bibitem [{\citenamefont {Zhong}\ \emph {et~al.}(2020)\citenamefont {Zhong},
  \citenamefont {Wang}, \citenamefont {Deng}, \citenamefont {Chen},
  \citenamefont {Peng}, \citenamefont {Luo}, \citenamefont {Qin}, \citenamefont
  {Wu}, \citenamefont {Ding}, \citenamefont {Hu} \emph
  {et~al.}}]{zhong2020quantum}%
  \BibitemOpen
  \bibfield  {author} {\bibinfo {author} {\bibfnamefont {H.-S.}\ \bibnamefont
  {Zhong}}, \bibinfo {author} {\bibfnamefont {H.}~\bibnamefont {Wang}},
  \bibinfo {author} {\bibfnamefont {Y.-H.}\ \bibnamefont {Deng}}, \bibinfo
  {author} {\bibfnamefont {M.-C.}\ \bibnamefont {Chen}}, \bibinfo {author}
  {\bibfnamefont {L.-C.}\ \bibnamefont {Peng}}, \bibinfo {author}
  {\bibfnamefont {Y.-H.}\ \bibnamefont {Luo}}, \bibinfo {author} {\bibfnamefont
  {J.}~\bibnamefont {Qin}}, \bibinfo {author} {\bibfnamefont {D.}~\bibnamefont
  {Wu}}, \bibinfo {author} {\bibfnamefont {X.}~\bibnamefont {Ding}}, \bibinfo
  {author} {\bibfnamefont {Y.}~\bibnamefont {Hu}},  \emph {et~al.},\
  }\href@noop {} {\bibfield  {journal} {\bibinfo  {journal} {Science}\ }\textbf
  {\bibinfo {volume} {370}},\ \bibinfo {pages} {1460} (\bibinfo {year}
  {2020})}\BibitemShut {NoStop}%
\bibitem [{\citenamefont {Jurcevic}\ \emph {et~al.}(2021)\citenamefont
  {Jurcevic}, \citenamefont {Javadi-Abhari}, \citenamefont {Bishop},
  \citenamefont {Lauer}, \citenamefont {Borgorin}, \citenamefont {Brink},
  \citenamefont {Capelluto}, \citenamefont {Gunluk}, \citenamefont {Itoko},
  \citenamefont {Kanazawa} \emph {et~al.}}]{jurcevic2021demonstration}%
  \BibitemOpen
  \bibfield  {author} {\bibinfo {author} {\bibfnamefont {P.}~\bibnamefont
  {Jurcevic}}, \bibinfo {author} {\bibfnamefont {A.}~\bibnamefont
  {Javadi-Abhari}}, \bibinfo {author} {\bibfnamefont {L.~S.}\ \bibnamefont
  {Bishop}}, \bibinfo {author} {\bibfnamefont {I.}~\bibnamefont {Lauer}},
  \bibinfo {author} {\bibfnamefont {D.}~\bibnamefont {Borgorin}}, \bibinfo
  {author} {\bibfnamefont {M.}~\bibnamefont {Brink}}, \bibinfo {author}
  {\bibfnamefont {L.}~\bibnamefont {Capelluto}}, \bibinfo {author}
  {\bibfnamefont {O.}~\bibnamefont {Gunluk}}, \bibinfo {author} {\bibfnamefont
  {T.}~\bibnamefont {Itoko}}, \bibinfo {author} {\bibfnamefont
  {N.}~\bibnamefont {Kanazawa}},  \emph {et~al.},\ }\href@noop {} {\bibfield
  {journal} {\bibinfo  {journal} {Quantum Science and Technology}\ } (\bibinfo
  {year} {2021})}\BibitemShut {NoStop}%
\bibitem [{\citenamefont {Sch{\"o}nhage}(1979)}]{schonhage1979power}%
  \BibitemOpen
  \bibfield  {author} {\bibinfo {author} {\bibfnamefont {A.}~\bibnamefont
  {Sch{\"o}nhage}},\ }in\ \href@noop {} {\emph {\bibinfo {booktitle}
  {International Colloquium on Automata, Languages, and Programming}}}\
  (\bibinfo {organization} {Springer},\ \bibinfo {year} {1979})\ pp.\ \bibinfo
  {pages} {520--529}\BibitemShut {NoStop}%
\bibitem [{\citenamefont {Landauer}(1996)}]{landauer1996physical}%
  \BibitemOpen
  \bibfield  {author} {\bibinfo {author} {\bibfnamefont {R.}~\bibnamefont
  {Landauer}},\ }\href@noop {} {\bibfield  {journal} {\bibinfo  {journal}
  {Physics letters A}\ }\textbf {\bibinfo {volume} {217}},\ \bibinfo {pages}
  {188} (\bibinfo {year} {1996})}\BibitemShut {NoStop}%
\bibitem [{\citenamefont {Aaronson}(2013)}]{aaronson2013quantum}%
  \BibitemOpen
  \bibfield  {author} {\bibinfo {author} {\bibfnamefont {S.}~\bibnamefont
  {Aaronson}},\ }\href@noop {} {\emph {\bibinfo {title} {Quantum computing
  since Democritus}}}\ (\bibinfo  {publisher} {Cambridge University Press},\
  \bibinfo {year} {2013})\BibitemShut {NoStop}%
\bibitem [{\citenamefont {Shor}(1995)}]{shor1995scheme}%
  \BibitemOpen
  \bibfield  {author} {\bibinfo {author} {\bibfnamefont {P.~W.}\ \bibnamefont
  {Shor}},\ }\href@noop {} {\bibfield  {journal} {\bibinfo  {journal} {Physical
  review A}\ }\textbf {\bibinfo {volume} {52}},\ \bibinfo {pages} {R2493}
  (\bibinfo {year} {1995})}\BibitemShut {NoStop}%
\bibitem [{\citenamefont {Kitaev}(1997)}]{kitaev1997quantum}%
  \BibitemOpen
  \bibfield  {author} {\bibinfo {author} {\bibfnamefont {A.~Y.}\ \bibnamefont
  {Kitaev}},\ }\href@noop {} {\bibfield  {journal} {\bibinfo  {journal}
  {Uspekhi Matematicheskikh Nauk}\ }\textbf {\bibinfo {volume} {52}},\ \bibinfo
  {pages} {53} (\bibinfo {year} {1997})}\BibitemShut {NoStop}%
\bibitem [{\citenamefont {Kitaev}(2003)}]{kitaev2003fault}%
  \BibitemOpen
  \bibfield  {author} {\bibinfo {author} {\bibfnamefont {A.~Y.}\ \bibnamefont
  {Kitaev}},\ }\href@noop {} {\bibfield  {journal} {\bibinfo  {journal} {Annals
  of Physics}\ }\textbf {\bibinfo {volume} {303}},\ \bibinfo {pages} {2}
  (\bibinfo {year} {2003})}\BibitemShut {NoStop}%
\bibitem [{\citenamefont {Freedman}\ and\ \citenamefont
  {Meyer}(2001)}]{freedman2001projective}%
  \BibitemOpen
  \bibfield  {author} {\bibinfo {author} {\bibfnamefont {M.~H.}\ \bibnamefont
  {Freedman}}\ and\ \bibinfo {author} {\bibfnamefont {D.~A.}\ \bibnamefont
  {Meyer}},\ }\href@noop {} {\bibfield  {journal} {\bibinfo  {journal}
  {Foundations of Computational Mathematics}\ }\textbf {\bibinfo {volume}
  {1}},\ \bibinfo {pages} {325} (\bibinfo {year} {2001})}\BibitemShut {NoStop}%
\bibitem [{\citenamefont {Bravyi}\ and\ \citenamefont
  {Kitaev}(1998)}]{bravyi1998quantum}%
  \BibitemOpen
  \bibfield  {author} {\bibinfo {author} {\bibfnamefont {S.~B.}\ \bibnamefont
  {Bravyi}}\ and\ \bibinfo {author} {\bibfnamefont {A.~Y.}\ \bibnamefont
  {Kitaev}},\ }\href@noop {} {\  (\bibinfo {year} {1998})},\ \Eprint
  {http://arxiv.org/abs/quant-ph/9811052} {arXiv:quant-ph/9811052 [quant-ph]}
  \BibitemShut {NoStop}%
\bibitem [{\citenamefont {Chen}\ \emph {et~al.}(2021)\citenamefont {Chen},
  \citenamefont {Satzinger}, \citenamefont {Atalaya}, \citenamefont {Korotkov},
  \citenamefont {Dunsworth}, \citenamefont {Sank}, \citenamefont {Quintana},
  \citenamefont {McEwen}, \citenamefont {Barends}, \citenamefont {Klimov} \emph
  {et~al.}}]{chen2021exponential}%
  \BibitemOpen
  \bibfield  {author} {\bibinfo {author} {\bibfnamefont {Z.}~\bibnamefont
  {Chen}}, \bibinfo {author} {\bibfnamefont {K.~J.}\ \bibnamefont {Satzinger}},
  \bibinfo {author} {\bibfnamefont {J.}~\bibnamefont {Atalaya}}, \bibinfo
  {author} {\bibfnamefont {A.~N.}\ \bibnamefont {Korotkov}}, \bibinfo {author}
  {\bibfnamefont {A.}~\bibnamefont {Dunsworth}}, \bibinfo {author}
  {\bibfnamefont {D.}~\bibnamefont {Sank}}, \bibinfo {author} {\bibfnamefont
  {C.}~\bibnamefont {Quintana}}, \bibinfo {author} {\bibfnamefont
  {M.}~\bibnamefont {McEwen}}, \bibinfo {author} {\bibfnamefont
  {R.}~\bibnamefont {Barends}}, \bibinfo {author} {\bibfnamefont {P.~V.}\
  \bibnamefont {Klimov}},  \emph {et~al.},\ }\href@noop {} {\bibfield
  {journal} {\bibinfo  {journal} {arXiv preprint arXiv:2102.06132}\ } (\bibinfo
  {year} {2021})},\ \Eprint {http://arxiv.org/abs/2102.06132} {arXiv:2102.06132
  [quant-ph]} \BibitemShut {NoStop}%
\bibitem [{\citenamefont {Calderbank}\ and\ \citenamefont
  {Shor}(1996{\natexlab{a}})}]{calderbank1996good}%
  \BibitemOpen
  \bibfield  {author} {\bibinfo {author} {\bibfnamefont {A.~R.}\ \bibnamefont
  {Calderbank}}\ and\ \bibinfo {author} {\bibfnamefont {P.~W.}\ \bibnamefont
  {Shor}},\ }\href@noop {} {\bibfield  {journal} {\bibinfo  {journal} {Physical
  Review A}\ }\textbf {\bibinfo {volume} {54}},\ \bibinfo {pages} {1098}
  (\bibinfo {year} {1996}{\natexlab{a}})}\BibitemShut {NoStop}%
\bibitem [{\citenamefont {Ashikhmin}\ \emph {et~al.}(2001)\citenamefont
  {Ashikhmin}, \citenamefont {Litsyn},\ and\ \citenamefont
  {Tsfasman}}]{ashikhmin2001asymptotically}%
  \BibitemOpen
  \bibfield  {author} {\bibinfo {author} {\bibfnamefont {A.}~\bibnamefont
  {Ashikhmin}}, \bibinfo {author} {\bibfnamefont {S.}~\bibnamefont {Litsyn}}, \
  and\ \bibinfo {author} {\bibfnamefont {M.~A.}\ \bibnamefont {Tsfasman}},\
  }\href@noop {} {\bibfield  {journal} {\bibinfo  {journal} {Physical Review
  A}\ }\textbf {\bibinfo {volume} {63}},\ \bibinfo {pages} {032311} (\bibinfo
  {year} {2001})}\BibitemShut {NoStop}%
\bibitem [{\citenamefont {Gallager}(1962)}]{gallager1962low}%
  \BibitemOpen
  \bibfield  {author} {\bibinfo {author} {\bibfnamefont {R.}~\bibnamefont
  {Gallager}},\ }\href@noop {} {\bibfield  {journal} {\bibinfo  {journal} {IRE
  Transactions on information theory}\ }\textbf {\bibinfo {volume} {8}},\
  \bibinfo {pages} {21} (\bibinfo {year} {1962})}\BibitemShut {NoStop}%
\bibitem [{\citenamefont {MacKay}\ and\ \citenamefont
  {Neal}(1996)}]{mackay1996near}%
  \BibitemOpen
  \bibfield  {author} {\bibinfo {author} {\bibfnamefont {D.~J.}\ \bibnamefont
  {MacKay}}\ and\ \bibinfo {author} {\bibfnamefont {R.~M.}\ \bibnamefont
  {Neal}},\ }\href@noop {} {\bibfield  {journal} {\bibinfo  {journal}
  {Electronics letters}\ }\textbf {\bibinfo {volume} {32}},\ \bibinfo {pages}
  {1645} (\bibinfo {year} {1996})}\BibitemShut {NoStop}%
\bibitem [{\citenamefont {Iliev}\ \emph {et~al.}(2008)\citenamefont {Iliev},
  \citenamefont {Hristov}, \citenamefont {Zahariev},\ and\ \citenamefont
  {Iliev}}]{iliev2008application}%
  \BibitemOpen
  \bibfield  {author} {\bibinfo {author} {\bibfnamefont {T.~B.}\ \bibnamefont
  {Iliev}}, \bibinfo {author} {\bibfnamefont {G.~V.}\ \bibnamefont {Hristov}},
  \bibinfo {author} {\bibfnamefont {P.~Z.}\ \bibnamefont {Zahariev}}, \ and\
  \bibinfo {author} {\bibfnamefont {M.~P.}\ \bibnamefont {Iliev}},\ }in\
  \href@noop {} {\emph {\bibinfo {booktitle} {Novel Algorithms and Techniques
  In Telecommunications, Automation and Industrial Electronics}}}\ (\bibinfo
  {publisher} {Springer},\ \bibinfo {year} {2008})\ pp.\ \bibinfo {pages}
  {532--536}\BibitemShut {NoStop}%
\bibitem [{\citenamefont {Gottesman}(2014)}]{gottesman2014overhead}%
  \BibitemOpen
  \bibfield  {author} {\bibinfo {author} {\bibfnamefont {D.}~\bibnamefont
  {Gottesman}},\ }\href {\doibase 10.26421/QIC14.15-16} {\bibfield  {journal}
  {\bibinfo  {journal} {Quantum Information and Computation}\ ,\ \bibinfo
  {pages} {1338}} (\bibinfo {year} {2014})}\BibitemShut {NoStop}%
\bibitem [{\citenamefont {Knill}\ \emph {et~al.}(1998)\citenamefont {Knill},
  \citenamefont {Laflamme},\ and\ \citenamefont {Zurek}}]{knill1998resilient}%
  \BibitemOpen
  \bibfield  {author} {\bibinfo {author} {\bibfnamefont {E.}~\bibnamefont
  {Knill}}, \bibinfo {author} {\bibfnamefont {R.}~\bibnamefont {Laflamme}}, \
  and\ \bibinfo {author} {\bibfnamefont {W.~H.}\ \bibnamefont {Zurek}},\
  }\href@noop {} {\bibfield  {journal} {\bibinfo  {journal} {Proceedings of the
  Royal Society of London. Series A: Mathematical, Physical and Engineering
  Sciences}\ }\textbf {\bibinfo {volume} {454}},\ \bibinfo {pages} {365}
  (\bibinfo {year} {1998})}\BibitemShut {NoStop}%
\bibitem [{\citenamefont {Aharonov}\ and\ \citenamefont
  {Ben-Or}(2008)}]{aharonov2008fault}%
  \BibitemOpen
  \bibfield  {author} {\bibinfo {author} {\bibfnamefont {D.}~\bibnamefont
  {Aharonov}}\ and\ \bibinfo {author} {\bibfnamefont {M.}~\bibnamefont
  {Ben-Or}},\ }\href {\doibase 10.1137/S0097539799359385} {\bibfield  {journal}
  {\bibinfo  {journal} {SIAM Journal on Computing}\ } (\bibinfo {year}
  {2008}),\ 10.1137/S0097539799359385}\BibitemShut {NoStop}%
\bibitem [{\citenamefont {Wilde}(2013)}]{wilde2013quantum}%
  \BibitemOpen
  \bibfield  {author} {\bibinfo {author} {\bibfnamefont {M.~M.}\ \bibnamefont
  {Wilde}},\ }\href@noop {} {\emph {\bibinfo {title} {Quantum information
  theory}}}\ (\bibinfo  {publisher} {Cambridge University Press},\ \bibinfo
  {year} {2013})\BibitemShut {NoStop}%
\bibitem [{\citenamefont {Gyongyosi}\ \emph {et~al.}(2018)\citenamefont
  {Gyongyosi}, \citenamefont {Imre},\ and\ \citenamefont
  {Nguyen}}]{gyongyosi2018survey}%
  \BibitemOpen
  \bibfield  {author} {\bibinfo {author} {\bibfnamefont {L.}~\bibnamefont
  {Gyongyosi}}, \bibinfo {author} {\bibfnamefont {S.}~\bibnamefont {Imre}}, \
  and\ \bibinfo {author} {\bibfnamefont {H.~V.}\ \bibnamefont {Nguyen}},\
  }\href@noop {} {\bibfield  {journal} {\bibinfo  {journal} {IEEE
  Communications Surveys \& Tutorials}\ }\textbf {\bibinfo {volume} {20}},\
  \bibinfo {pages} {1149} (\bibinfo {year} {2018})}\BibitemShut {NoStop}%
\bibitem [{\citenamefont {Delfosse}\ and\ \citenamefont
  {Z{\'e}mor}(2012)}]{delfosse2012upper}%
  \BibitemOpen
  \bibfield  {author} {\bibinfo {author} {\bibfnamefont {N.}~\bibnamefont
  {Delfosse}}\ and\ \bibinfo {author} {\bibfnamefont {G.}~\bibnamefont
  {Z{\'e}mor}},\ }\href@noop {} {\  (\bibinfo {year} {2012})},\ \Eprint
  {http://arxiv.org/abs/1205.7036} {arXiv:1205.7036 [quant-ph]} \BibitemShut
  {NoStop}%
\bibitem [{\citenamefont {Richardson}\ and\ \citenamefont
  {Urbanke}(2008)}]{richardson2008modern}%
  \BibitemOpen
  \bibfield  {author} {\bibinfo {author} {\bibfnamefont {T.}~\bibnamefont
  {Richardson}}\ and\ \bibinfo {author} {\bibfnamefont {R.}~\bibnamefont
  {Urbanke}},\ }\href@noop {} {\emph {\bibinfo {title} {Modern coding
  theory}}}\ (\bibinfo  {publisher} {Cambridge University Press},\ \bibinfo
  {year} {2008})\BibitemShut {NoStop}%
\bibitem [{\citenamefont {MacKay}\ \emph {et~al.}(2004)\citenamefont {MacKay},
  \citenamefont {Mitchison},\ and\ \citenamefont
  {McFadden}}]{mackay2004sparse}%
  \BibitemOpen
  \bibfield  {author} {\bibinfo {author} {\bibfnamefont {D.~J.}\ \bibnamefont
  {MacKay}}, \bibinfo {author} {\bibfnamefont {G.}~\bibnamefont {Mitchison}}, \
  and\ \bibinfo {author} {\bibfnamefont {P.~L.}\ \bibnamefont {McFadden}},\
  }\href@noop {} {\bibfield  {journal} {\bibinfo  {journal} {IEEE Transactions
  on Information Theory}\ }\textbf {\bibinfo {volume} {50}},\ \bibinfo {pages}
  {2315} (\bibinfo {year} {2004})}\BibitemShut {NoStop}%
\bibitem [{\citenamefont {Lidar}\ and\ \citenamefont
  {Brun}(2013)}]{lidar2013quantum}%
  \BibitemOpen
  \bibfield  {author} {\bibinfo {author} {\bibfnamefont {D.~A.}\ \bibnamefont
  {Lidar}}\ and\ \bibinfo {author} {\bibfnamefont {T.~A.}\ \bibnamefont
  {Brun}},\ }\href@noop {} {\emph {\bibinfo {title} {Quantum error
  correction}}}\ (\bibinfo  {publisher} {Cambridge university press},\ \bibinfo
  {year} {2013})\BibitemShut {NoStop}%
\bibitem [{\citenamefont {Dennis}\ \emph {et~al.}(2002)\citenamefont {Dennis},
  \citenamefont {Kitaev}, \citenamefont {Landahl},\ and\ \citenamefont
  {Preskill}}]{dennis2002topological}%
  \BibitemOpen
  \bibfield  {author} {\bibinfo {author} {\bibfnamefont {E.}~\bibnamefont
  {Dennis}}, \bibinfo {author} {\bibfnamefont {A.}~\bibnamefont {Kitaev}},
  \bibinfo {author} {\bibfnamefont {A.}~\bibnamefont {Landahl}}, \ and\
  \bibinfo {author} {\bibfnamefont {J.}~\bibnamefont {Preskill}},\ }\href@noop
  {} {\bibfield  {journal} {\bibinfo  {journal} {Journal of Mathematical
  Physics}\ }\textbf {\bibinfo {volume} {43}},\ \bibinfo {pages} {4452}
  (\bibinfo {year} {2002})}\BibitemShut {NoStop}%
\bibitem [{\citenamefont {Kubica}(2018)}]{kubica2018abcs}%
  \BibitemOpen
  \bibfield  {author} {\bibinfo {author} {\bibfnamefont {A.~M.}\ \bibnamefont
  {Kubica}},\ }\emph {\bibinfo {title} {The ABCs of the color code: A study of
  topological quantum codes as toy models for fault-tolerant quantum
  computation and quantum phases of matter}},\ \href@noop {} {Ph.D. thesis},\
  \bibinfo  {school} {California Institute of Technology} (\bibinfo {year}
  {2018})\BibitemShut {NoStop}%
\bibitem [{\citenamefont {Terhal}(2015)}]{terhal2015quantum}%
  \BibitemOpen
  \bibfield  {author} {\bibinfo {author} {\bibfnamefont {B.~M.}\ \bibnamefont
  {Terhal}},\ }\href@noop {} {\bibfield  {journal} {\bibinfo  {journal}
  {Reviews of Modern Physics}\ }\textbf {\bibinfo {volume} {87}},\ \bibinfo
  {pages} {307} (\bibinfo {year} {2015})}\BibitemShut {NoStop}%
\bibitem [{\citenamefont {Preskill}(1998)}]{preskill1998lecture}%
  \BibitemOpen
  \bibfield  {author} {\bibinfo {author} {\bibfnamefont {J.}~\bibnamefont
  {Preskill}},\ }\href@noop {} {\bibfield  {journal} {\bibinfo  {journal}
  {California Institute of Technology}\ }\textbf {\bibinfo {volume} {16}},\
  \bibinfo {pages} {10} (\bibinfo {year} {1998})}\BibitemShut {NoStop}%
\bibitem [{\citenamefont {Gottesman}(1997)}]{gottesman1997stabilizer}%
  \BibitemOpen
  \bibfield  {author} {\bibinfo {author} {\bibfnamefont {D.}~\bibnamefont
  {Gottesman}},\ }\emph {\bibinfo {title} {Stabilizer codes and quantum error
  correction}},\ \href@noop {} {Ph.D. thesis},\ \bibinfo  {school} {Caltech}
  (\bibinfo {year} {1997}),\ \Eprint {http://arxiv.org/abs/quant-ph/9705052}
  {arXiv:quant-ph/9705052 [quant-ph]} \BibitemShut {NoStop}%
\bibitem [{\citenamefont {Calderbank}\ \emph {et~al.}(1998)\citenamefont
  {Calderbank}, \citenamefont {Rains}, \citenamefont {Shor},\ and\
  \citenamefont {Sloane}}]{calderbank1998quantum}%
  \BibitemOpen
  \bibfield  {author} {\bibinfo {author} {\bibfnamefont {A.~R.}\ \bibnamefont
  {Calderbank}}, \bibinfo {author} {\bibfnamefont {E.~M.}\ \bibnamefont
  {Rains}}, \bibinfo {author} {\bibfnamefont {P.}~\bibnamefont {Shor}}, \ and\
  \bibinfo {author} {\bibfnamefont {N.~J.}\ \bibnamefont {Sloane}},\
  }\href@noop {} {\bibfield  {journal} {\bibinfo  {journal} {IEEE Transactions
  on Information Theory}\ }\textbf {\bibinfo {volume} {44}},\ \bibinfo {pages}
  {1369} (\bibinfo {year} {1998})}\BibitemShut {NoStop}%
\bibitem [{\citenamefont {Bravyi}\ \emph
  {et~al.}(2010{\natexlab{a}})\citenamefont {Bravyi}, \citenamefont {Terhal},\
  and\ \citenamefont {Leemhuis}}]{bravyi2010majorana}%
  \BibitemOpen
  \bibfield  {author} {\bibinfo {author} {\bibfnamefont {S.}~\bibnamefont
  {Bravyi}}, \bibinfo {author} {\bibfnamefont {B.~M.}\ \bibnamefont {Terhal}},
  \ and\ \bibinfo {author} {\bibfnamefont {B.}~\bibnamefont {Leemhuis}},\
  }\href@noop {} {\bibfield  {journal} {\bibinfo  {journal} {New Journal of
  Physics}\ }\textbf {\bibinfo {volume} {12}},\ \bibinfo {pages} {083039}
  (\bibinfo {year} {2010}{\natexlab{a}})}\BibitemShut {NoStop}%
\bibitem [{\citenamefont {Bomb\'in}\ and\ \citenamefont
  {Martin-Delgado}(2007)}]{bombin2007homological}%
  \BibitemOpen
  \bibfield  {author} {\bibinfo {author} {\bibfnamefont {H.}~\bibnamefont
  {Bomb\'in}}\ and\ \bibinfo {author} {\bibfnamefont {M.~A.}\ \bibnamefont
  {Martin-Delgado}},\ }\href@noop {} {\bibfield  {journal} {\bibinfo  {journal}
  {Journal of mathematical physics}\ }\textbf {\bibinfo {volume} {48}},\
  \bibinfo {pages} {052105} (\bibinfo {year} {2007})}\BibitemShut {NoStop}%
\bibitem [{\citenamefont {Hatcher}(2002)}]{hatcher2002algebraic}%
  \BibitemOpen
  \bibfield  {author} {\bibinfo {author} {\bibfnamefont {A.}~\bibnamefont
  {Hatcher}},\ }\href@noop {} {\emph {\bibinfo {title} {Algebraic Topology}}}\
  (\bibinfo  {publisher} {Cambridge University Press},\ \bibinfo {year}
  {2002})\BibitemShut {NoStop}%
\bibitem [{\citenamefont {Breuckmann}(2017)}]{breuckmann2017homological}%
  \BibitemOpen
  \bibfield  {author} {\bibinfo {author} {\bibfnamefont {N.~P.}\ \bibnamefont
  {Breuckmann}},\ }\emph {\bibinfo {title} {Homological quantum codes beyond
  the toric code}},\ \href {https://d-nb.info/1162900415/34} {Ph.D. thesis},\
  \bibinfo  {school} {RWTH Aachen University} (\bibinfo {year}
  {2017})\BibitemShut {NoStop}%
\bibitem [{\citenamefont {Nakahara}(2003)}]{nakahara2003geometry}%
  \BibitemOpen
  \bibfield  {author} {\bibinfo {author} {\bibfnamefont {M.}~\bibnamefont
  {Nakahara}},\ }\href@noop {} {\emph {\bibinfo {title} {Geometry, topology and
  physics}}}\ (\bibinfo  {publisher} {CRC press},\ \bibinfo {year}
  {2003})\BibitemShut {NoStop}%
\bibitem [{\citenamefont {Breuckmann}\ and\ \citenamefont
  {Terhal}(2016)}]{breuckmann2016constructions}%
  \BibitemOpen
  \bibfield  {author} {\bibinfo {author} {\bibfnamefont {N.~P.}\ \bibnamefont
  {Breuckmann}}\ and\ \bibinfo {author} {\bibfnamefont {B.~M.}\ \bibnamefont
  {Terhal}},\ }\href@noop {} {\bibfield  {journal} {\bibinfo  {journal} {IEEE
  transactions on Information Theory}\ }\textbf {\bibinfo {volume} {62}},\
  \bibinfo {pages} {3731} (\bibinfo {year} {2016})}\BibitemShut {NoStop}%
\bibitem [{\citenamefont {Breuckmann}\ \emph {et~al.}(2017)\citenamefont
  {Breuckmann}, \citenamefont {Vuillot}, \citenamefont {Campbell},
  \citenamefont {Krishna},\ and\ \citenamefont
  {Terhal}}]{breuckmann2017hyperbolic}%
  \BibitemOpen
  \bibfield  {author} {\bibinfo {author} {\bibfnamefont {N.~P.}\ \bibnamefont
  {Breuckmann}}, \bibinfo {author} {\bibfnamefont {C.}~\bibnamefont {Vuillot}},
  \bibinfo {author} {\bibfnamefont {E.}~\bibnamefont {Campbell}}, \bibinfo
  {author} {\bibfnamefont {A.}~\bibnamefont {Krishna}}, \ and\ \bibinfo
  {author} {\bibfnamefont {B.~M.}\ \bibnamefont {Terhal}},\ }\href@noop {}
  {\bibfield  {journal} {\bibinfo  {journal} {Quantum Science and Technology}\
  }\textbf {\bibinfo {volume} {2}},\ \bibinfo {pages} {035007} (\bibinfo {year}
  {2017})}\BibitemShut {NoStop}%
\bibitem [{\citenamefont {Kim}(2007)}]{kim2007quantum}%
  \BibitemOpen
  \bibfield  {author} {\bibinfo {author} {\bibfnamefont {I.~H.}\ \bibnamefont
  {Kim}},\ }\emph {\bibinfo {title} {Quantum codes on Hurwitz surfaces}},\
  \href {https://dspace.mit.edu/handle/1721.1/40917} {\bibinfo {type}
  {{Bachelor's Thesis}}},\ \bibinfo  {school} {Massachusetts Institute of
  Technology} (\bibinfo {year} {2007})\BibitemShut {NoStop}%
\bibitem [{\citenamefont {Z{\'e}mor}(2009)}]{zemor2009cayley}%
  \BibitemOpen
  \bibfield  {author} {\bibinfo {author} {\bibfnamefont {G.}~\bibnamefont
  {Z{\'e}mor}},\ }in\ \href@noop {} {\emph {\bibinfo {booktitle} {International
  Conference on Coding and Cryptology}}}\ (\bibinfo {organization} {Springer},\
  \bibinfo {year} {2009})\ pp.\ \bibinfo {pages} {259--273}\BibitemShut
  {NoStop}%
\bibitem [{\citenamefont {Delfosse}(2013)}]{delfosse2013tradeoffs}%
  \BibitemOpen
  \bibfield  {author} {\bibinfo {author} {\bibfnamefont {N.}~\bibnamefont
  {Delfosse}},\ }in\ \href@noop {} {\emph {\bibinfo {booktitle} {2013 IEEE
  International Symposium on Information Theory}}}\ (\bibinfo {organization}
  {IEEE},\ \bibinfo {year} {2013})\ pp.\ \bibinfo {pages}
  {917--921}\BibitemShut {NoStop}%
\bibitem [{\citenamefont {Kovalev}\ and\ \citenamefont
  {Pryadko}(2013)}]{kovalev2013fault}%
  \BibitemOpen
  \bibfield  {author} {\bibinfo {author} {\bibfnamefont {A.~A.}\ \bibnamefont
  {Kovalev}}\ and\ \bibinfo {author} {\bibfnamefont {L.~P.}\ \bibnamefont
  {Pryadko}},\ }\href@noop {} {\bibfield  {journal} {\bibinfo  {journal}
  {Physical Review A}\ }\textbf {\bibinfo {volume} {87}},\ \bibinfo {pages}
  {020304} (\bibinfo {year} {2013})}\BibitemShut {NoStop}%
\bibitem [{\citenamefont {Delfosse}\ and\ \citenamefont
  {Nickerson}(2017)}]{delfosse2017almost}%
  \BibitemOpen
  \bibfield  {author} {\bibinfo {author} {\bibfnamefont {N.}~\bibnamefont
  {Delfosse}}\ and\ \bibinfo {author} {\bibfnamefont {N.~H.}\ \bibnamefont
  {Nickerson}},\ }\href@noop {} {\  (\bibinfo {year} {2017})},\ \Eprint
  {http://arxiv.org/abs/1709.06218} {arXiv:1709.06218 [quant-ph]} \BibitemShut
  {NoStop}%
\bibitem [{\citenamefont {Higgott}\ and\ \citenamefont
  {Breuckmann}(2020)}]{higgott2020subsystem}%
  \BibitemOpen
  \bibfield  {author} {\bibinfo {author} {\bibfnamefont {O.}~\bibnamefont
  {Higgott}}\ and\ \bibinfo {author} {\bibfnamefont {N.~P.}\ \bibnamefont
  {Breuckmann}},\ }\href@noop {} {\  (\bibinfo {year} {2020})},\ \Eprint
  {http://arxiv.org/abs/2010.09626} {arXiv:2010.09626 [quant-ph]} \BibitemShut
  {NoStop}%
\bibitem [{\citenamefont {Soares~Jr}\ and\ \citenamefont
  {Da~Silva}(2018)}]{soares2018hyperbolic}%
  \BibitemOpen
  \bibfield  {author} {\bibinfo {author} {\bibfnamefont {W.~S.}\ \bibnamefont
  {Soares~Jr}}\ and\ \bibinfo {author} {\bibfnamefont {E.~B.}\ \bibnamefont
  {Da~Silva}},\ }\href@noop {} {\  (\bibinfo {year} {2018})},\ \Eprint
  {http://arxiv.org/abs/1804.06382} {arXiv:1804.06382 [quant-ph]} \BibitemShut
  {NoStop}%
\bibitem [{\citenamefont {Vuillot}\ and\ \citenamefont
  {Breuckmann}(2019)}]{vuillot2019quantum}%
  \BibitemOpen
  \bibfield  {author} {\bibinfo {author} {\bibfnamefont {C.}~\bibnamefont
  {Vuillot}}\ and\ \bibinfo {author} {\bibfnamefont {N.~P.}\ \bibnamefont
  {Breuckmann}},\ }\href@noop {} {\  (\bibinfo {year} {2019})},\ \Eprint
  {http://arxiv.org/abs/1906.11394} {arXiv:1906.11394 [quant-ph]} \BibitemShut
  {NoStop}%
\bibitem [{\citenamefont {Guth}\ and\ \citenamefont
  {Lubotzky}(2014)}]{guth2014quantum}%
  \BibitemOpen
  \bibfield  {author} {\bibinfo {author} {\bibfnamefont {L.}~\bibnamefont
  {Guth}}\ and\ \bibinfo {author} {\bibfnamefont {A.}~\bibnamefont
  {Lubotzky}},\ }\href@noop {} {\bibfield  {journal} {\bibinfo  {journal}
  {Journal of Mathematical Physics}\ }\textbf {\bibinfo {volume} {55}},\
  \bibinfo {pages} {082202} (\bibinfo {year} {2014})}\BibitemShut {NoStop}%
\bibitem [{\citenamefont {Hastings}(2014)}]{hastings2013decoding}%
  \BibitemOpen
  \bibfield  {author} {\bibinfo {author} {\bibfnamefont {M.~B.}\ \bibnamefont
  {Hastings}},\ }\href {\doibase 10.26421/QIC14.13-14} {\bibfield  {journal}
  {\bibinfo  {journal} {Quantum Information and Computation}\ } (\bibinfo
  {year} {2014}),\ 10.26421/QIC14.13-14}\BibitemShut {NoStop}%
\bibitem [{\citenamefont {Londe}\ and\ \citenamefont
  {Leverrier}(2019)}]{londe2017golden}%
  \BibitemOpen
  \bibfield  {author} {\bibinfo {author} {\bibfnamefont {V.}~\bibnamefont
  {Londe}}\ and\ \bibinfo {author} {\bibfnamefont {A.}~\bibnamefont
  {Leverrier}},\ }\href {\doibase 10.26421/QIC19.5-6-1} {\bibfield  {journal}
  {\bibinfo  {journal} {Quantum Information and Computation}\ ,\ \bibinfo
  {pages} {0361}} (\bibinfo {year} {2019})}\BibitemShut {NoStop}%
\bibitem [{\citenamefont {Breuckmann}\ and\ \citenamefont
  {Londe}(2020)}]{breuckmann2020single}%
  \BibitemOpen
  \bibfield  {author} {\bibinfo {author} {\bibfnamefont {N.~P.}\ \bibnamefont
  {Breuckmann}}\ and\ \bibinfo {author} {\bibfnamefont {V.}~\bibnamefont
  {Londe}},\ }\href@noop {} {\  (\bibinfo {year} {2020})},\ \Eprint
  {http://arxiv.org/abs/2001.03568} {arXiv:2001.03568 [quant-ph]} \BibitemShut
  {NoStop}%
\bibitem [{\citenamefont {Freedman}\ \emph {et~al.}(2002)\citenamefont
  {Freedman}, \citenamefont {Meyer},\ and\ \citenamefont
  {Luo}}]{freedman2002z2}%
  \BibitemOpen
  \bibfield  {author} {\bibinfo {author} {\bibfnamefont {M.~H.}\ \bibnamefont
  {Freedman}}, \bibinfo {author} {\bibfnamefont {D.~A.}\ \bibnamefont {Meyer}},
  \ and\ \bibinfo {author} {\bibfnamefont {F.}~\bibnamefont {Luo}},\
  }\href@noop {} {\bibfield  {journal} {\bibinfo  {journal} {Mathematics of
  quantum computation, Chapman \& Hall/CRC}\ ,\ \bibinfo {pages} {287}}
  (\bibinfo {year} {2002})}\BibitemShut {NoStop}%
\bibitem [{Note1()}]{Note1}%
  \BibitemOpen
  \bibinfo {note} {{Note that there is an unfortunate typo in~\cite
  {freedman2002z2}, claiming distance scaling as $\Omega (\protect \sqrt
  [2]{\protect \qopname \relax o{log}(n)}\protect \sqrt {n})$.}}\BibitemShut
  {Stop}%
\bibitem [{\citenamefont {Fetaya}(2011)}]{fetaya2011homological}%
  \BibitemOpen
  \bibfield  {author} {\bibinfo {author} {\bibfnamefont {E.}~\bibnamefont
  {Fetaya}},\ }\emph {\bibinfo {title} {Homological error correcting codes and
  systolic geometry}},\ \href@noop {} {Master's thesis},\ \bibinfo  {school}
  {The Hebrew University of Jerusalem} (\bibinfo {year} {2011}),\ \Eprint
  {http://arxiv.org/abs/1108.2886} {arXiv:1108.2886 [math.DG]} \BibitemShut
  {NoStop}%
\bibitem [{\citenamefont {Haah}(2016)}]{haah2016algebraic}%
  \BibitemOpen
  \bibfield  {author} {\bibinfo {author} {\bibfnamefont {J.}~\bibnamefont
  {Haah}},\ }\href@noop {} {\bibfield  {journal} {\bibinfo  {journal} {Revista
  colombiana de matematicas}\ }\textbf {\bibinfo {volume} {50}},\ \bibinfo
  {pages} {299} (\bibinfo {year} {2016})}\BibitemShut {NoStop}%
\bibitem [{\citenamefont {Haah}(2011)}]{haah2011local}%
  \BibitemOpen
  \bibfield  {author} {\bibinfo {author} {\bibfnamefont {J.}~\bibnamefont
  {Haah}},\ }\href@noop {} {\bibfield  {journal} {\bibinfo  {journal} {Physical
  Review A}\ }\textbf {\bibinfo {volume} {83}},\ \bibinfo {pages} {042330}
  (\bibinfo {year} {2011})}\BibitemShut {NoStop}%
\bibitem [{\citenamefont {Panteleev}\ and\ \citenamefont
  {Kalachev}(2020)}]{panteleev2020quantum}%
  \BibitemOpen
  \bibfield  {author} {\bibinfo {author} {\bibfnamefont {P.}~\bibnamefont
  {Panteleev}}\ and\ \bibinfo {author} {\bibfnamefont {G.}~\bibnamefont
  {Kalachev}},\ }\href@noop {} {\enquote {\bibinfo {title} {Quantum ldpc codes
  with almost linear minimum distance},}\ } (\bibinfo {year} {2020}),\ \Eprint
  {http://arxiv.org/abs/2012.04068} {arXiv:2012.04068 [cs.IT]} \BibitemShut
  {NoStop}%
\bibitem [{\citenamefont {Brown}\ \emph {et~al.}(2016)\citenamefont {Brown},
  \citenamefont {Loss}, \citenamefont {Pachos}, \citenamefont {Self},\ and\
  \citenamefont {Wootton}}]{brown2016quantum}%
  \BibitemOpen
  \bibfield  {author} {\bibinfo {author} {\bibfnamefont {B.~J.}\ \bibnamefont
  {Brown}}, \bibinfo {author} {\bibfnamefont {D.}~\bibnamefont {Loss}},
  \bibinfo {author} {\bibfnamefont {J.~K.}\ \bibnamefont {Pachos}}, \bibinfo
  {author} {\bibfnamefont {C.~N.}\ \bibnamefont {Self}}, \ and\ \bibinfo
  {author} {\bibfnamefont {J.~R.}\ \bibnamefont {Wootton}},\ }\href@noop {}
  {\bibfield  {journal} {\bibinfo  {journal} {Reviews of Modern Physics}\
  }\textbf {\bibinfo {volume} {88}},\ \bibinfo {pages} {045005} (\bibinfo
  {year} {2016})}\BibitemShut {NoStop}%
\bibitem [{\citenamefont {Fetaya}(2012)}]{fetaya2012bounding}%
  \BibitemOpen
  \bibfield  {author} {\bibinfo {author} {\bibfnamefont {E.}~\bibnamefont
  {Fetaya}},\ }\href@noop {} {\bibfield  {journal} {\bibinfo  {journal}
  {Journal of mathematical physics}\ }\textbf {\bibinfo {volume} {53}},\
  \bibinfo {pages} {062202} (\bibinfo {year} {2012})}\BibitemShut {NoStop}%
\bibitem [{\citenamefont {Bravyi}\ \emph
  {et~al.}(2010{\natexlab{b}})\citenamefont {Bravyi}, \citenamefont {Poulin},\
  and\ \citenamefont {Terhal}}]{bravyi2010tradeoffs}%
  \BibitemOpen
  \bibfield  {author} {\bibinfo {author} {\bibfnamefont {S.}~\bibnamefont
  {Bravyi}}, \bibinfo {author} {\bibfnamefont {D.}~\bibnamefont {Poulin}}, \
  and\ \bibinfo {author} {\bibfnamefont {B.}~\bibnamefont {Terhal}},\
  }\href@noop {} {\bibfield  {journal} {\bibinfo  {journal} {Physical review
  letters}\ }\textbf {\bibinfo {volume} {104}},\ \bibinfo {pages} {050503}
  (\bibinfo {year} {2010}{\natexlab{b}})}\BibitemShut {NoStop}%
\bibitem [{\citenamefont {Devakul}\ and\ \citenamefont
  {Williamson}(2021)}]{devakul2021fractalizing}%
  \BibitemOpen
  \bibfield  {author} {\bibinfo {author} {\bibfnamefont {T.}~\bibnamefont
  {Devakul}}\ and\ \bibinfo {author} {\bibfnamefont {D.~J.}\ \bibnamefont
  {Williamson}},\ }\href@noop {} {\bibfield  {journal} {\bibinfo  {journal}
  {Quantum}\ }\textbf {\bibinfo {volume} {5}},\ \bibinfo {pages} {438}
  (\bibinfo {year} {2021})}\BibitemShut {NoStop}%
\bibitem [{\citenamefont {Katz}(2007)}]{katz2007systolic}%
  \BibitemOpen
  \bibfield  {author} {\bibinfo {author} {\bibfnamefont {M.~G.}\ \bibnamefont
  {Katz}},\ }\href@noop {} {\emph {\bibinfo {title} {Systolic geometry and
  topology}}},\ \bibinfo {number} {137}\ (\bibinfo  {publisher} {American
  Mathematical Soc.},\ \bibinfo {year} {2007})\BibitemShut {NoStop}%
\bibitem [{\citenamefont {Tillich}\ and\ \citenamefont
  {Z{\'e}mor}(2013)}]{tillich2013quantum}%
  \BibitemOpen
  \bibfield  {author} {\bibinfo {author} {\bibfnamefont {J.-P.}\ \bibnamefont
  {Tillich}}\ and\ \bibinfo {author} {\bibfnamefont {G.}~\bibnamefont
  {Z{\'e}mor}},\ }\href@noop {} {\bibfield  {journal} {\bibinfo  {journal}
  {IEEE Transactions on Information Theory}\ }\textbf {\bibinfo {volume}
  {60}},\ \bibinfo {pages} {1193} (\bibinfo {year} {2013})}\BibitemShut
  {NoStop}%
\bibitem [{\citenamefont {Bravyi}\ and\ \citenamefont
  {Hastings}(2014)}]{bravyi2014homological}%
  \BibitemOpen
  \bibfield  {author} {\bibinfo {author} {\bibfnamefont {S.}~\bibnamefont
  {Bravyi}}\ and\ \bibinfo {author} {\bibfnamefont {M.~B.}\ \bibnamefont
  {Hastings}},\ }in\ \href@noop {} {\emph {\bibinfo {booktitle} {Proceedings of
  the forty-sixth annual ACM symposium on Theory of computing}}}\ (\bibinfo
  {year} {2014})\ pp.\ \bibinfo {pages} {273--282}\BibitemShut {NoStop}%
\bibitem [{\citenamefont {Hastings}(2016{\natexlab{a}})}]{hastings2016weight}%
  \BibitemOpen
  \bibfield  {author} {\bibinfo {author} {\bibfnamefont {M.~B.}\ \bibnamefont
  {Hastings}},\ }\href@noop {} {\  (\bibinfo {year} {2016}{\natexlab{a}})},\
  \Eprint {http://arxiv.org/abs/1611.03790} {arXiv:1611.03790 [quant-ph]}
  \BibitemShut {NoStop}%
\bibitem [{\citenamefont {Evra}\ \emph {et~al.}(2020)\citenamefont {Evra},
  \citenamefont {Kaufman},\ and\ \citenamefont {Zémor}}]{evra_decodable_2020}%
  \BibitemOpen
  \bibfield  {author} {\bibinfo {author} {\bibfnamefont {S.}~\bibnamefont
  {Evra}}, \bibinfo {author} {\bibfnamefont {T.}~\bibnamefont {Kaufman}}, \
  and\ \bibinfo {author} {\bibfnamefont {G.}~\bibnamefont {Zémor}},\ }\href
  {http://arxiv.org/abs/2004.07935} {\bibfield  {journal} {\bibinfo  {journal}
  {arXiv:2004.07935 [quant-ph]}\ } (\bibinfo {year} {2020})},\ \bibinfo {note}
  {arXiv: 2004.07935}\BibitemShut {NoStop}%
\bibitem [{\citenamefont {Kaufman}\ and\ \citenamefont
  {Tessler}(2020)}]{kaufman2020new}%
  \BibitemOpen
  \bibfield  {author} {\bibinfo {author} {\bibfnamefont {T.}~\bibnamefont
  {Kaufman}}\ and\ \bibinfo {author} {\bibfnamefont {R.~J.}\ \bibnamefont
  {Tessler}},\ }\href@noop {} {\enquote {\bibinfo {title} {New cosystolic
  expanders from tensors imply explicit quantum ldpc codes with
  $\omega(\sqrt{n}\log^kn)$ distance},}\ } (\bibinfo {year} {2020}),\ \Eprint
  {http://arxiv.org/abs/2008.09495} {arXiv:2008.09495 [quant-ph]} \BibitemShut
  {NoStop}%
\bibitem [{\citenamefont {Hastings}\ \emph {et~al.}(2020)\citenamefont
  {Hastings}, \citenamefont {Haah},\ and\ \citenamefont
  {O'Donnell}}]{hastings2020fiber}%
  \BibitemOpen
  \bibfield  {author} {\bibinfo {author} {\bibfnamefont {M.~B.}\ \bibnamefont
  {Hastings}}, \bibinfo {author} {\bibfnamefont {J.}~\bibnamefont {Haah}}, \
  and\ \bibinfo {author} {\bibfnamefont {R.}~\bibnamefont {O'Donnell}},\
  }\href@noop {} {\enquote {\bibinfo {title} {Fiber bundle codes: Breaking the
  $n^{1/2} \operatorname{polylog}(n)$ barrier for quantum ldpc codes},}\ }
  (\bibinfo {year} {2020}),\ \Eprint {http://arxiv.org/abs/2009.03921}
  {arXiv:2009.03921 [quant-ph]} \BibitemShut {NoStop}%
\bibitem [{\citenamefont {Panteleev}\ and\ \citenamefont
  {Kalachev}(2019)}]{panteleev2019degenerate}%
  \BibitemOpen
  \bibfield  {author} {\bibinfo {author} {\bibfnamefont {P.}~\bibnamefont
  {Panteleev}}\ and\ \bibinfo {author} {\bibfnamefont {G.}~\bibnamefont
  {Kalachev}},\ }\href@noop {} {\enquote {\bibinfo {title} {Degenerate quantum
  ldpc codes with good finite length performance},}\ } (\bibinfo {year}
  {2019}),\ \Eprint {http://arxiv.org/abs/1904.02703} {arXiv:1904.02703
  [quant-ph]} \BibitemShut {NoStop}%
\bibitem [{\citenamefont {Breuckmann}\ and\ \citenamefont
  {Eberhardt}(2020)}]{breuckmann2020balanced}%
  \BibitemOpen
  \bibfield  {author} {\bibinfo {author} {\bibfnamefont {N.~P.}\ \bibnamefont
  {Breuckmann}}\ and\ \bibinfo {author} {\bibfnamefont {J.~N.}\ \bibnamefont
  {Eberhardt}},\ }\href@noop {} {\enquote {\bibinfo {title} {Balanced product
  quantum codes},}\ } (\bibinfo {year} {2020}),\ \Eprint
  {http://arxiv.org/abs/2012.09271} {arXiv:2012.09271 [quant-ph]} \BibitemShut
  {NoStop}%
\bibitem [{\citenamefont {Weibel}(1994)}]{weibel_introduction_1994}%
  \BibitemOpen
  \bibfield  {author} {\bibinfo {author} {\bibfnamefont {C.~A.}\ \bibnamefont
  {Weibel}},\ }\href {\doibase 10.1017/CBO9781139644136} {\emph {\bibinfo
  {title} {An {Introduction} to {Homological} {Algebra}}}},\ Cambridge
  {Studies} in {Advanced} {Mathematics}\ (\bibinfo  {publisher} {Cambridge
  University Press},\ \bibinfo {address} {Cambridge},\ \bibinfo {year}
  {1994})\BibitemShut {NoStop}%
\bibitem [{\citenamefont {Leverrier}\ \emph
  {et~al.}(2015{\natexlab{a}})\citenamefont {Leverrier}, \citenamefont
  {Tillich},\ and\ \citenamefont {Zémor}}]{leverrier_quantum_2015}%
  \BibitemOpen
  \bibfield  {author} {\bibinfo {author} {\bibfnamefont {A.}~\bibnamefont
  {Leverrier}}, \bibinfo {author} {\bibfnamefont {J.}~\bibnamefont {Tillich}},
  \ and\ \bibinfo {author} {\bibfnamefont {G.}~\bibnamefont {Zémor}},\ }in\
  \href {\doibase 10.1109/FOCS.2015.55} {\emph {\bibinfo {booktitle} {2015
  {IEEE} 56th {Annual} {Symposium} on {Foundations} of {Computer} {Science}}}}\
  (\bibinfo {year} {2015})\ pp.\ \bibinfo {pages} {810--824},\ \bibinfo {note}
  {iSSN: 0272-5428}\BibitemShut {NoStop}%
\bibitem [{\citenamefont {Fawzi}\ \emph
  {et~al.}(2018{\natexlab{a}})\citenamefont {Fawzi}, \citenamefont
  {Grospellier},\ and\ \citenamefont {Leverrier}}]{fawzi_constant_2018}%
  \BibitemOpen
  \bibfield  {author} {\bibinfo {author} {\bibfnamefont {O.}~\bibnamefont
  {Fawzi}}, \bibinfo {author} {\bibfnamefont {A.}~\bibnamefont {Grospellier}},
  \ and\ \bibinfo {author} {\bibfnamefont {A.}~\bibnamefont {Leverrier}},\ }in\
  \href {\doibase 10.1109/FOCS.2018.00076} {\emph {\bibinfo {booktitle} {2018
  {IEEE} 59th {Annual} {Symposium} on {Foundations} of {Computer} {Science}
  ({FOCS})}}}\ (\bibinfo {year} {2018})\ pp.\ \bibinfo {pages} {743--754},\
  \bibinfo {note} {iSSN: 2575-8454}\BibitemShut {NoStop}%
\bibitem [{\citenamefont {Sipser}\ and\ \citenamefont
  {Spielman}(1996)}]{sipser_expander_1996}%
  \BibitemOpen
  \bibfield  {author} {\bibinfo {author} {\bibfnamefont {M.}~\bibnamefont
  {Sipser}}\ and\ \bibinfo {author} {\bibfnamefont {D.~A.}\ \bibnamefont
  {Spielman}},\ }\href {\doibase 10.1109/18.556667} {\bibfield  {journal}
  {\bibinfo  {journal} {IEEE Transactions on Information Theory}\ }\textbf
  {\bibinfo {volume} {42}},\ \bibinfo {pages} {1710} (\bibinfo {year}
  {1996})},\ \bibinfo {note} {conference Name: IEEE Transactions on Information
  Theory}\BibitemShut {NoStop}%
\bibitem [{\citenamefont {Kovalev}\ and\ \citenamefont
  {Pryadko}(2012)}]{kovalev_improved_2012}%
  \BibitemOpen
  \bibfield  {author} {\bibinfo {author} {\bibfnamefont {A.~A.}\ \bibnamefont
  {Kovalev}}\ and\ \bibinfo {author} {\bibfnamefont {L.~P.}\ \bibnamefont
  {Pryadko}},\ }in\ \href {\doibase 10.1109/ISIT.2012.6284206} {\emph {\bibinfo
  {booktitle} {2012 {IEEE} {International} {Symposium} on {Information}
  {Theory} {Proceedings}}}}\ (\bibinfo {year} {2012})\ pp.\ \bibinfo {pages}
  {348--352},\ \bibinfo {note} {iSSN: 2157-8117}\BibitemShut {NoStop}%
\bibitem [{\citenamefont {Zeng}\ and\ \citenamefont
  {Pryadko}(2019)}]{zeng_higher-dimensional_2019}%
  \BibitemOpen
  \bibfield  {author} {\bibinfo {author} {\bibfnamefont {W.}~\bibnamefont
  {Zeng}}\ and\ \bibinfo {author} {\bibfnamefont {L.~P.}\ \bibnamefont
  {Pryadko}},\ }\href {\doibase 10.1103/PhysRevLett.122.230501} {\bibfield
  {journal} {\bibinfo  {journal} {Physical Review Letters}\ }\textbf {\bibinfo
  {volume} {122}},\ \bibinfo {pages} {230501} (\bibinfo {year} {2019})},\
  \bibinfo {note} {publisher: American Physical Society}\BibitemShut {NoStop}%
\bibitem [{\citenamefont {Audoux}\ and\ \citenamefont
  {Couvreur}(2019)}]{audoux_tensor_2019}%
  \BibitemOpen
  \bibfield  {author} {\bibinfo {author} {\bibfnamefont {B.}~\bibnamefont
  {Audoux}}\ and\ \bibinfo {author} {\bibfnamefont {A.}~\bibnamefont
  {Couvreur}},\ }\href {\doibase 10.4171/AIHPD/71} {\bibfield  {journal}
  {\bibinfo  {journal} {Annales de l'Institut Henri Poincaré (D)
  Combinatorics, Physics and their Interactions}\ }\textbf {\bibinfo {volume}
  {6}},\ \bibinfo {pages} {239} (\bibinfo {year} {2019})},\ \bibinfo {note}
  {publisher: European Mathematical Society}\BibitemShut {NoStop}%
\bibitem [{\citenamefont {Lubotzky}(2017)}]{lubotzky_high_2017}%
  \BibitemOpen
  \bibfield  {author} {\bibinfo {author} {\bibfnamefont {A.}~\bibnamefont
  {Lubotzky}},\ }\href {http://arxiv.org/abs/1712.02526} {\bibfield  {journal}
  {\bibinfo  {journal} {arXiv:1712.02526 [math]}\ } (\bibinfo {year} {2017})},\
  \bibinfo {note} {arXiv: 1712.02526}\BibitemShut {NoStop}%
\bibitem [{\citenamefont {Lubotzky}\ \emph {et~al.}(1988)\citenamefont
  {Lubotzky}, \citenamefont {Phillips},\ and\ \citenamefont
  {Sarnak}}]{lubotzky_ramanujan_1988}%
  \BibitemOpen
  \bibfield  {author} {\bibinfo {author} {\bibfnamefont {A.}~\bibnamefont
  {Lubotzky}}, \bibinfo {author} {\bibfnamefont {R.}~\bibnamefont {Phillips}},
  \ and\ \bibinfo {author} {\bibfnamefont {P.}~\bibnamefont {Sarnak}},\ }\href
  {\doibase 10.1007/BF02126799} {\bibfield  {journal} {\bibinfo  {journal}
  {Combinatorica}\ }\textbf {\bibinfo {volume} {8}},\ \bibinfo {pages} {261}
  (\bibinfo {year} {1988})}\BibitemShut {NoStop}%
\bibitem [{\citenamefont {Leverrier}\ \emph {et~al.}(2020)\citenamefont
  {Leverrier}, \citenamefont {Apers},\ and\ \citenamefont
  {Vuillot}}]{leverrier2020quantum}%
  \BibitemOpen
  \bibfield  {author} {\bibinfo {author} {\bibfnamefont {A.}~\bibnamefont
  {Leverrier}}, \bibinfo {author} {\bibfnamefont {S.}~\bibnamefont {Apers}}, \
  and\ \bibinfo {author} {\bibfnamefont {C.}~\bibnamefont {Vuillot}},\
  }\href@noop {} {\  (\bibinfo {year} {2020})},\ \Eprint
  {http://arxiv.org/abs/2011.09746} {arXiv:2011.09746 [quant-ph]} \BibitemShut
  {NoStop}%
\bibitem [{\citenamefont {Maurice}(2014)}]{maurice2014codes}%
  \BibitemOpen
  \bibfield  {author} {\bibinfo {author} {\bibfnamefont {D.}~\bibnamefont
  {Maurice}},\ }\emph {\bibinfo {title} {Codes correcteurs quantiques pouvant
  se d{\'e}coder it{\'e}rativement}},\ \href@noop {} {Ph.D. thesis},\ \bibinfo
  {school} {Universit{\'e} Pierre et Marie Curie-Paris VI} (\bibinfo {year}
  {2014})\BibitemShut {NoStop}%
\bibitem [{\citenamefont {Chamon}(2005)}]{chamon2005quantum}%
  \BibitemOpen
  \bibfield  {author} {\bibinfo {author} {\bibfnamefont {C.}~\bibnamefont
  {Chamon}},\ }\href@noop {} {\bibfield  {journal} {\bibinfo  {journal}
  {Physical review letters}\ }\textbf {\bibinfo {volume} {94}},\ \bibinfo
  {pages} {040402} (\bibinfo {year} {2005})}\BibitemShut {NoStop}%
\bibitem [{\citenamefont {Bravyi}\ \emph {et~al.}(2011)\citenamefont {Bravyi},
  \citenamefont {Leemhuis},\ and\ \citenamefont
  {Terhal}}]{bravyi2011topological}%
  \BibitemOpen
  \bibfield  {author} {\bibinfo {author} {\bibfnamefont {S.}~\bibnamefont
  {Bravyi}}, \bibinfo {author} {\bibfnamefont {B.}~\bibnamefont {Leemhuis}}, \
  and\ \bibinfo {author} {\bibfnamefont {B.~M.}\ \bibnamefont {Terhal}},\
  }\href@noop {} {\bibfield  {journal} {\bibinfo  {journal} {Annals of
  Physics}\ }\textbf {\bibinfo {volume} {326}},\ \bibinfo {pages} {839}
  (\bibinfo {year} {2011})}\BibitemShut {NoStop}%
\bibitem [{\citenamefont {Bonilla-Ataides}\ \emph {et~al.}(2020)\citenamefont
  {Bonilla-Ataides}, \citenamefont {Tuckett}, \citenamefont {Bartlett},
  \citenamefont {Flammia},\ and\ \citenamefont {Brown}}]{bonilla2020xzzx}%
  \BibitemOpen
  \bibfield  {author} {\bibinfo {author} {\bibfnamefont {J.~P.}\ \bibnamefont
  {Bonilla-Ataides}}, \bibinfo {author} {\bibfnamefont {D.~K.}\ \bibnamefont
  {Tuckett}}, \bibinfo {author} {\bibfnamefont {S.~D.}\ \bibnamefont
  {Bartlett}}, \bibinfo {author} {\bibfnamefont {S.~T.}\ \bibnamefont
  {Flammia}}, \ and\ \bibinfo {author} {\bibfnamefont {B.~J.}\ \bibnamefont
  {Brown}},\ }\href@noop {} {\  (\bibinfo {year} {2020})},\ \Eprint
  {http://arxiv.org/abs/2009.07851} {arXiv:2009.07851 [quant-ph]} \BibitemShut
  {NoStop}%
\bibitem [{\citenamefont {Hastings}(2016{\natexlab{b}})}]{hastings2016quantum}%
  \BibitemOpen
  \bibfield  {author} {\bibinfo {author} {\bibfnamefont {M.~B.}\ \bibnamefont
  {Hastings}},\ }\href@noop {} {\  (\bibinfo {year} {2016}{\natexlab{b}})},\
  \Eprint {http://arxiv.org/abs/1608.05089} {arXiv:1608.05089 [quant-ph]}
  \BibitemShut {NoStop}%
\bibitem [{\citenamefont {Bravyi}(2011)}]{bravyi2011subsystem}%
  \BibitemOpen
  \bibfield  {author} {\bibinfo {author} {\bibfnamefont {S.}~\bibnamefont
  {Bravyi}},\ }\href@noop {} {\bibfield  {journal} {\bibinfo  {journal}
  {Physical Review A}\ }\textbf {\bibinfo {volume} {83}},\ \bibinfo {pages}
  {012320} (\bibinfo {year} {2011})}\BibitemShut {NoStop}%
\bibitem [{\citenamefont {Yoder}(2019)}]{yoder_bbs}%
  \BibitemOpen
  \bibfield  {author} {\bibinfo {author} {\bibfnamefont {T.~J.}\ \bibnamefont
  {Yoder}},\ }\href {\doibase 10.1103/PhysRevA.99.052333} {\bibfield  {journal}
  {\bibinfo  {journal} {Phys. Rev. A}\ }\textbf {\bibinfo {volume} {99}},\
  \bibinfo {pages} {052333} (\bibinfo {year} {2019})}\BibitemShut {NoStop}%
\bibitem [{\citenamefont {Li}\ and\ \citenamefont
  {Yoder}(2020)}]{li2020numerical}%
  \BibitemOpen
  \bibfield  {author} {\bibinfo {author} {\bibfnamefont {M.}~\bibnamefont
  {Li}}\ and\ \bibinfo {author} {\bibfnamefont {T.~J.}\ \bibnamefont {Yoder}},\
  }in\ \href@noop {} {\emph {\bibinfo {booktitle} {2020 IEEE International
  Conference on Quantum Computing and Engineering (QCE)}}}\ (\bibinfo
  {organization} {IEEE},\ \bibinfo {year} {2020})\ pp.\ \bibinfo {pages}
  {109--119}\BibitemShut {NoStop}%
\bibitem [{\citenamefont {Bacon}\ \emph {et~al.}(2017)\citenamefont {Bacon},
  \citenamefont {Flammia}, \citenamefont {Harrow},\ and\ \citenamefont
  {Shi}}]{bacon2017sparse}%
  \BibitemOpen
  \bibfield  {author} {\bibinfo {author} {\bibfnamefont {D.}~\bibnamefont
  {Bacon}}, \bibinfo {author} {\bibfnamefont {S.~T.}\ \bibnamefont {Flammia}},
  \bibinfo {author} {\bibfnamefont {A.~W.}\ \bibnamefont {Harrow}}, \ and\
  \bibinfo {author} {\bibfnamefont {J.}~\bibnamefont {Shi}},\ }\href@noop {}
  {\bibfield  {journal} {\bibinfo  {journal} {IEEE Transactions on Information
  Theory}\ }\textbf {\bibinfo {volume} {63}},\ \bibinfo {pages} {2464}
  (\bibinfo {year} {2017})}\BibitemShut {NoStop}%
\bibitem [{\citenamefont {Calderbank}\ and\ \citenamefont
  {Shor}(1996{\natexlab{b}})}]{goodquantumcodes}%
  \BibitemOpen
  \bibfield  {author} {\bibinfo {author} {\bibfnamefont {A.~R.}\ \bibnamefont
  {Calderbank}}\ and\ \bibinfo {author} {\bibfnamefont {P.~W.}\ \bibnamefont
  {Shor}},\ }\href {\doibase 10.1103/PhysRevA.54.1098} {\bibfield  {journal}
  {\bibinfo  {journal} {Phys. Rev. A}\ }\textbf {\bibinfo {volume} {54}},\
  \bibinfo {pages} {1098} (\bibinfo {year} {1996}{\natexlab{b}})}\BibitemShut
  {NoStop}%
\bibitem [{\citenamefont {Bohdanowicz}\ \emph {et~al.}(2019)\citenamefont
  {Bohdanowicz}, \citenamefont {Crosson}, \citenamefont {Nirkhe},\ and\
  \citenamefont {Yuen}}]{bohdanowicz2019good}%
  \BibitemOpen
  \bibfield  {author} {\bibinfo {author} {\bibfnamefont {T.~C.}\ \bibnamefont
  {Bohdanowicz}}, \bibinfo {author} {\bibfnamefont {E.}~\bibnamefont
  {Crosson}}, \bibinfo {author} {\bibfnamefont {C.}~\bibnamefont {Nirkhe}}, \
  and\ \bibinfo {author} {\bibfnamefont {H.}~\bibnamefont {Yuen}},\ }in\
  \href@noop {} {\emph {\bibinfo {booktitle} {Proceedings of the 51st Annual
  ACM SIGACT Symposium on Theory of Computing}}}\ (\bibinfo {year} {2019})\
  pp.\ \bibinfo {pages} {481--490}\BibitemShut {NoStop}%
\bibitem [{\citenamefont {Brown}\ and\ \citenamefont
  {Fawzi}(2013)}]{brown2013short}%
  \BibitemOpen
  \bibfield  {author} {\bibinfo {author} {\bibfnamefont {W.}~\bibnamefont
  {Brown}}\ and\ \bibinfo {author} {\bibfnamefont {O.}~\bibnamefont {Fawzi}},\
  }in\ \href@noop {} {\emph {\bibinfo {booktitle} {2013 IEEE International
  Symposium on Information Theory}}}\ (\bibinfo {organization} {IEEE},\
  \bibinfo {year} {2013})\ pp.\ \bibinfo {pages} {346--350}\BibitemShut
  {NoStop}%
\bibitem [{\citenamefont {Breuckmann}\ and\ \citenamefont
  {Terhal}(2014)}]{breuckmann2014space}%
  \BibitemOpen
  \bibfield  {author} {\bibinfo {author} {\bibfnamefont {N.~P.}\ \bibnamefont
  {Breuckmann}}\ and\ \bibinfo {author} {\bibfnamefont {B.~M.}\ \bibnamefont
  {Terhal}},\ }\href@noop {} {\bibfield  {journal} {\bibinfo  {journal}
  {Journal of Physics A: Mathematical and Theoretical}\ }\textbf {\bibinfo
  {volume} {47}},\ \bibinfo {pages} {195304} (\bibinfo {year}
  {2014})}\BibitemShut {NoStop}%
\bibitem [{\citenamefont {Movassagh}\ and\ \citenamefont
  {Ouyang}(2020)}]{movassagh2020constructing}%
  \BibitemOpen
  \bibfield  {author} {\bibinfo {author} {\bibfnamefont {R.}~\bibnamefont
  {Movassagh}}\ and\ \bibinfo {author} {\bibfnamefont {Y.}~\bibnamefont
  {Ouyang}},\ }\href@noop {} {\  (\bibinfo {year} {2020})},\ \Eprint
  {http://arxiv.org/abs/2012.01453} {arXiv:2012.01453 [quant-ph]} \BibitemShut
  {NoStop}%
\bibitem [{\citenamefont {Knill}(2005)}]{knill2005quantum}%
  \BibitemOpen
  \bibfield  {author} {\bibinfo {author} {\bibfnamefont {E.}~\bibnamefont
  {Knill}},\ }\href@noop {} {\bibfield  {journal} {\bibinfo  {journal}
  {Nature}\ }\textbf {\bibinfo {volume} {434}},\ \bibinfo {pages} {39}
  (\bibinfo {year} {2005})}\BibitemShut {NoStop}%
\bibitem [{\citenamefont {Fawzi}\ \emph
  {et~al.}(2018{\natexlab{b}})\citenamefont {Fawzi}, \citenamefont
  {Grospellier},\ and\ \citenamefont {Leverrier}}]{fawzi2018constant}%
  \BibitemOpen
  \bibfield  {author} {\bibinfo {author} {\bibfnamefont {O.}~\bibnamefont
  {Fawzi}}, \bibinfo {author} {\bibfnamefont {A.}~\bibnamefont {Grospellier}},
  \ and\ \bibinfo {author} {\bibfnamefont {A.}~\bibnamefont {Leverrier}},\ }in\
  \href@noop {} {\emph {\bibinfo {booktitle} {2018 IEEE 59th Annual Symposium
  on Foundations of Computer Science (FOCS)}}}\ (\bibinfo {organization}
  {IEEE},\ \bibinfo {year} {2018})\ pp.\ \bibinfo {pages}
  {743--754}\BibitemShut {NoStop}%
\bibitem [{\citenamefont {Leverrier}\ \emph
  {et~al.}(2015{\natexlab{b}})\citenamefont {Leverrier}, \citenamefont
  {Tillich},\ and\ \citenamefont {Z{\'e}mor}}]{leverrier2015quantum}%
  \BibitemOpen
  \bibfield  {author} {\bibinfo {author} {\bibfnamefont {A.}~\bibnamefont
  {Leverrier}}, \bibinfo {author} {\bibfnamefont {J.-P.}\ \bibnamefont
  {Tillich}}, \ and\ \bibinfo {author} {\bibfnamefont {G.}~\bibnamefont
  {Z{\'e}mor}},\ }in\ \href@noop {} {\emph {\bibinfo {booktitle} {2015 IEEE
  56th Annual Symposium on Foundations of Computer Science}}}\ (\bibinfo
  {organization} {IEEE},\ \bibinfo {year} {2015})\ pp.\ \bibinfo {pages}
  {810--824}\BibitemShut {NoStop}%
\bibitem [{\citenamefont {Campbell}\ \emph {et~al.}(2017)\citenamefont
  {Campbell}, \citenamefont {Terhal},\ and\ \citenamefont
  {Vuillot}}]{campbell2017roads}%
  \BibitemOpen
  \bibfield  {author} {\bibinfo {author} {\bibfnamefont {E.~T.}\ \bibnamefont
  {Campbell}}, \bibinfo {author} {\bibfnamefont {B.~M.}\ \bibnamefont
  {Terhal}}, \ and\ \bibinfo {author} {\bibfnamefont {C.}~\bibnamefont
  {Vuillot}},\ }\href@noop {} {\bibfield  {journal} {\bibinfo  {journal}
  {Nature}\ }\textbf {\bibinfo {volume} {549}},\ \bibinfo {pages} {172}
  (\bibinfo {year} {2017})}\BibitemShut {NoStop}%
\bibitem [{\citenamefont {Bravyi}\ and\ \citenamefont
  {K{\"o}nig}(2013)}]{bravyi2013classification}%
  \BibitemOpen
  \bibfield  {author} {\bibinfo {author} {\bibfnamefont {S.}~\bibnamefont
  {Bravyi}}\ and\ \bibinfo {author} {\bibfnamefont {R.}~\bibnamefont
  {K{\"o}nig}},\ }\href@noop {} {\bibfield  {journal} {\bibinfo  {journal}
  {Physical review letters}\ }\textbf {\bibinfo {volume} {110}},\ \bibinfo
  {pages} {170503} (\bibinfo {year} {2013})}\BibitemShut {NoStop}%
\bibitem [{\citenamefont {Pastawski}\ and\ \citenamefont
  {Yoshida}(2015)}]{pastawski2015fault}%
  \BibitemOpen
  \bibfield  {author} {\bibinfo {author} {\bibfnamefont {F.}~\bibnamefont
  {Pastawski}}\ and\ \bibinfo {author} {\bibfnamefont {B.}~\bibnamefont
  {Yoshida}},\ }\href@noop {} {\bibfield  {journal} {\bibinfo  {journal}
  {Physical Review A}\ }\textbf {\bibinfo {volume} {91}},\ \bibinfo {pages}
  {012305} (\bibinfo {year} {2015})}\BibitemShut {NoStop}%
\bibitem [{\citenamefont {Krishna}\ and\ \citenamefont
  {Poulin}(2019)}]{krishna2019fault}%
  \BibitemOpen
  \bibfield  {author} {\bibinfo {author} {\bibfnamefont {A.}~\bibnamefont
  {Krishna}}\ and\ \bibinfo {author} {\bibfnamefont {D.}~\bibnamefont
  {Poulin}},\ }\href@noop {} {\  (\bibinfo {year} {2019})},\ \Eprint
  {http://arxiv.org/abs/1909.07424} {arXiv:1909.07424 [quant-ph]} \BibitemShut
  {NoStop}%
\bibitem [{\citenamefont {Burton}\ and\ \citenamefont
  {Browne}(2020)}]{burton2020limitations}%
  \BibitemOpen
  \bibfield  {author} {\bibinfo {author} {\bibfnamefont {S.}~\bibnamefont
  {Burton}}\ and\ \bibinfo {author} {\bibfnamefont {D.}~\bibnamefont
  {Browne}},\ }\href@noop {} {\  (\bibinfo {year} {2020})},\ \Eprint
  {http://arxiv.org/abs/2012.05842} {arXiv:2012.05842 [quant-ph]} \BibitemShut
  {NoStop}%
\bibitem [{\citenamefont {Jochym-O'Connor}(2019)}]{jochym2019fault}%
  \BibitemOpen
  \bibfield  {author} {\bibinfo {author} {\bibfnamefont {T.}~\bibnamefont
  {Jochym-O'Connor}},\ }\href@noop {} {\bibfield  {journal} {\bibinfo
  {journal} {Quantum}\ }\textbf {\bibinfo {volume} {3}},\ \bibinfo {pages}
  {120} (\bibinfo {year} {2019})}\BibitemShut {NoStop}%
\bibitem [{\citenamefont {Iyer}\ and\ \citenamefont
  {Poulin}(2015)}]{iyer2015hardness}%
  \BibitemOpen
  \bibfield  {author} {\bibinfo {author} {\bibfnamefont {P.}~\bibnamefont
  {Iyer}}\ and\ \bibinfo {author} {\bibfnamefont {D.}~\bibnamefont {Poulin}},\
  }\href@noop {} {\bibfield  {journal} {\bibinfo  {journal} {IEEE Transactions
  on Information Theory}\ }\textbf {\bibinfo {volume} {61}},\ \bibinfo {pages}
  {5209} (\bibinfo {year} {2015})}\BibitemShut {NoStop}%
\bibitem [{\citenamefont {Das}\ \emph {et~al.}(2020)\citenamefont {Das},
  \citenamefont {Pattison}, \citenamefont {Manne}, \citenamefont {Carmean},
  \citenamefont {Svore}, \citenamefont {Qureshi},\ and\ \citenamefont
  {Delfosse}}]{das2020scalable}%
  \BibitemOpen
  \bibfield  {author} {\bibinfo {author} {\bibfnamefont {P.}~\bibnamefont
  {Das}}, \bibinfo {author} {\bibfnamefont {C.~A.}\ \bibnamefont {Pattison}},
  \bibinfo {author} {\bibfnamefont {S.}~\bibnamefont {Manne}}, \bibinfo
  {author} {\bibfnamefont {D.}~\bibnamefont {Carmean}}, \bibinfo {author}
  {\bibfnamefont {K.}~\bibnamefont {Svore}}, \bibinfo {author} {\bibfnamefont
  {M.}~\bibnamefont {Qureshi}}, \ and\ \bibinfo {author} {\bibfnamefont
  {N.}~\bibnamefont {Delfosse}},\ }\href@noop {} {\  (\bibinfo {year}
  {2020})},\ \Eprint {http://arxiv.org/abs/2001.06598} {arXiv:2001.06598
  [quant-ph]} \BibitemShut {NoStop}%
\bibitem [{\citenamefont {Leifer}\ and\ \citenamefont
  {Poulin}(2008)}]{LEIFER20081899}%
  \BibitemOpen
  \bibfield  {author} {\bibinfo {author} {\bibfnamefont {M.}~\bibnamefont
  {Leifer}}\ and\ \bibinfo {author} {\bibfnamefont {D.}~\bibnamefont
  {Poulin}},\ }\href {\doibase https://doi.org/10.1016/j.aop.2007.10.001}
  {\bibfield  {journal} {\bibinfo  {journal} {Annals of Physics}\ }\textbf
  {\bibinfo {volume} {323}},\ \bibinfo {pages} {1899} (\bibinfo {year}
  {2008})}\BibitemShut {NoStop}%
\bibitem [{\citenamefont {Poulin}\ and\ \citenamefont
  {Bilgin}(2008)}]{poulin2008belief}%
  \BibitemOpen
  \bibfield  {author} {\bibinfo {author} {\bibfnamefont {D.}~\bibnamefont
  {Poulin}}\ and\ \bibinfo {author} {\bibfnamefont {E.}~\bibnamefont
  {Bilgin}},\ }\href {\doibase 10.1103/PhysRevA.77.052318} {\bibfield
  {journal} {\bibinfo  {journal} {Phys. Rev. A}\ }\textbf {\bibinfo {volume}
  {77}},\ \bibinfo {pages} {052318} (\bibinfo {year} {2008})}\BibitemShut
  {NoStop}%
\bibitem [{\citenamefont {Poulin}\ and\ \citenamefont
  {Chung}(2008)}]{poulin2008iterative}%
  \BibitemOpen
  \bibfield  {author} {\bibinfo {author} {\bibfnamefont {D.}~\bibnamefont
  {Poulin}}\ and\ \bibinfo {author} {\bibfnamefont {Y.}~\bibnamefont {Chung}},\
  }\href {\doibase 10.26421/QIC8.10} {\bibfield  {journal} {\bibinfo  {journal}
  {Quantum Information and Computation}\ ,\ \bibinfo {pages} {0987}} (\bibinfo
  {year} {2008})}\BibitemShut {NoStop}%
\bibitem [{\citenamefont {{Wang}}\ \emph {et~al.}(2012)\citenamefont {{Wang}},
  \citenamefont {{Sanders}}, \citenamefont {{Bai}},\ and\ \citenamefont
  {{Wang}}}]{6145510}%
  \BibitemOpen
  \bibfield  {author} {\bibinfo {author} {\bibfnamefont {Y.}~\bibnamefont
  {{Wang}}}, \bibinfo {author} {\bibfnamefont {B.~C.}\ \bibnamefont
  {{Sanders}}}, \bibinfo {author} {\bibfnamefont {B.}~\bibnamefont {{Bai}}}, \
  and\ \bibinfo {author} {\bibfnamefont {X.}~\bibnamefont {{Wang}}},\ }\href
  {\doibase 10.1109/TIT.2011.2169534} {\bibfield  {journal} {\bibinfo
  {journal} {IEEE Transactions on Information Theory}\ }\textbf {\bibinfo
  {volume} {58}},\ \bibinfo {pages} {1231} (\bibinfo {year}
  {2012})}\BibitemShut {NoStop}%
\bibitem [{\citenamefont {Kuo}\ and\ \citenamefont
  {Lai}(2020)}]{kuo2020refined}%
  \BibitemOpen
  \bibfield  {author} {\bibinfo {author} {\bibfnamefont {K.-Y.}\ \bibnamefont
  {Kuo}}\ and\ \bibinfo {author} {\bibfnamefont {C.-Y.}\ \bibnamefont {Lai}},\
  }\href@noop {} {\bibfield  {journal} {\bibinfo  {journal} {IEEE Journal on
  Selected Areas in Information Theory}\ }\textbf {\bibinfo {volume} {1}},\
  \bibinfo {pages} {487} (\bibinfo {year} {2020})}\BibitemShut {NoStop}%
\bibitem [{\citenamefont {Raveendran}\ \emph {et~al.}(2019)\citenamefont
  {Raveendran}, \citenamefont {Bahrami},\ and\ \citenamefont
  {Vasic}}]{raveendran2019syndrome}%
  \BibitemOpen
  \bibfield  {author} {\bibinfo {author} {\bibfnamefont {N.}~\bibnamefont
  {Raveendran}}, \bibinfo {author} {\bibfnamefont {M.}~\bibnamefont {Bahrami}},
  \ and\ \bibinfo {author} {\bibfnamefont {B.}~\bibnamefont {Vasic}},\ }in\
  \href@noop {} {\emph {\bibinfo {booktitle} {ICC 2019-2019 IEEE International
  Conference on Communications (ICC)}}}\ (\bibinfo {organization} {IEEE},\
  \bibinfo {year} {2019})\ pp.\ \bibinfo {pages} {1--6}\BibitemShut {NoStop}%
\bibitem [{\citenamefont {Duclos-Cianci}\ and\ \citenamefont
  {Poulin}(2010)}]{duclos2010fast}%
  \BibitemOpen
  \bibfield  {author} {\bibinfo {author} {\bibfnamefont {G.}~\bibnamefont
  {Duclos-Cianci}}\ and\ \bibinfo {author} {\bibfnamefont {D.}~\bibnamefont
  {Poulin}},\ }\href {\doibase 10.1103/PhysRevLett.104.050504} {\bibfield
  {journal} {\bibinfo  {journal} {Phys. Rev. Lett.}\ }\textbf {\bibinfo
  {volume} {104}},\ \bibinfo {pages} {050504} (\bibinfo {year}
  {2010})}\BibitemShut {NoStop}%
\bibitem [{\citenamefont {Grospellier}\ and\ \citenamefont
  {Krishna}(2018)}]{grospellier2018numerical}%
  \BibitemOpen
  \bibfield  {author} {\bibinfo {author} {\bibfnamefont {A.}~\bibnamefont
  {Grospellier}}\ and\ \bibinfo {author} {\bibfnamefont {A.}~\bibnamefont
  {Krishna}},\ }\href@noop {} {\  (\bibinfo {year} {2018})},\ \Eprint
  {http://arxiv.org/abs/1810.03681} {arXiv:1810.03681 [quant-ph]} \BibitemShut
  {NoStop}%
\bibitem [{\citenamefont {Grospellier}\ \emph {et~al.}(2020)\citenamefont
  {Grospellier}, \citenamefont {Grou{\`e}s}, \citenamefont {Krishna},\ and\
  \citenamefont {Leverrier}}]{grospellier2020combining}%
  \BibitemOpen
  \bibfield  {author} {\bibinfo {author} {\bibfnamefont {A.}~\bibnamefont
  {Grospellier}}, \bibinfo {author} {\bibfnamefont {L.}~\bibnamefont
  {Grou{\`e}s}}, \bibinfo {author} {\bibfnamefont {A.}~\bibnamefont {Krishna}},
  \ and\ \bibinfo {author} {\bibfnamefont {A.}~\bibnamefont {Leverrier}},\
  }\href@noop {} {\  (\bibinfo {year} {2020})},\ \Eprint
  {http://arxiv.org/abs/2004.11199} {arXiv:2004.11199 [quant-ph]} \BibitemShut
  {NoStop}%
\bibitem [{\citenamefont {Roffe}\ \emph {et~al.}(2020)\citenamefont {Roffe},
  \citenamefont {White}, \citenamefont {Burton},\ and\ \citenamefont
  {Campbell}}]{roffe2020decoding}%
  \BibitemOpen
  \bibfield  {author} {\bibinfo {author} {\bibfnamefont {J.}~\bibnamefont
  {Roffe}}, \bibinfo {author} {\bibfnamefont {D.~R.}\ \bibnamefont {White}},
  \bibinfo {author} {\bibfnamefont {S.}~\bibnamefont {Burton}}, \ and\ \bibinfo
  {author} {\bibfnamefont {E.~T.}\ \bibnamefont {Campbell}},\ }\href@noop {} {\
   (\bibinfo {year} {2020})},\ \Eprint {http://arxiv.org/abs/2005.07016}
  {arXiv:2005.07016 [quant-ph]} \BibitemShut {NoStop}%
\bibitem [{\citenamefont {{Liang}}\ \emph {et~al.}(2011)\citenamefont
  {{Liang}}, \citenamefont {{Cheng}}, \citenamefont {{Lai}}, \citenamefont
  {{Chen}},\ and\ \citenamefont {{Chen}}}]{liang2011hardware}%
  \BibitemOpen
  \bibfield  {author} {\bibinfo {author} {\bibfnamefont {C.}~\bibnamefont
  {{Liang}}}, \bibinfo {author} {\bibfnamefont {C.}~\bibnamefont {{Cheng}}},
  \bibinfo {author} {\bibfnamefont {Y.}~\bibnamefont {{Lai}}}, \bibinfo
  {author} {\bibfnamefont {L.}~\bibnamefont {{Chen}}}, \ and\ \bibinfo {author}
  {\bibfnamefont {H.~H.}\ \bibnamefont {{Chen}}},\ }\href {\doibase
  10.1109/TCSVT.2011.2125570} {\bibfield  {journal} {\bibinfo  {journal} {IEEE
  Transactions on Circuits and Systems for Video Technology}\ }\textbf
  {\bibinfo {volume} {21}},\ \bibinfo {pages} {525} (\bibinfo {year}
  {2011})}\BibitemShut {NoStop}%
\bibitem [{\citenamefont {Chen}\ and\ \citenamefont
  {Fossorier}(2002)}]{chen2002near}%
  \BibitemOpen
  \bibfield  {author} {\bibinfo {author} {\bibfnamefont {J.}~\bibnamefont
  {Chen}}\ and\ \bibinfo {author} {\bibfnamefont {M.~P.}\ \bibnamefont
  {Fossorier}},\ }\href@noop {} {\bibfield  {journal} {\bibinfo  {journal}
  {IEEE Transactions on communications}\ }\textbf {\bibinfo {volume} {50}},\
  \bibinfo {pages} {406} (\bibinfo {year} {2002})}\BibitemShut {NoStop}%
\bibitem [{\citenamefont {Delfosse}\ \emph {et~al.}(2021)\citenamefont
  {Delfosse}, \citenamefont {Londe},\ and\ \citenamefont
  {Beverland}}]{delfosse2021toward}%
  \BibitemOpen
  \bibfield  {author} {\bibinfo {author} {\bibfnamefont {N.}~\bibnamefont
  {Delfosse}}, \bibinfo {author} {\bibfnamefont {V.}~\bibnamefont {Londe}}, \
  and\ \bibinfo {author} {\bibfnamefont {M.}~\bibnamefont {Beverland}},\
  }\href@noop {} {\  (\bibinfo {year} {2021})},\ \Eprint
  {http://arxiv.org/abs/2103.08049} {arXiv:2103.08049 [quant-ph]} \BibitemShut
  {NoStop}%
\bibitem [{\citenamefont {Delfosse}\ and\ \citenamefont
  {Hastings}(2020)}]{delfosse2020union}%
  \BibitemOpen
  \bibfield  {author} {\bibinfo {author} {\bibfnamefont {N.}~\bibnamefont
  {Delfosse}}\ and\ \bibinfo {author} {\bibfnamefont {M.~B.}\ \bibnamefont
  {Hastings}},\ }\href@noop {} {\  (\bibinfo {year} {2020})},\ \Eprint
  {http://arxiv.org/abs/2009.14226} {arXiv:2009.14226 [quant-ph]} \BibitemShut
  {NoStop}%
\bibitem [{\citenamefont {Kovalev}\ \emph {et~al.}(2018)\citenamefont
  {Kovalev}, \citenamefont {Prabhakar}, \citenamefont {Dumer},\ and\
  \citenamefont {Pryadko}}]{kovalev2018numerical}%
  \BibitemOpen
  \bibfield  {author} {\bibinfo {author} {\bibfnamefont {A.~A.}\ \bibnamefont
  {Kovalev}}, \bibinfo {author} {\bibfnamefont {S.}~\bibnamefont {Prabhakar}},
  \bibinfo {author} {\bibfnamefont {I.}~\bibnamefont {Dumer}}, \ and\ \bibinfo
  {author} {\bibfnamefont {L.~P.}\ \bibnamefont {Pryadko}},\ }\href {\doibase
  10.1103/PhysRevA.97.062320} {\bibfield  {journal} {\bibinfo  {journal} {Phys.
  Rev. A}\ }\textbf {\bibinfo {volume} {97}},\ \bibinfo {pages} {062320}
  (\bibinfo {year} {2018})}\BibitemShut {NoStop}%
\bibitem [{\citenamefont {Bomb{\'\i}n}(2015)}]{bombin2015single}%
  \BibitemOpen
  \bibfield  {author} {\bibinfo {author} {\bibfnamefont {H.}~\bibnamefont
  {Bomb{\'\i}n}},\ }\href@noop {} {\bibfield  {journal} {\bibinfo  {journal}
  {Physical Review X}\ }\textbf {\bibinfo {volume} {5}},\ \bibinfo {pages}
  {031043} (\bibinfo {year} {2015})}\BibitemShut {NoStop}%
\bibitem [{\citenamefont {Grospellier}\ \emph {et~al.}(2021)\citenamefont
  {Grospellier}, \citenamefont {Grou{\`e}s}, \citenamefont {Krishna},\ and\
  \citenamefont {Leverrier}}]{grospellier2021combining}%
  \BibitemOpen
  \bibfield  {author} {\bibinfo {author} {\bibfnamefont {A.}~\bibnamefont
  {Grospellier}}, \bibinfo {author} {\bibfnamefont {L.}~\bibnamefont
  {Grou{\`e}s}}, \bibinfo {author} {\bibfnamefont {A.}~\bibnamefont {Krishna}},
  \ and\ \bibinfo {author} {\bibfnamefont {A.}~\bibnamefont {Leverrier}},\
  }\href@noop {} {\bibfield  {journal} {\bibinfo  {journal} {Quantum}\ }\textbf
  {\bibinfo {volume} {5}},\ \bibinfo {pages} {432} (\bibinfo {year}
  {2021})}\BibitemShut {NoStop}%
\bibitem [{\citenamefont {Quintavalle}\ \emph {et~al.}(2020)\citenamefont
  {Quintavalle}, \citenamefont {Vasmer}, \citenamefont {Roffe},\ and\
  \citenamefont {Campbell}}]{quintavalle2020single}%
  \BibitemOpen
  \bibfield  {author} {\bibinfo {author} {\bibfnamefont {A.~O.}\ \bibnamefont
  {Quintavalle}}, \bibinfo {author} {\bibfnamefont {M.}~\bibnamefont {Vasmer}},
  \bibinfo {author} {\bibfnamefont {J.}~\bibnamefont {Roffe}}, \ and\ \bibinfo
  {author} {\bibfnamefont {E.~T.}\ \bibnamefont {Campbell}},\ }\href@noop {} {\
   (\bibinfo {year} {2020})},\ \Eprint {http://arxiv.org/abs/2009.11790}
  {arXiv:2009.11790 [quant-ph]} \BibitemShut {NoStop}%
\bibitem [{\citenamefont {Hastings}(2021)}]{hastingsreduction}%
  \BibitemOpen
  \bibfield  {author} {\bibinfo {author} {\bibfnamefont {M.~B.}\ \bibnamefont
  {Hastings}},\ }\href@noop {} {\  (\bibinfo {year} {2021})},\ \Eprint
  {http://arxiv.org/abs/2102.10030} {arXiv:2102.10030 [quant-ph]} \BibitemShut
  {NoStop}%
\bibitem [{\citenamefont {Koll{\'a}r}\ \emph {et~al.}(2019)\citenamefont
  {Koll{\'a}r}, \citenamefont {Fitzpatrick},\ and\ \citenamefont
  {Houck}}]{kollar2019hyperbolic}%
  \BibitemOpen
  \bibfield  {author} {\bibinfo {author} {\bibfnamefont {A.~J.}\ \bibnamefont
  {Koll{\'a}r}}, \bibinfo {author} {\bibfnamefont {M.}~\bibnamefont
  {Fitzpatrick}}, \ and\ \bibinfo {author} {\bibfnamefont {A.~A.}\ \bibnamefont
  {Houck}},\ }\href@noop {} {\bibfield  {journal} {\bibinfo  {journal}
  {Nature}\ }\textbf {\bibinfo {volume} {571}},\ \bibinfo {pages} {45}
  (\bibinfo {year} {2019})}\BibitemShut {NoStop}%
\bibitem [{\citenamefont {Bilski}(1992)}]{bilski1992embedding}%
  \BibitemOpen
  \bibfield  {author} {\bibinfo {author} {\bibfnamefont {T.}~\bibnamefont
  {Bilski}},\ }\href@noop {} {\bibfield  {journal} {\bibinfo  {journal} {IEE
  Proceedings E (Computers and Digital Techniques)}\ }\textbf {\bibinfo
  {volume} {139}},\ \bibinfo {pages} {134} (\bibinfo {year}
  {1992})}\BibitemShut {NoStop}%
\bibitem [{\citenamefont {Dujmovi{\'c}}\ \emph {et~al.}(2016)\citenamefont
  {Dujmovi{\'c}}, \citenamefont {Sidiropoulos},\ and\ \citenamefont
  {Wood}}]{dujmovic2015layouts}%
  \BibitemOpen
  \bibfield  {author} {\bibinfo {author} {\bibfnamefont {V.}~\bibnamefont
  {Dujmovi{\'c}}}, \bibinfo {author} {\bibfnamefont {A.}~\bibnamefont
  {Sidiropoulos}}, \ and\ \bibinfo {author} {\bibfnamefont {D.~R.}\
  \bibnamefont {Wood}},\ }\href@noop {} {\bibfield  {journal} {\bibinfo
  {journal} {Chicago J. Theoretical Computer Science}\ }\textbf {\bibinfo
  {volume} {16}} (\bibinfo {year} {2016})}\BibitemShut {NoStop}%
\bibitem [{qub()}]{qubitzoo}%
  \BibitemOpen
  \href@noop {} {\enquote {\bibinfo {title} {{Qubit Zoo}},}\ }\bibinfo
  {howpublished} {\url{https://www.qubitzoo.com/}},\ \bibinfo {note} {accessed:
  2021-02-02}\BibitemShut {NoStop}%
\bibitem [{\citenamefont {Fowler}\ \emph {et~al.}(2012)\citenamefont {Fowler},
  \citenamefont {Mariantoni}, \citenamefont {Martinis},\ and\ \citenamefont
  {Cleland}}]{fowler2012surface}%
  \BibitemOpen
  \bibfield  {author} {\bibinfo {author} {\bibfnamefont {A.~G.}\ \bibnamefont
  {Fowler}}, \bibinfo {author} {\bibfnamefont {M.}~\bibnamefont {Mariantoni}},
  \bibinfo {author} {\bibfnamefont {J.~M.}\ \bibnamefont {Martinis}}, \ and\
  \bibinfo {author} {\bibfnamefont {A.~N.}\ \bibnamefont {Cleland}},\
  }\href@noop {} {\bibfield  {journal} {\bibinfo  {journal} {Physical Review
  A}\ }\textbf {\bibinfo {volume} {86}},\ \bibinfo {pages} {032324} (\bibinfo
  {year} {2012})}\BibitemShut {NoStop}%
\bibitem [{\citenamefont {Monroe}\ and\ \citenamefont
  {Kim}(2013)}]{monroe2013scaling}%
  \BibitemOpen
  \bibfield  {author} {\bibinfo {author} {\bibfnamefont {C.}~\bibnamefont
  {Monroe}}\ and\ \bibinfo {author} {\bibfnamefont {J.}~\bibnamefont {Kim}},\
  }\href@noop {} {\bibfield  {journal} {\bibinfo  {journal} {Science}\ }\textbf
  {\bibinfo {volume} {339}},\ \bibinfo {pages} {1164} (\bibinfo {year}
  {2013})}\BibitemShut {NoStop}%
\bibitem [{\citenamefont {Nigmatullin}\ \emph {et~al.}(2016)\citenamefont
  {Nigmatullin}, \citenamefont {Ballance}, \citenamefont {De~Beaudrap},\ and\
  \citenamefont {Benjamin}}]{nigmatullin2016minimally}%
  \BibitemOpen
  \bibfield  {author} {\bibinfo {author} {\bibfnamefont {R.}~\bibnamefont
  {Nigmatullin}}, \bibinfo {author} {\bibfnamefont {C.~J.}\ \bibnamefont
  {Ballance}}, \bibinfo {author} {\bibfnamefont {N.}~\bibnamefont
  {De~Beaudrap}}, \ and\ \bibinfo {author} {\bibfnamefont {S.~C.}\ \bibnamefont
  {Benjamin}},\ }\href@noop {} {\bibfield  {journal} {\bibinfo  {journal} {New
  Journal of Physics}\ }\textbf {\bibinfo {volume} {18}},\ \bibinfo {pages}
  {103028} (\bibinfo {year} {2016})}\BibitemShut {NoStop}%
\bibitem [{\citenamefont {Nickerson}\ \emph {et~al.}(2014)\citenamefont
  {Nickerson}, \citenamefont {Fitzsimons},\ and\ \citenamefont
  {Benjamin}}]{nickerson2014freely}%
  \BibitemOpen
  \bibfield  {author} {\bibinfo {author} {\bibfnamefont {N.~H.}\ \bibnamefont
  {Nickerson}}, \bibinfo {author} {\bibfnamefont {J.~F.}\ \bibnamefont
  {Fitzsimons}}, \ and\ \bibinfo {author} {\bibfnamefont {S.~C.}\ \bibnamefont
  {Benjamin}},\ }\href {\doibase 10.1103/PhysRevX.4.041041} {\bibfield
  {journal} {\bibinfo  {journal} {Phys. Rev. X}\ }\textbf {\bibinfo {volume}
  {4}},\ \bibinfo {pages} {041041} (\bibinfo {year} {2014})}\BibitemShut
  {NoStop}%
\bibitem [{\citenamefont {Rudolph}(2017)}]{rudolph2017optimistic}%
  \BibitemOpen
  \bibfield  {author} {\bibinfo {author} {\bibfnamefont {T.}~\bibnamefont
  {Rudolph}},\ }\href@noop {} {\bibfield  {journal} {\bibinfo  {journal} {APL
  Photonics}\ }\textbf {\bibinfo {volume} {2}},\ \bibinfo {pages} {030901}
  (\bibinfo {year} {2017})}\BibitemShut {NoStop}%
\bibitem [{\citenamefont {Pellizzari}\ \emph {et~al.}(1995)\citenamefont
  {Pellizzari}, \citenamefont {Gardiner}, \citenamefont {Cirac},\ and\
  \citenamefont {Zoller}}]{PhysRevLett.75.3788}%
  \BibitemOpen
  \bibfield  {author} {\bibinfo {author} {\bibfnamefont {T.}~\bibnamefont
  {Pellizzari}}, \bibinfo {author} {\bibfnamefont {S.~A.}\ \bibnamefont
  {Gardiner}}, \bibinfo {author} {\bibfnamefont {J.~I.}\ \bibnamefont {Cirac}},
  \ and\ \bibinfo {author} {\bibfnamefont {P.}~\bibnamefont {Zoller}},\ }\href
  {\doibase 10.1103/PhysRevLett.75.3788} {\bibfield  {journal} {\bibinfo
  {journal} {Phys. Rev. Lett.}\ }\textbf {\bibinfo {volume} {75}},\ \bibinfo
  {pages} {3788} (\bibinfo {year} {1995})}\BibitemShut {NoStop}%
\bibitem [{\citenamefont {Wade}\ \emph {et~al.}(2016)\citenamefont {Wade},
  \citenamefont {Mattioli},\ and\ \citenamefont
  {M\o{}lmer}}]{PhysRevA.94.053830}%
  \BibitemOpen
  \bibfield  {author} {\bibinfo {author} {\bibfnamefont {A.~C.~J.}\
  \bibnamefont {Wade}}, \bibinfo {author} {\bibfnamefont {M.}~\bibnamefont
  {Mattioli}}, \ and\ \bibinfo {author} {\bibfnamefont {K.}~\bibnamefont
  {M\o{}lmer}},\ }\href {\doibase 10.1103/PhysRevA.94.053830} {\bibfield
  {journal} {\bibinfo  {journal} {Phys. Rev. A}\ }\textbf {\bibinfo {volume}
  {94}},\ \bibinfo {pages} {053830} (\bibinfo {year} {2016})}\BibitemShut
  {NoStop}%
\bibitem [{\citenamefont {Conrad}\ \emph {et~al.}(2018)\citenamefont {Conrad},
  \citenamefont {Chamberland}, \citenamefont {Breuckmann},\ and\ \citenamefont
  {Terhal}}]{conrad2018small}%
  \BibitemOpen
  \bibfield  {author} {\bibinfo {author} {\bibfnamefont {J.}~\bibnamefont
  {Conrad}}, \bibinfo {author} {\bibfnamefont {C.}~\bibnamefont {Chamberland}},
  \bibinfo {author} {\bibfnamefont {N.~P.}\ \bibnamefont {Breuckmann}}, \ and\
  \bibinfo {author} {\bibfnamefont {B.~M.}\ \bibnamefont {Terhal}},\
  }\href@noop {} {\bibfield  {journal} {\bibinfo  {journal} {Philosophical
  Transactions of the Royal Society A: Mathematical, Physical and Engineering
  Sciences}\ }\textbf {\bibinfo {volume} {376}},\ \bibinfo {pages} {20170323}
  (\bibinfo {year} {2018})}\BibitemShut {NoStop}%
\bibitem [{\citenamefont {Kitaev}\ \emph {et~al.}(2002)\citenamefont {Kitaev},
  \citenamefont {Shen}, \citenamefont {Vyalyi},\ and\ \citenamefont
  {Vyalyi}}]{kitaev2002classical}%
  \BibitemOpen
  \bibfield  {author} {\bibinfo {author} {\bibfnamefont {A.~Y.}\ \bibnamefont
  {Kitaev}}, \bibinfo {author} {\bibfnamefont {A.}~\bibnamefont {Shen}},
  \bibinfo {author} {\bibfnamefont {M.~N.}\ \bibnamefont {Vyalyi}}, \ and\
  \bibinfo {author} {\bibfnamefont {M.~N.}\ \bibnamefont {Vyalyi}},\
  }\href@noop {} {\emph {\bibinfo {title} {Classical and quantum
  computation}}},\ \bibinfo {number} {47}\ (\bibinfo  {publisher} {American
  Mathematical Soc.},\ \bibinfo {year} {2002})\BibitemShut {NoStop}%
\bibitem [{\citenamefont {Aharonov}\ and\ \citenamefont
  {Naveh}(2002)}]{aharonov2002quantum}%
  \BibitemOpen
  \bibfield  {author} {\bibinfo {author} {\bibfnamefont {D.}~\bibnamefont
  {Aharonov}}\ and\ \bibinfo {author} {\bibfnamefont {T.}~\bibnamefont
  {Naveh}},\ }\href@noop {} {\  (\bibinfo {year} {2002})},\ \Eprint
  {http://arxiv.org/abs/quant-ph/0210077} {arXiv:quant-ph/0210077 [quant-ph]}
  \BibitemShut {NoStop}%
\bibitem [{\citenamefont {Arora}\ and\ \citenamefont
  {Safra}(1998)}]{arora1998probabilistic}%
  \BibitemOpen
  \bibfield  {author} {\bibinfo {author} {\bibfnamefont {S.}~\bibnamefont
  {Arora}}\ and\ \bibinfo {author} {\bibfnamefont {S.}~\bibnamefont {Safra}},\
  }\href@noop {} {\bibfield  {journal} {\bibinfo  {journal} {Journal of the ACM
  (JACM)}\ }\textbf {\bibinfo {volume} {45}},\ \bibinfo {pages} {70} (\bibinfo
  {year} {1998})}\BibitemShut {NoStop}%
\bibitem [{\citenamefont {Dinur}(2007)}]{dinur2007pcp}%
  \BibitemOpen
  \bibfield  {author} {\bibinfo {author} {\bibfnamefont {I.}~\bibnamefont
  {Dinur}},\ }\href@noop {} {\bibfield  {journal} {\bibinfo  {journal} {Journal
  of the ACM (JACM)}\ }\textbf {\bibinfo {volume} {54}},\ \bibinfo {pages} {12}
  (\bibinfo {year} {2007})}\BibitemShut {NoStop}%
\bibitem [{\citenamefont {Hastings}(2013)}]{hastings2012trivial}%
  \BibitemOpen
  \bibfield  {author} {\bibinfo {author} {\bibfnamefont {M.~B.}\ \bibnamefont
  {Hastings}},\ }\href {\doibase 10.26421/QIC13.5-6} {\bibfield  {journal}
  {\bibinfo  {journal} {Quantum Information and Computation}\ } (\bibinfo
  {year} {2013}),\ 10.26421/QIC13.5-6}\BibitemShut {NoStop}%
\bibitem [{\citenamefont {Freedman}\ and\ \citenamefont
  {Hastings}(2013)}]{freedman2013quantum}%
  \BibitemOpen
  \bibfield  {author} {\bibinfo {author} {\bibfnamefont {M.~H.}\ \bibnamefont
  {Freedman}}\ and\ \bibinfo {author} {\bibfnamefont {M.~B.}\ \bibnamefont
  {Hastings}},\ }\href@noop {} {\  (\bibinfo {year} {2013})},\ \Eprint
  {http://arxiv.org/abs/1301.1363} {arXiv:1301.1363 [quant-ph]} \BibitemShut
  {NoStop}%
\bibitem [{\citenamefont {Anshu}\ and\ \citenamefont
  {Nirkhe}(2020)}]{anshu2020circuit}%
  \BibitemOpen
  \bibfield  {author} {\bibinfo {author} {\bibfnamefont {A.}~\bibnamefont
  {Anshu}}\ and\ \bibinfo {author} {\bibfnamefont {C.}~\bibnamefont {Nirkhe}},\
  }\href@noop {} {\  (\bibinfo {year} {2020})},\ \Eprint
  {http://arxiv.org/abs/2011.02044} {arXiv:2011.02044 [quant-ph]} \BibitemShut
  {NoStop}%
\bibitem [{\citenamefont {Eldar}\ and\ \citenamefont
  {Harrow}(2017)}]{eldar_local_2017}%
  \BibitemOpen
  \bibfield  {author} {\bibinfo {author} {\bibfnamefont {L.}~\bibnamefont
  {Eldar}}\ and\ \bibinfo {author} {\bibfnamefont {A.~W.}\ \bibnamefont
  {Harrow}},\ }in\ \href {\doibase 10.1109/FOCS.2017.46} {\emph {\bibinfo
  {booktitle} {2017 {IEEE} 58th {Annual} {Symposium} on {Foundations} of
  {Computer} {Science} ({FOCS})}}}\ (\bibinfo {year} {2017})\ pp.\ \bibinfo
  {pages} {427--438},\ \bibinfo {note} {iSSN: 0272-5428}\BibitemShut {NoStop}%
\bibitem [{\citenamefont {Aharonov}\ \emph {et~al.}(2013)\citenamefont
  {Aharonov}, \citenamefont {Arad},\ and\ \citenamefont
  {Vidick}}]{aharonov2013guest}%
  \BibitemOpen
  \bibfield  {author} {\bibinfo {author} {\bibfnamefont {D.}~\bibnamefont
  {Aharonov}}, \bibinfo {author} {\bibfnamefont {I.}~\bibnamefont {Arad}}, \
  and\ \bibinfo {author} {\bibfnamefont {T.}~\bibnamefont {Vidick}},\
  }\href@noop {} {\bibfield  {journal} {\bibinfo  {journal} {Acm sigact news}\
  }\textbf {\bibinfo {volume} {44}},\ \bibinfo {pages} {47} (\bibinfo {year}
  {2013})}\BibitemShut {NoStop}%
\bibitem [{\citenamefont {Freedman}\ and\ \citenamefont
  {Hastings}(2020)}]{freedman2020building}%
  \BibitemOpen
  \bibfield  {author} {\bibinfo {author} {\bibfnamefont {M.}~\bibnamefont
  {Freedman}}\ and\ \bibinfo {author} {\bibfnamefont {M.~B.}\ \bibnamefont
  {Hastings}},\ }\href@noop {} {\  (\bibinfo {year} {2020})},\ \Eprint
  {http://arxiv.org/abs/2012.02249} {arXiv:2012.02249 [quant-ph]} \BibitemShut
  {NoStop}%
\bibitem [{\citenamefont {Agarwal}\ \emph {et~al.}(2019)\citenamefont
  {Agarwal}, \citenamefont {Chandrasekaran}, \citenamefont {Kolla},\ and\
  \citenamefont {Madan}}]{agarwal_expansion_2019}%
  \BibitemOpen
  \bibfield  {author} {\bibinfo {author} {\bibfnamefont {N.}~\bibnamefont
  {Agarwal}}, \bibinfo {author} {\bibfnamefont {K.}~\bibnamefont
  {Chandrasekaran}}, \bibinfo {author} {\bibfnamefont {A.}~\bibnamefont
  {Kolla}}, \ and\ \bibinfo {author} {\bibfnamefont {V.}~\bibnamefont
  {Madan}},\ }\href {\doibase 10.1137/17M1141047} {\bibfield  {journal}
  {\bibinfo  {journal} {SIAM Journal on Discrete Mathematics}\ }\textbf
  {\bibinfo {volume} {33}},\ \bibinfo {pages} {1338} (\bibinfo {year}
  {2019})},\ \bibinfo {note} {publisher: Society for Industrial and Applied
  Mathematics}\BibitemShut {NoStop}%
\end{thebibliography}%


\end{document}